\documentclass[a4paper,11pt]{article}
\usepackage{jheppub}
\hypersetup{hypertexnames=false}
\usepackage[T1]{fontenc}
\usepackage[utf8]{inputenc}
\usepackage{multirow}
\usepackage{braket}
\usepackage{tabularx}
\usepackage{longtable}
\usepackage{graphicx}
\usepackage{relsize}
\usepackage{slashed}
\usepackage{booktabs}
\usepackage[normalem]{ulem}
\usepackage{rotating}
\usepackage{bigstrut}
\usepackage{makecell}
\usepackage[dvipsnames]{xcolor}
\usepackage{amsmath}
\usepackage{amssymb,bm}
\usepackage{amsthm}
\allowdisplaybreaks[4]
\usepackage{mathrsfs}
\usepackage{array}
\usepackage[all]{xy}
\usepackage{euscript}
\usepackage{enumerate}
\usepackage{enumitem}
\usepackage{slashed}
\usepackage{float}
\usepackage{mathtools}
\usepackage{soul}
\usepackage{orcidlink}
\usepackage{tikz}
\allowdisplaybreaks[4]
\newcommand{\D}[0]{\mathrm{d}}

\newcommand{\ee}[0]{e^+e^-}

\newcommand{\chIII}[0]{$D^0D^{*-}\pi^+$}

\newcommand{\beq}{\begin{equation}}
\newcommand{\eeq}{\end{equation}}
\renewcommand{\vec}{\bm}
\newcommand{\dash}{--}
\newcommand{\stot}{s_{\rm tot}}
\newcommand{\Pzero}{{\cal P}^{\gamma^*}_A}
\newcommand{\Palpha}{{\cal P}^{\gamma^*}_\alpha}

\definecolor{poscolor}{RGB} {252,188,190}
\definecolor{negcolor}{RGB} {168,168,234}
\usepackage{colortbl}

\usepackage{relsize}
\def\babar{\mbox{\slshape B\kern-0.1em{\smaller A}\kern-0.1em
 B\kern-0.1em{\smaller A\kern-0.2em R}}}

\usetikzlibrary{decorations.markings}
\usetikzlibrary{snakes}
\tikzset{
 photon/.style={decorate, decoration={snake}, draw=black},
 electron/.style={draw=black, postaction={decorate},
 decoration={markings,mark=at position .55 with {\arrow[draw=black]{>}}}},
 gluon/.style={decorate, draw=magenta,
 decoration={coil,amplitude=3pt, segment length=4pt}},
 scalar/.style={dashed,line width=.6pt, postaction={decorate}}
}

\newcommand{\Np}{N_\text{p}}
\newcommand{\Ne}{N_\text{e}}
\newcommand{\Nexp}{N_\text{exp}}

\graphicspath{{Figs.dir/}}

\title{Vector charmonium(-like) states in the energy range of 4.1\dash4.6~GeV}

\author[a,b]{Xiang-Kun Dong\orcidlink{0000-0001-6392-7143}}
\author[c]{Vadim Baru\orcidlink{0000-0001-6472-1008}}
\author[d]{Leon~von~Detten\orcidlink{0000-0001-7636-6840}}
\author[e,f,g]{Feng-Kun~Guo\orcidlink{0000-0002-2919-2064}}
\author[h]{Christoph~Hanhart\orcidlink{0000-0002-3509-2473}}
\author[a,b]{Teng Ji\orcidlink{0000-0003-0366-1042}}
\author[a,b,h,i]{Ulf-G. Mei{\ss}ner\orcidlink{0000-0003-1254-442X}}
\author[j]{Alexey Nefediev\orcidlink{0000-0002-9988-9430}}

\affiliation[a]{Helmholtz Institut f\"{u}r Strahlen- und Kernphysik and Cluster of Excellence  ``Color meets Flavor'', Universit\"{a}t Bonn,\\ D-53115 Bonn, Germany}
\affiliation[b]{Bethe Center for Theoretical Physics, Universit\"{a}t Bonn,\\ D-53115 Bonn, Germany}
\affiliation[c]{Institut f\"ur Theoretische Physik II, Ruhr-Universit\"at Bochum,\\ D-44780 Bochum, Germany}
\affiliation[d]{Institute for Advanced Simulation (IAS-4), Forschungszentrum J\"ulich,\\ D-52425 J\"ulich, Germany}
\affiliation[e]{Institute of Theoretical Physics, Chinese Academy of Sciences,\\ Beijing 100190, China}
\affiliation[f]{School of Physical Sciences, University of Chinese Academy of Sciences,\\ Beijing 100049, China}
\affiliation[g]{Southern Center for Nuclear-Science Theory (SCNT), Institute of Modern Physics, Chinese Academy of Sciences,\\ Huizhou 516000, China}
\affiliation[h]{Institute for Advanced Simulation (IAS-4) and Cluster of Excellence ``Color meets Flavor'', Forschungszentrum J\"ulich,\\ D-52425 J\"ulich, Germany}
\affiliation[i]{Peng Huanwu Collaborative Center for Research and Education, International Institute for Interdisciplinary and Frontiers, Beihang University,\\ Beijing 100191, China}
\affiliation[j]{Helmholtz Institut f\"{u}r Strahlen- und Kernphysik, Universit\"{a}t Bonn,\\ D-53115 Bonn, Germany}

\emailAdd{xiangkun@hiskp.uni-bonn.de}
\emailAdd{vadim.baru@tp2.rub.de}
\emailAdd{Leon.qcd.78@proton.me}
\emailAdd{fkguo@itp.ac.cn}
\emailAdd{c.hanhart@fz-juelich.de}
\emailAdd{teng@hiskp.uni-bonn.de}
\emailAdd{meissner@hiskp.uni-bonn.de}
\emailAdd{nefediev@hiskp.uni-bonn.de}

\abstract{
The spectrum of vector charmonium(-like) states in the 4.1\dash4.6~GeV energy region exhibits a long-standing tension between inclusive and exclusive measurements. While the inclusive $R$-value indicates the presence of only vector charmonia that are usually interpreted as conventional states such as $\psi(4160)$ and $\psi(4415)$, exclusive $\ee$ cross sections reveal additional structures whose parameters strongly depend on the observed final states when fitted with Breit--Wigner functions. This puzzling pattern suggests that coupled-channel and threshold effects play an essential role.
In this work, we develop a unified coupled-channel framework relevant for the $1^{--}$ resonances in the considered energy region. The framework incorporates the $S$-wave open-charm channels $D\bar{D}_1$, $D^*\bar{D}_1$, and $D^*\bar{D}_2^*$, constrained by heavy-quark spin symmetry, optional bare poles corresponding to quark-model states which may be associated with the $\psi(4160)$ and the $\psi(4415)$, and final-state interaction in the $Z_c$ channels relevant for the three-body final states.
We employ several benchmark models to perform simultaneous fits to the BESIII data for the cross sections of $\ee\to J/\psi\pi^+\pi^-$, $h_c\pi^+\pi^-$, $D\bar{D}^*\pi$, $D^*\bar{D}^*\pi$, $J/\psi\eta$, and $\chi_{c0}\omega$, together with the available invariant-mass distributions that exhibit the $Z_c(3900)$ and $Z_c(4020)$ structures. The models differ in the number of bare seed states
and the fitting strategy. We show that even the purely dynamical scheme without bare charmonia captures the gross features of all the analyzed distributions. We therefore conclude that the nontrivial behavior of the measured line shapes in the studied energy range can be understood in terms of strong coupled-channel effects with dynamically generated poles. The inclusion of bare compact states leads to a quantitative improvement of the fit quality but does not alter this conclusion. Finally we discuss what can be learned about possibly existing heavy-quark spin partners of the exotic $J^{PC}=1^{--}$ states based on our study.
}
\begin{document}

\maketitle

\section{Introduction}
\label{sec:intro}

Understanding the spectrum of charmonia above the open-charm threshold remains one of the central challenges in heavy-hadron spectroscopy. Since the discovery of the $X(3872)$~\cite{Belle:2003nnu}, a large number of charmonium-like states have been observed, whose properties cannot be naturally accommodated within the conventional quark-model classification~\cite{GellMann:1964nj,Zweig:1964jf,Godfrey:1985xj}. For general reviews of the experimental and theoretical status of the $XYZ$ states, see, for example, Refs.~\cite{Lebed:2016hpi,Esposito:2016noz,Hosaka:2016pey,Ali:2017jda,Guo:2017jvc,Olsen:2017bmm,Guo:2019twa,Brambilla:2019esw,Dong:2021juy,Chen:2022asf,Meng:2022ozq,Liu:2024uxn, Chen:2024eaq,Wang:2025sic,Wang:2025dur,Dai:2026fkg,Wang:2025clb,Bai:2026atm}.

The vector sector plays a distinguished role, since states with $J^{PC}=1^{--}$ can be produced directly in $\ee$ annihilations and therefore provide a particularly clean arena for studying line shapes, threshold effects, and pole structures.
The main difficulties in the theoretical description of the spectrum of vector charmonium are twofold: on the one hand, this family is very rich, and, on the other hand, the data that need to be analyzed cover a broad energy range of nearly 1~GeV above the open-charm threshold at around 3.7~GeV. This corresponds to typical momenta in the system of two charmed mesons of up to 1.0\dash1.5~GeV, which cannot be accommodated within any reliable hadronic effective field theory (EFT). Moreover, the large number of charmonium and charmonium-like states and hadronic thresholds coexisting in this broad energy range leads to a large number of parameters in any realistic model aimed at a simultaneous description of the entire set of experimental data. Such parameters cannot firmly be constrained by the existing data.
A possible way out is to employ model approaches, for example, based on specific $K$-matrix parameterizations \cite{Uglov:2016orr,Husken:2022yik,Husken:2024hmi}. However, in many realizations, these approaches violate analyticity. 

In this work, we investigate the possible interplay between states emerging from two-hadron scattering and additional pole terms. To keep the problem tractable, we restrict the analysis to the most relevant data sets and employ complex-valued fit parameters, at the cost of sacrificing exact unitarity to some extent, while carefully preserving the analytic structure of the amplitudes. Furthermore, we focus on the energy range from 4.1 to 4.6~GeV, where a substantial body of experimental data is available. In particular, we refrain from including in our combined fit the data on the exclusive $D_{(s)}^{(*)}\bar{D}_{(s)}^{(*)}$ channels \cite{Belle:2006hvs,BaBar:2006qlj,
CLEO:2008ojp,Dong:2017tpt,BaBar:2009elc,Belle:2010fwv,BaBar:2010plp,Belle:2017grj,BESIII:2021yvc, BESIII:2023wsc, BESIII:2024zdh, BESIII:2024ths}; see Refs.~\cite{Xue:2017xpu,Nakamura:2023obk} for works including those channels.
These channels do not exhibit clear structures associated, for example, with the $\psi(4230)$ and couple to the relevant $J^{PC}=1^{--}$ states only through odd partial waves. This suggests that the coupling of the $\psi(4230)$ and its possible spin partners to such two-body channels is suppressed. The data in these channels are mainly localized at lower energies, with only the tails of their line shapes extending into the energy range studied in this work. 
The tails are insufficient to constrain the multiple additional fitting parameters required to include these extra open-charm channels. 
Thus, since the analyzed set of experimental data is not complete, we are forced to treat some parameters of our fit as complex-valued quantities, whose imaginary parts effectively mimic the contributions from all neglected open channels.
In fact, it is basically impossible to include all channels that can sizably contribute to the decays of vector charmonium-like states. In particular, hadronic molecules formed by two charmed mesons with a relatively large binding energy would be able to decay into purely light flavor hadrons without the Okubo-Zweig-Iizuka suppression~\cite{BESIII:2023gqy}. Such channels cannot be included explicitly in a reliable manner within a coupled-channel framework.

\begin{figure}[t]
\centering
\includegraphics[width=\linewidth]{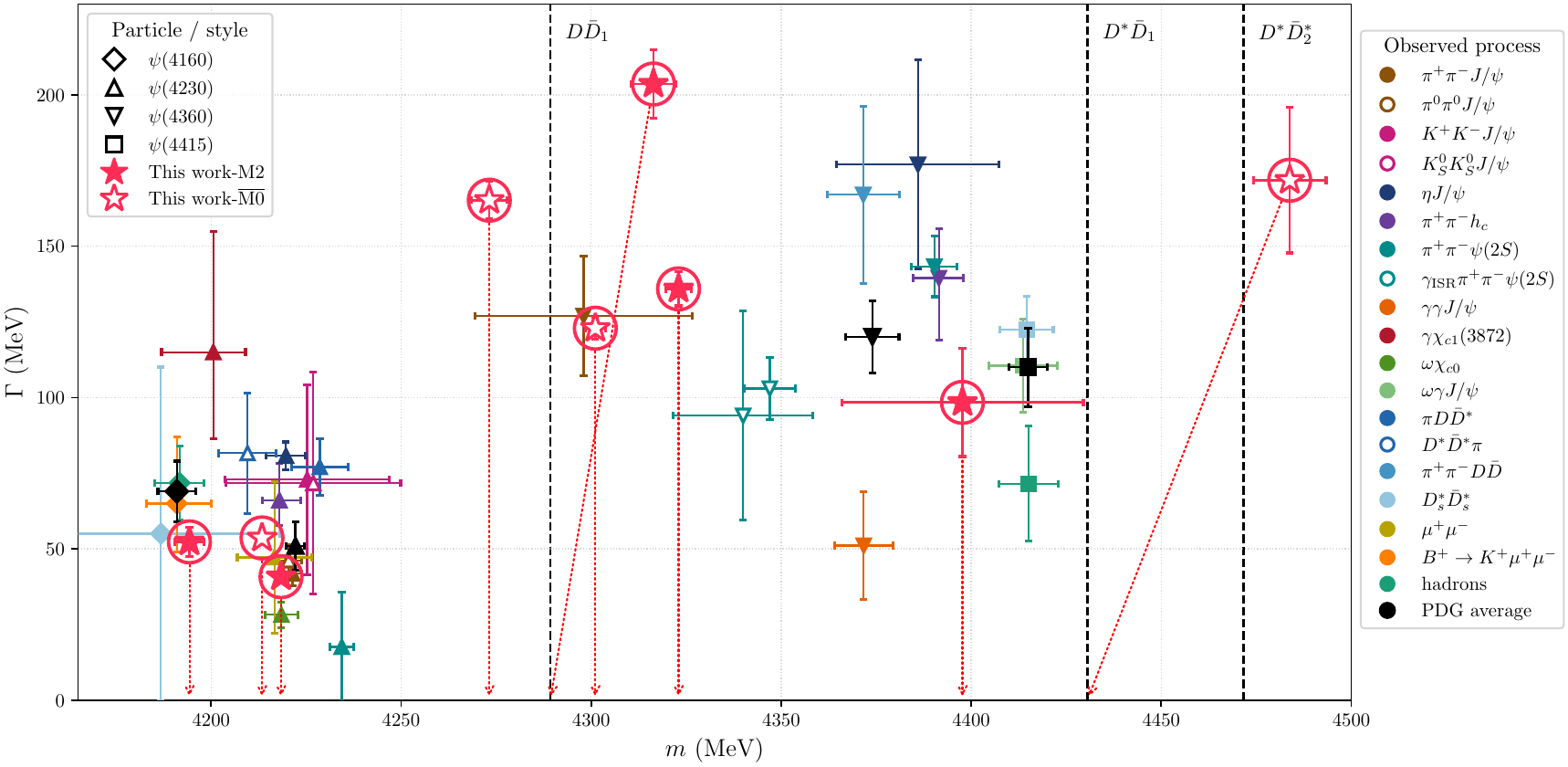}
\caption{Masses and widths of the vector $\psi$ states from experimental analyses of individual channels used in the Particle Data Group (PDG) average~\cite{ParticleDataGroup:2024cfk}. All observed processes are from $e^+e^-$ annihilations except for the $B^+\to K^+\mu^+\mu^-$. The parameters of $\psi(4160)$ (diamond), $\psi(4230)$ (upward triangle), $\psi(4360)$ (downward triangle), and $\psi(4415)$ (square) are taken from Refs.~\cite{BESIII:2023wsc,LHCb:2013ywr,BES:2007zwq}, \cite{BESIII:2023tll,BESIII:2022kcv,BESIII:2023cmv,BESIII:2022qal,BESIII:2022joj,BESIII:2021njb,Ablikim:2020jrn,BESIII:2020oph,BESIII:2019gjc,BESIII:2018iea,BESIII:2019qvy,BESIII:2016adj}, \cite{BESIII:2023tll,BESIII:2022quc,BESIII:2022qal,BESIII:2021njb,BESIII:2021yal,BESIII:2016adj,Belle:2014wyt,BaBar:2012hpr,BESIII:2026qhe}, and \cite{BESIII:2024jzg,BESIII:2023wsc,BES:2007zwq}, respectively. Red stars show the results of the combined fit obtained in this work in coupled-channel models M2 (solid) and $\overline{\rm M0}$ (open) as defined in Section~\ref{sec:models}. The dotted lines with arrows indicate the shortest paths from the poles to the physical axis. If these trajectories are not perpendicular to the physical axis, this indicates that the corresponding poles are not directly connected to it, but rather reside on a hidden Riemann sheet. The black dashed vertical lines indicate the thresholds of the two-body open-charm channels $D\bar{D}_1$, $D^*\bar{D}_1$, and $D^*\bar{D}_2^*$.}
 \label{fig:polecom}
\end{figure}

Now, after this general introduction and disclaimer, we come to a discussion of particular vector states in the spectrum of charmonium and their experimental status.
The state now known as $\psi(4230)$ was first reported by the \babar\ Collaboration as $Y(4260)$ in the initial state radiation (ISR) process $\ee\to\gamma_{\rm ISR} J/\psi\pi^+\pi^-$~\cite{BaBar:2005hhc} and later confirmed by the CLEO and Belle Collaborations~\cite{CLEO:2006tct,Belle:2007dxy}. With the high-statistics energy scan of BESIII, it became clear that the structure around 4.2\dash4.3 GeV departed clearly from a simple isolated Breit-Wigner resonance. Later on, multiple signals associated with $\psi(4230)$ were observed in a variety of hidden-charm and open-charm channels, including $J/\psi\pi^+\pi^-$, $h_c\pi^+\pi^-$, $D\bar{D}^*\pi$, $D^*\bar{D}^*\pi$, $J/\psi\eta$, $\chi_{c0}\omega$, and several others~\cite{BESIII:2016adj,BESIII:2018iea,BESIII:2019qvy,BESIII:2019gjc,Ablikim:2020jrn,BESIII:2020oph,BESIII:2021njb,BESIII:2022kcv,BESIII:2022qal,BESIII:2022joj,BESIII:2023tll,BESIII:2023cmv,BESIII:2025qkn}. However, the masses and widths extracted in different channels scatter significantly, and the asymmetric line shape in the $J/\psi\pi^+\pi^-$ channel is often parameterized experimentally by introducing an additional Breit-Wigner resonance near 4.32~GeV~\cite{BESIII:2016bnd,BESIII:2022qal}. At the same time, the higher-energy region contains the structures commonly labeled as $\psi(4320)$, $\psi(4360)$, and $\psi(4415)$, whose mutual relation is likewise highly channel-dependent~\cite{BESIII:2023tll,BESIII:2022quc,BESIII:2022qal,BESIII:2021njb,BESIII:2021yal,BESIII:2016adj,Belle:2014wyt,BaBar:2012hpr,BESIII:2026qhe, BESIII:2024jzg,BESIII:2023wsc,BES:2007zwq}, as shown in Figure~\ref{fig:polecom}. This situation suggests that the observed enhancements in the line shapes should not be interpreted, as na{\"i}vely done, as a set of independent states extracted using a sum of Breit-Wigner amplitudes.

Several qualitative features make the $\psi(4230)$ region particularly intriguing. In particular, the line shapes differ significantly from channel to channel, the hidden- and open-charm production rates exhibit striking contrasts, and the state (originally called $Y(4260)$) is closely connected to the emergence of the charged structures $Z_c(3900)$ and $Z_c(4020)$ in various channels~\cite{BESIII:2013ris,BESIII:2013ouc,BESIII:2015pqw,BESIII:2013mhi}. In addition, the inclusive $R$-value does not display a prominent conventional resonance peak at the position of the $\psi(4230)$~\cite{Osterheld:1986hw,BES:2001ckj,CLEO:2008ojp}.
Taken together, these features indicate that the experimentally observed line shapes encode nontrivial coupled-channel dynamics~\cite{Nakamura:2023obk,vonDetten:2024eie}, rather than providing a direct one-to-one correspondence with the underlying pole content.

The peculiar situation just discussed has motivated a wide range of theoretical interpretations. Soon after the discovery of the $Y(4260)$, this state was interpreted in terms of hybrid charmonium, baryonium, hadro-charmonium, compact multiquark, and hadronic-molecule scenarios, as well as interference effects~\cite{Close:2005iz,Kou:2005gt,Zhu:2005hp,Llanes-Estrada:2005qvr,Qiao:2005av,Dubynskiy:2008mq,Ding:2008gr,Kalashnikova:2008qr,Chen:2010nv}. Among these, the hadronic-molecule and coupled-channel interpretations have attracted particular attention since the $D\bar{D}_1$ threshold lies not far from 4.26~GeV, which is the mass value of the original determination, and naturally gives rise to strong $S$-wave threshold effects~\cite{Wang:2013cya,Cleven:2013mka,Wang:2013kra,Cleven:2015era,Qin:2016spb}. Here and in the following, $D_1$ denotes the narrow $D_1(2420)$ resonance.
A recent comprehensive analysis of the energy region from 4.2 to 4.35~GeV has shown that a single state with pronounced $D\bar{D}_1$ molecular features, together with interference effects and the nearby $D\bar{D}_1$ threshold, can describe a broad set of exclusive data in this window~\cite{vonDetten:2024eie}.
On the other hand, Ref.~\cite{Nakamura:2023obk} claims the need of an additional pole around 4.32~GeV.

In the present work, we extend the study of Ref.~\cite{vonDetten:2024eie} by broadening the investigated energy range from 4.2\dash4.35~GeV to 4.1\dash4.6~GeV and by investigating whether the pertinent states can be understood as dynamically generated from $\{D,D^*\}$-$\{\bar D_1,\bar D_2^*\}$ scattering.\footnote{The meson pairs are understood as linear combinations with their charge-conjugated counterparts that form states with negative $C$ parity; see Eq.~\eqref{elastic2} below for details.} This at the same
time acknowledges that,
in this broader energy interval, additional $S$-wave open-charm channels, most notably $D^*\bar{D}_1$ and $D^*\bar{D}_2^*$, become relevant in
addition to $D\bar{D}_1$.
The three channels are related by
heavy-quark spin symmetry (HQSS)~\cite{Guo:2017jvc,Ji:2022blw,Peng:2022nrj, Zhang:2025gmm}. In addition,
the three-body final states require an explicit treatment of the $Z_c$-related final-state interactions (FSIs) in the subsystems $J/\psi\pi$, $h_c\pi$, $D\bar{D}^*$, and $D^*\bar{D}^*$~\cite{Wang:2013cya,Albaladejo:2015lob,Chen:2023def,Du:2022jjv,Ye:2026azi,Yu:2024sqv,Yan:2023bwt,Danilkin:2020kce,He:2017lhy,Chen:2026fnz,Chen:2011xk}.
Therefore, a realistic description of the vector charmonium-like structures in this region should incorporate at least three key ingredients: conventional charmonium seeds, coupled-channel rescattering among the dominant open-charm channels, and channel-dependent FSIs in the experimentally observed decay modes with three-body final states.
In this work, we construct such a framework and perform a simultaneous fit to the BESIII data for the cross sections of $\ee\to J/\psi\pi^+\pi^-$, $h_c\pi^+\pi^-$, $D\bar{D}^*\pi$, $D^*\bar{D}^*\pi$, $J/\psi\eta$, and $\chi_{c0}\omega$, together with the available invariant-mass distributions involving the $Z_c(3900)$ and $Z_c(4020)$ states. Our goal is to identify a consistent dynamical interpretation of the entire data set and clarify how many poles (physical states) are actually required in the energy region considered. In particular, we investigate to what extent the apparent proliferation of vector structures can be understood as a combined consequence of nearby open-charm thresholds, rescattering effects, and channel-dependent production and decay mechanisms. Addressing these questions is necessary for clarifying the status of the vector charmonium(-like) spectrum and for establishing a more reliable connection between experimental line shapes and the underlying degrees of freedom in quantum chromodynamics (QCD).

The paper is organized as follows. In Section~\ref{sec:qm}, we give a brief introduction to the vector charmonia in the framework of quark models. In Section~\ref{sec:data}, we specify and motivate the choice of the experimental input to be used in the analysis. Section~\ref{sec:formalism} contains the details of the employed theoretical coupled-channel framework. We present our main results in Section~\ref{sec:results} and discuss and summarize them in Section~\ref{sec:summary}. Additional details of the formalism, explicit expressions for the production amplitudes, predicted distributions in complementary channels, and the parameter values obtained in the best fits are collected in the appendices. 

\section{Vector charmonium in the quark model: Inspiration and puzzles}
\label{sec:qm}

We start with a brief introduction to quark model calculations of the charmonium spectrum. Charmonia are mesons composed of a charm quark and its antiquark.
The first hadron in this family, now known as the $J/\psi$, was observed during the November revolution in 1974 \cite{E598:1974sol,SLAC-SP-017:1974ind}.
Subsequently, discoveries of excited vector charmonia followed rapidly: $\psi(3686)$~\cite{Abrams:1974yy}, $\psi(3770)$ \cite{Goldhaber:1976xn}, $\psi(4040)$ \cite{Rapidis:1977cv}, $\psi(4160)$ \cite{DASP:1978dns}, $\psi(4415)$~\cite{Siegrist:1976br}, and others. Within the quark model framework, these mesons can generally be understood as radial excitations of predominantly $S$- or $D$-wave quark--antiquark states; see, for example, Refs.~\cite{Eichten:1978tg,Eichten:1979ms,Godfrey:1985xj,Ebert:2002pp,Barnes:2005pb,Li:2009zu,Ferretti:2013faa}. In particular, it is commonly believed that the two lightest vector charmonia, $J/\psi$ and $\psi(3686)$, are genuine $S$-wave quark model states, so that $\psi(3686)$ is conventionally denoted as $\psi(2S)$ \cite{ParticleDataGroup:2024cfk}. However, already the next excitation, the $\psi(3770)$, lies above the open-charm threshold, making its assignment as a pure $D$-wave quark model state questionable. Thus one may need to account for $S$-$D$ mixing through 
charmed particle pairs to explain puzzling features in the decays of the $\psi(3770)$~\cite{Rosner:2001nm}. 

The problem of the proper identification becomes even more severe for higher states, where the effects of multiple hadronic thresholds must be strong. Furthermore, as mentioned in the Introduction, a number of additional vector states have been reported by various experimental collaborations (see Figure~\ref{fig:polecom} for a visual compilation of these results) in the energy range between 4.1 and 4.6~GeV, and these states cannot be naturally accommodated within the na{\"i}ve quark model framework. Therefore, understanding the spectrum of vector charmonia requires analyses employing advanced coupled-channel schemes that incorporate both bare compact seeds inspired by the quark model and open- and hidden-charm channels with thresholds lying in the relevant energy region.
Examples of such analyses can be found in Refs.~\cite{Uglov:2016orr,Husken:2024hmi,Nakamura:2023obk,vonDetten:2024eie}. However, in Ref.~\cite{Uglov:2016orr}, only a limited set of experimental data on $e^+e^-$ annihilation into pairs of open-charm mesons was analyzed, and the coupled-channel model based on the $K$-matrix approach did not fully implement the analytic structure of the amplitude. In Ref.~\cite{Husken:2024hmi}, a more advanced $K$-matrix approach was employed, so that the resulting amplitude was suitable for a pole search, but only the lower part of the spectrum below 4.2~GeV was addressed. 
In Ref.~\cite{vonDetten:2024eie}, a unitary scheme was employed to study vector charmonium and charmonium-like states in the energy interval 4.2\dash4.35~GeV,
however, only the $D_1\bar D$ channel was kept dynamically.
This study led
to the conclusion that one compact quark model state, $\psi(4160)$, coexisted with a dynamically generated pole around 4.23~GeV, associated with the $\psi(4230)$ state.
In Ref.~\cite{Nakamura:2023obk}, a global multi-parameter coupled-channel analysis of the world data on $e^+e^- \to c\bar{c}$ was performed, approximately respecting three-body unitarity, and several poles located near the open-charm thresholds were identified. However, in this global fit, some contact-interaction matrix elements among the HQSS-partner channels $D\bar{D}_1$, $D^*\bar{D}_1$, and $D^*\bar{D}_2^*$ were set to zero, at odds with the requirements of HQSS (see Table~XIII in Ref.~\cite{Nakamura:2023obk}). In particular, no direct short-range coupling between the $D_1 \bar{D}$ channel and the heavier $D_1 \bar{D}^*$ and $D_2^* \bar{D}^*$ ones was included. A consistent treatment of these HQSS-related coupled-channel transitions is one of the main foci of the present analysis.
Accordingly,
our study is based on the ideas
underlying
Ref.~\cite{vonDetten:2024eie},
but we enlarge the covered energy range up to 4.6~GeV and broaden
both the set of experimental data analyzed and the coupled-channel framework employed; see Sects.~\ref{sec:data} and \ref{sec:formalism} below for a detailed discussion.

\section{Experimental Input}
\label{sec:data}

In this section, we specify the data sets included in the fit. The selection is guided by the following considerations. We include both hidden-charm and open-charm channels and retain both total cross sections and invariant-mass distributions. These data constrain the line shapes in the 4.1\dash4.6~GeV energy range as well as the $Z_c$-related FSIs. The selected data sets include:
\begin{itemize}
\item Exclusive cross sections for $e^+e^-$ annihilation into $J/\psi\pi^+\pi^-$ \cite{BESIII:2022qal}, $h_c\pi^+\pi^-$ \cite{BESIII:2025bce}, $D^0\bar{D}^{*-}\pi^+$ + c.c. \cite{BESIII:2018iea}, $D^{*0}\bar{D}^{*-}\pi^+$ + c.c. \cite{BESIII:2023cmv}, $J/\psi\eta$ \cite{BESIII:2023tll}, and $\chi_{c0}\omega$ \cite{BESIII:2014rja,BESIII:2015som,BESIII:2019gjc}.
 \item Invariant mass distributions for the $Z_c$ structures observed in $J/\psi\pi^\pm$ \cite{BESIII:2025qkn}, $h_c\pi^\pm$ \cite{BESIII:2013ouc}, $(D\bar{D}^*)^\pm$ \cite{BESIII:2015pqw}, and $(D^*\bar{D}^*)^\pm$ \cite{BESIII:2013mhi} from the corresponding three-body final states.
\end{itemize}
Several observed channels that also exhibit the $\psi(4230)$ structure are excluded from the current analysis since 
\begin{itemize}
 \item $J/\psi\pi^0\pi^0$ \cite{BESIII:2020oph} does not provide significant information beyond that contained in the $J/\psi\pi^+\pi^-$ mode
and has much lower statistics;
\item $\psi(2S)\pi\pi$ \cite{BESIII:2021njb,BESIII:2026qhe} requires a dedicated treatment of the $\psi(2S)\pi$ distribution simultaneously with the $\pi\pi$ FSI \cite{Guo:2019twa,Molnar:2019uos}, which lies beyond the scope of the present research;
\item the description of $J/\psi K^+K^-$ and $J/\psi K_{\rm S}^0K_{\rm S}^0$ requires the $\pi\pi$-$K\bar K$ FSIs~\cite{Danilkin:2020kce,Ermolina:2026jgp} not included in the present framework;
\item $\gamma X(3872)$ \cite{BESIII:2019qvy}, $J/\psi\eta'$ \cite{BESIII:2019nmu}, and $h_c\eta$~\cite{BESIII:2024yqi} suffer from low experimental statistics;
\item the $\mu^+\mu^-$ measurement \cite{Ablikim:2020jrn} appears to be in tension with the measured $R$ value and raises theoretical consistency issues~\cite{Farrar:2023zmj}. 
\end{itemize}
Nevertheless, as will be demonstrated below, the retained set already covers the dominant hidden-charm, open-charm, and $Z_c$-sensitive observables sufficiently to constrain the pole positions
of the states in the energy range of interest.

\section{Coupled-Channel Framework}
\label{sec:formalism}

As explained above, the experimental data analyzed in this work are total and differential cross sections measured in $e^+e^-$ collisions. We therefore need to establish a comprehensive theoretical framework that incorporates the dynamics of coupled channels and the production mechanisms from virtual photons. Our aim is to retain a minimal set of degrees of freedom capable of capturing the dominant threshold effects in the $1^{--}$ sector while maintaining direct contact with the available BESIII data in both hidden-charm and open-charm final states. In this section, we present the coupled-channel scheme used in the combined analysis of the selected data set, aiming at a consistent description of the spectrum.

\subsection{Degrees of freedom}
\label{sec:degrees}

The employed coupled-channel formalism incorporates three distinct categories of channels:
\begin{enumerate}[leftmargin=*]
\item $\Np$ bare poles (compact states) tagged by capital Latin letters $A$ and $B$. The inclusion of such poles is motivated by the data and follows the previous analysis of the lower part of the spectrum in Ref.~\cite{vonDetten:2024eie}. Here we consider up to two bare poles,
\beq
\begin{split}
&\mbox{bare pole 1 leading to the } \psi(4160),\\
&\mbox{bare pole 2 leading to the } \psi(4415),
\end{split}
\label{bare}
\eeq
so we have $\Np=2$ at most. One of the primary goals of this analysis is to study the minimal bare-pole content of the developed coupled-channel scheme that is consistent with the existing data. Thus, we build fits with $\Np=0$ (no bare poles), $\Np=1$ (with only $\psi(4160)$ included), and $\Np=2$ (with both $\psi(4160)$ and $\psi(4415)$ included).
\item $\Ne$ dynamically coupled open-charm two-body $S$-wave channels. In what follows these channels will be referred to as ``two-body elastic channels'' for brevity and tagged by Greek letters from the beginning of the alphabet, that is, $\alpha$ and $\beta$ (note that Greek letters $\mu$ and $\nu$ are reserved for four-dimensional Lorentz indices). In the studied broad energy range, the $J^{PC}=1^{--}$ states can couple to several such channels related through HQSS. In this analysis, we include (charge-conjugate channels are implied; see Eq.~\eqref{elastic2} below for details)
\beq
\begin{split}
&D\bar{D}_1~\mbox{as two-body elastic channel 1},\\
&D^*\bar{D}_1~\mbox{as two-body elastic channel 2},\\
&D^*\bar{D}_2^*~\mbox{as two-body elastic channel 3},
\end{split}
\label{elastic}
\eeq
so we have $\Ne=3$. The convention for the charge conjugation transformation with the operator $\hat C$ is chosen to be
\begin{align}
\hat C\{D,D^*,D_1,D^*_2\}=\{\bar{D}, -\bar{D}^*,\bar{D}_1,-\bar{D}^{*}_2\}.
\end{align}
In this convention, a two-meson state $\ket{M_1\bar{M}_2}_C$, with a given $C$-parity, is built as
\beq
\ket{M_1\bar{M}_2}_C=\frac{1}{\sqrt{2}}\Bigl(\ket{M_1\bar{M}_2}+(-1)^{J-J_1-J_2}CC_1C_2\ket{\bar{M}_1M_2}\Bigr),
\label{M1M2C}
\eeq
where $C_1$ and $C_2$ are the charge parities of the mesons $M_1$ and $M_2$, respectively, $J_1$ and $J_2$ are their total spins, and $J$ is the total spin of the two-meson system. Then the physical states with the quantum numbers $J^{PC}=1^{--}$ composed of two charmed mesons above are constructed as~\cite{Ji:2022blw}
\begin{equation}
 \begin{aligned}
&\ket{D\bar{D}_1}_{J^{PC}=1^{--}}= \frac{1}{\sqrt2}(\ket{D\bar{D}_1}-\ket{\bar{D}D_1}),\\
&\ket{D^*\bar{D}_1}_{J^{PC}=1^{--}}=\frac{1}{\sqrt2}(\ket{D^*\bar{D}_1}-\ket{\bar{D}^*D_1}),\\
&\ket{D^*\bar{D}_2^*}_{J^{PC}=1^{--}}=\frac{1}{\sqrt2}(\ket{D^*\bar{D}^*_2}-\ket{\bar{D}^*D^*_2}).
\end{aligned}
\label{elastic2}
\end{equation}
Finally, the isoscalar flavor combination entering the above $C$-parity eigenstates is, for example, for the $D\bar{D}_1$ pair,
\beq
\ket{D\bar{D}_1}_{I=0}=\frac1{\sqrt{2}}(\ket{D^+D_1^-}+\ket{D^0\bar{D}_1^0}),
\label{DDiso}
\eeq
and similarly for the other elastic channels. In what follows, we refer to the two-body elastic channels in Eq.~\eqref{elastic} for simplicity, implicitly assuming the proper $C$-parity and isospin states defined in Eqs.~\eqref{elastic2} and \eqref{DDiso}.
\item $\Nexp=6$ channels corresponding to the experimentally observed final states discussed in Section~\ref{sec:data}. We refer to these channels as ``observation channels'' and tag them by Latin letters from the beginning of the alphabet, that is, $a$, $b$, and $c$ (note that Latin letters $i$, $j$, $k$, and $l$ are reserved for three-dimensional Lorentz indices). The observation channels are categorized as having three-body (hidden-charm or open-charm) final states,
\beq
\begin{split}
&J/\psi\pi\pi~\mbox{as observation channel 1},\\
&h_c\pi\pi~\mbox{as observation channel 2},\\
&D\bar{D}^*\pi~\mbox{as observation channel 3},\\
&D^*\bar{D}^*\pi~\mbox{as observation channel 4},
\end{split}
\label{observation1}
\eeq
and those with two-body hidden-charm final states,
\beq
\begin{split}
&J/\psi\eta~\mbox{as observation channel 5},\\
&\chi_{c0}\omega~\mbox{as observation channel 6}.
\label{observation2}
\end{split}
\eeq
Besides contributing to the absorptive parts of the amplitudes and thereby affecting the pole positions, the observation channels in Eqs.~\eqref{observation1} and \eqref{observation2} constrain the production sources and, in the three-body channels above, contain channel-dependent FSIs as discussed in detail in Section~\ref{sec:fsi} below.
Note that the three-body open-charm channels, observation channels 3 and 4 in Eq.~\eqref{observation1}, can also be regarded as elastic, analogously to the two-body channels in Eq.~\eqref{elastic}. However, to avoid confusion, we do not refer to them as such, and therefore reserve the term ``elastic'' exclusively for the channels in Eq.~\eqref{elastic}.
\end{enumerate}

\subsection{Lippmann-Schwinger equation}
\label{sec:LSE}

We employ a standard relativistic framework, where the $S$-matrix is by convention related to the scattering multi-channel amplitude $T$ as
\begin{equation}
S=1-iT,
\end{equation}
and $T$ obeys the Lippmann-Schwinger equation (LSE)
\begin{equation}
T=V+VGT.
\label{LSEgen}
\end{equation}
In the most general form the potential matrix $V$ of dimension $\Np +\Ne +\Nexp $ includes transitions between the bare charmonium states in Eq.~\eqref{bare}, two-body elastic channels in Eq.~\eqref{elastic}, and observation channels in Eqs.~\eqref{observation1} and \eqref{observation2},
\beq
V=\begin{pmatrix}
V_{AB}&V_{A\beta}&V_{Ab}\\
V_{\alpha B}&V_{\alpha\beta}&V_{\alpha b}\\
V_{aB}&V_{a\beta}&V_{ab}
\end{pmatrix},
\label{Vfull}
\eeq
$$
A,B=1\ldots\Np;\quad \alpha,\beta=1\ldots\Ne;\quad a,b=1\ldots\Nexp.
$$
The diagonal matrix of the free Green's functions reads
\beq
G={\rm diag}(G_{AB},G_{\alpha\beta},G_{ab}).
\eeq

In the strict heavy quark limit, the leading-order (LO) contact interactions $V_{\alpha\beta}$ ($\alpha,\beta=1,2,3$) between the three two-body elastic channels in Eq.~\eqref{elastic}
can be expressed in terms of four contact terms~\cite{Guo:2017jvc} introduced in Eq.~\eqref{eq:FIjl} (the subscript ``0'' carried by all $F$'s indicates isospin $I=0$; see also Appendix~\ref{sec:HbarT_potential} for further details),
\begin{equation}\label{eq:VY}
\begin{aligned}
V_{11}& =\,\frac{1}{8}\left(-F_{01}^c+3 F_{01}^d+5(F_{02}^d-F_{02}^c)\right), \\
V_{22}& =\frac{1}{16}\left(7 F_{01}^c+11 F_{01}^d+5 (F_{02}^d-F_{02}^c)\right), \\
V_{33}& = \frac{1}{16}\left(-5 F_{01}^c+15 F_{01}^d+(F_{02}^d-F_{02}^c)\right),\\
V_{12} & =-\frac{1}{8 \sqrt{2}}\left(F_{01}^c+5 F_{01}^d-5(F_{02}^d-F_{02}^c)\right), \\
V_{13} & =\frac{1}{8} \sqrt{\frac{5}{2}}\left(-3 F_{01}^c+F_{01}^d-(F_{02}^d-F_{02}^c)\right), \\
V_{23} & =\frac{\sqrt{5}}{16}\left(5F_{01}^c+F_{01}^d-(F_{02}^d-F_{02}^c)\right).
\end{aligned}
\end{equation}
Note that the six potentials $V_{\alpha\beta}$ appearing here are controlled by only three independent parameters, since two of them always enter through the same linear combination, $F^d_{02}-F^c_{02}$. Moreover, it is not consistent to set only a subset of the matrix elements listed above to zero.
As discussed in the Introduction, a large number of open channels cannot be considered explicitly in the present analysis. Under these circumstances, some parameters of the effective potentials must be treated as complex-valued. We therefore adopt the following procedure: First, we neglect direct transitions among the bare quark-model states, since their effect can be absorbed into a redefinition of the parameters of the bare compact states, and set $V_{AB}=0$ for $A,B=1\dots\Np$. Second, we emphasize that a full inclusion of the observation channels in the coupled-channel framework would introduce numerous additional parameters that cannot be reliably constrained by the available data. 
Therefore, we refrain from considering direct transitions between observation channels, which allows us to reduce the dimension of the potential matrix in Eq.~\eqref{Vfull} by introducing an effective potential $\bar{V}$ (Hereinafter, summation over repeated indices is implied unless stated otherwise),
\begin{equation}
\begin{aligned}
\bar{V}_{AB}=&V_{Aa}G_{ab}V_{bB},\\
\bar{V}_{\alpha B}=&V_{\alpha B}+V_{\alpha a}G_{ab}V_{bB},\\
\bar{V}_{A\beta}=&V_{A\beta}+V_{A a}G_{ab}V_{b\beta},\\
\bar{V}_{\alpha\beta}=&V_{\alpha\beta}+V_{\alpha a}G_{ab}V_{b\beta},
\end{aligned}
\label{Veff}
\end{equation}
that absorbs the effects of the observation channels \cite{Hanhart:2015cua,Guo:2016bjq,Ji:2025hjw}.
Then, rather than incorporating the observation channels explicitly into the coupled-channel framework, we employ the effective transition potentials in Eq.~\eqref{Veff} and neglect the real parts of the contributions from the observation channels, while accounting for their absorptive effects by treating the imaginary parts of the effective elastic potentials as constant fit parameters (see Section~\ref{sec:parameters} for a summary of all fit parameters). In this way, we accept a certain degree of unitarity violation and neglect the energy dependence associated with the observation channels, which is expected to generate only slowly varying higher-order corrections, since the corresponding thresholds lie far below the energy region studied here.

The resulting LSE reduces to
\begin{align}
T=\bar{V}+\bar{V} G T,
\label{LSEred}
\end{align}
with
\begin{align}
\bar{V}=\left(\begin{array}{cc}
\bar{V}_{AB}&\bar{V}_{A\beta}\\
\bar{V}_{\alpha B}&\bar{V}_{\alpha\beta}
\end{array}
\right), \quad
G=\mbox{diag}(G_{AB},G_{\alpha\beta}),
\label{eq:VGT0alpha}
\end{align}
where $\bar V_{\alpha\beta}$ are treated as complex valued quantities and
\beq
\begin{split}
&G_{AB}=\frac{\delta_{AB}}{E^2-m_{0,A}^2+i\,m_{0,A}^{}\Gamma_{0,A}^{}},\\[2mm]
&G_{\alpha\beta}=\delta_{\alpha\beta}G^{\rm DR}
(E,m_{1 \alpha},m_{2 \alpha }-{i}\,\Gamma_{2 \alpha}/2).
\end{split}
\label{Gs}
\eeq
Here, $E$ is the total energy in the center of mass frame, and $m_{0,A}$ and $\Gamma_{0,A}$ denote the bare mass and width of pole $A$, respectively, while $m_{1\alpha,2\alpha}$ and $\Gamma_{1\alpha,2\alpha}$ represent the masses and widths of the constituent particles in channel $\alpha$. The widths of the $D$ and $D^*$ mesons are negligibly small and are therefore omitted, whereas the widths of the $D_1^{}$ and $D_2^*$ states are treated as constants throughout the analysis. We employ a dimensional regularization (DR) scheme to tame the logarithmically divergent loop integrals, so the regularized two-body propagator reads
\beq
\begin{split}
G^{\rm DR}(E,m_1,m_2)
=&\ i\int\frac{d^4l}{(2\pi)^4}\frac{1}{[l^2-m_1^2+i\epsilon][(P-l)^2-m_2^2+i\epsilon]}\\
=&\ \frac1{16\pi^2}\bigg\{a(\mu)+\log\frac{m_1^2}{\mu^2}
+\frac{m_2^2-m_1^2+E^2}{2E^2} \log\frac{m_2^2}{m_1^2}\\
&+\frac{k}{E} \Big[
\log\left(2k E+E^2+\Delta\right) +
\log\left(2k E+E^2-\Delta\right)\\
&-\log\left(2k E-E^2+\Delta\right) -
\log\left(2k E-E^2-\Delta\right)
\Big]\bigg\},
\label{eq:GDR}
\end{split}
\eeq
where $m_1$ and $m_2$ are the complex masses of the constituents, $\Delta= m_1^2-m_2^2$, and the three-momentum
\beq
k=\frac{1}{2\sqrt{\stot}}\sqrt{\lambda(\stot,m_1^2,m_2^2)
}, 
\eeq
with $\stot=P^2=E^2$ the two-body invariant mass squared for the given two-body system, and
$\lambda(x,y,z)=x^2+y^2+z^2-2xy-2yz-2zx$ the K{\"a}ll{\'e}n triangle function. The dimensional regularization scale $\mu$ is set to $1~\mathrm{GeV}$. The subtraction constant $a(\mu)$ is determined by matching the DR loop function $G^{\rm DR}$ to the hard-cutoff (HC) loop function $G^{\rm HC}$ at the complex threshold~\cite{Oller:1998hw,Garcia-Recio:2010enl,Albaladejo:2015lob,Du:2022jjv}, with the cutoff $\Lambda=1~\mathrm{GeV}$, that is,
\begin{align}
 G^{\rm DR}(m_1+m_2,m_1,m_2)=G^{\rm HC}(m_1+m_2,m_1,m_2)\,,
\end{align}
where
\begin{align}
 G^{\rm HC}(E,m_1,m_2)=\int_{0}^{\Lambda} \frac{l^2\, d l}{4 \pi^2} \frac{1}{\omega_1 \omega_2} \frac{\omega_1+\omega_2}{E^2-\left(\omega_1+\omega_2\right)^2+i \epsilon}\,,
\end{align}
with $\omega_n=\sqrt{m_n^2+l^2}$ ($n=1,2$).

In what follows, the observation channels do not appear explicitly in the intermediate states of the reactions considered. We therefore define the amplitude $T$ as solution of the reduced LSE in Eq.~\eqref{LSEred}, with the Green functions and the potential matrix defined in Eq.~\eqref{eq:VGT0alpha}.

\subsection{Production amplitudes for bare states and two-body elastic channels}\label{sec:amplitudes}

The production of the various final states in electron--positron annihilation studied in this work proceeds through a single virtual photon. Since the $e^+e^-\to\gamma^*$ annihilation vertex and the photon propagator are fixed by QED, they are omitted for simplicity in the amplitudes discussed below and restored only when cross sections are calculated.

We start from the bare poles in Eq.~\eqref{bare} and two-body elastic channels in Eq.~\eqref{elastic} produced from the virtual photon $\gamma^*$, with the polarization of the latter denoted as ${\bm\epsilon}_{\gamma^*}$. The polarization information of the bare compact states
is contained in their polarization vectors, $\{{\bm\epsilon_A}\}$ ($A=1\ldots\Np$). For the two-body elastic channels in the final states this information is encoded in the set of the polarization vectors ($\alpha=1,2,3$),
\begin{align}
\epsilon^i_\alpha&=\left\{\epsilon^i_{D_1},\ \frac{i}{\sqrt2}\epsilon^{ijk}\epsilon_{D^*}^j\epsilon_{D_1}^k,\ {\sqrt{\frac35}}\epsilon_{D_2^*}^{ij}\epsilon_{D^*}^j\right\}.
\label{pols}
\end{align}
Denoting the tree-level production strengths of the bare states and two-body elastic channels from the virtual photon by $\Pzero$ and $\Palpha$, respectively, the corresponding dressed production amplitudes read
\beq
{\cal M}_{A,\alpha}=\left({\bm\epsilon}_{\gamma^*}\cdot{\bm \epsilon}_{A,\alpha}^*\right)U_{A,\alpha}
\label{Uampl},
\eeq
with
\begin{align}
(U_{A},U_{\alpha})=({\cal P}_{A}^{\gamma^*},{\cal P}_{\alpha}^{\gamma^*})(1+GT),
\end{align}
where, as mentioned above, $T$ is solution of the LSE in Eq.~\eqref{LSEred} and
the matrix of the propagators $G$ is specified in Eq.~\eqref{eq:VGT0alpha}.

\begin{figure}[t]
\centering
\includegraphics[width=\linewidth]{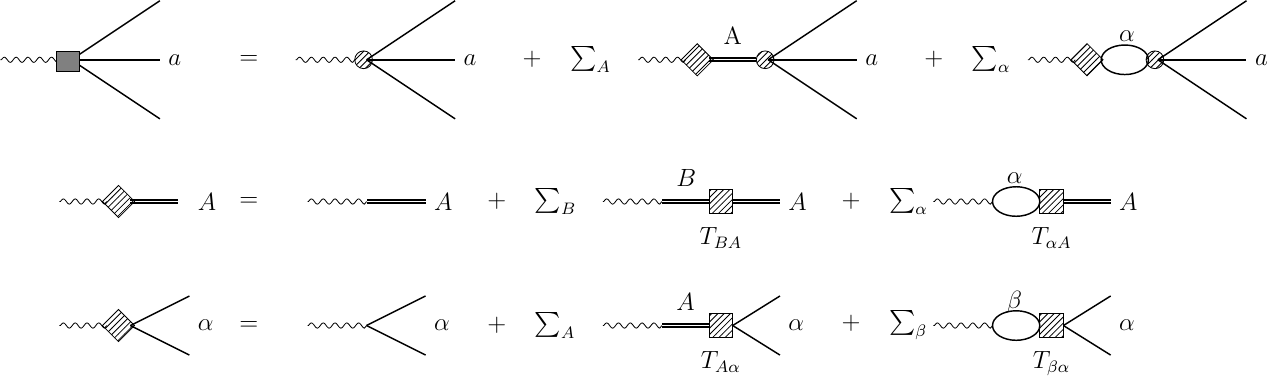}
\caption{Production mechanisms for the observation channels with three-body final states in Eq.~\eqref{observation1} from the virtual photon as detailed in Eq.~\eqref{M1ab}. Diagrams for the production of the two-body observation channels in Eq.~\eqref{observation2} look similar.
The shaded square box represents the full production amplitude in Eq.~\eqref{M1ab}. The hatched diamond represents the full transition amplitude, Eq.~\eqref{Uampl}, from the virtual photon to bare states $\psi$ in Eq.~\eqref{bare} or two-body elastic channels in Eq.~\eqref{elastic}, including all coupled-channel effects. The hatched circle represents the full transition amplitudes from virtual photon, bare states or two-body elastic channels to the observation channels. The full transition amplitudes include the effects of the interaction in the final state discussed in Section~\ref{sec:fsi}. The amplitude $T$, whose various components appear in the diagrams as hatched square boxes, is solution of the reduced LSE in Eq.~\eqref{LSEred}.}
\label{fig:Production}
\end{figure}

\subsection{Cross sections in the observation channels}
\label{sec:cross_sections_expressions_detail}

The total production amplitude for a particular observation channel in Eqs.~\eqref{observation1} and \eqref{observation2} can be written in the generic form (no summation in $a$ is implied on the right-hand side)
\beq
{\cal M}_a=\epsilon_{\gamma^*}^i{\cal M}_a^{ij}\epsilon^{j*}_a,\quad a=1\ldots\Nexp,
\label{Mas}
\eeq
where, as before, ${\bm\epsilon}_{\gamma^*}$ is the polarization vector of the photon and
\begin{align}
\epsilon^i_a=\Bigl\{&\epsilon^{i}_{\psi},\ -\frac{i}{\sqrt2}\epsilon^{ijk}\epsilon_{h_c}^j\epsilon_{L=1}^k,\ \epsilon^{i}_{D^{*-}},\ -\frac{i}{\sqrt2}\epsilon^{ijk}\epsilon_{D^{*0}}^j\epsilon_{D^{*-}}^k,\ -\frac{i}{\sqrt2}\epsilon^{ijk}\epsilon_\psi^j\epsilon_{L=1}^k,\ \epsilon_\omega^i\Bigr\},
\label{eas}
\end{align}
with ${\bm\epsilon}_X$ the polarization vector of the particle $X$ and, for the polarization vector of the state with the orbital angular momentum $L=1$,
\beq
\begin{split}
&\epsilon_{L=1}^i(m)=\sqrt{\frac{3}{4\pi}}\int d\Omega\; \hat{p}^iY_{1m}^*(\hat{\bm p}),\quad \sum_m \epsilon_{L=1}^i(m)\epsilon_{L=1}^{j*}(m)=\delta^{ij},
\end{split}
\eeq 
with $\hat{\bm p}$ the unit vector in the direction of the momentum.

Each production amplitude in Eq.~\eqref{Mas} can be presented as a sum of (i) the direct production term, (ii) the production through the bare charmonium poles, and (iii) the production through the two-body elastic intermediate states,
\beq
\begin{split}
\mathcal{M}_a=\mathcal{M}^{\gamma^*}_a
+\sum_A\mathcal{M}^A_a
+\sum_\alpha\mathcal{M}^\alpha_a.
\end{split}
\label{M1ab}
\eeq
We illustrate these three mechanisms in Figure~\ref{fig:Production}. The spin-averaged squared amplitude is then obtained as
\begin{align}
\overline{\left|\mathcal{M}_a\right|^2}=\frac{1}{E^4}\sum_{\rm spins} L^{ii'}\mathcal M_a^{ij}\mathcal M_a^{*i'j}
=\frac{e^2}{E^2}\bar{M}_a^{\dagger} \cdot {Q_a}\cdot \bar{M}_a,
\label{barMdef}
\end{align}
where
\beq
L^{ii'}=\frac{1}{2}e^2E^2\left(\delta^{ii'}-\delta^{i3}\delta^{i'3}\right)
\eeq
is the leptonic tensor. For each observation channel $a$, $Q_a$ is a kinematical matrix and $\bar{M}_a$ is a vector.
For the channels with only one tensor structure, both of these objects reduce to single-component functions. We provide explicit expressions for the $Q$'s and $\bar{M}$'s for all observation channels involved in Appendix~\ref{app:prodi}.

To proceed to the cross sections, we address the observation channels with the three-body final states first.
The kinematics of the corresponding processes can be defined in terms of the four-momenta $P^\mu_{}$, $p_1^\mu$, $p_2^\mu$, and $p_3^\mu$ for the virtual photon $\gamma^*$ and the final-state particles $a_1$, $a_2$, and $a_3$, respectively. For the four three-body observation channels in Eq.~\eqref{observation1} considered here, we have
\beq
\begin{pmatrix}
a_1\\ a_2\\a_3
\end{pmatrix}
=
\begin{pmatrix}
\pi^+\\ \pi^-\\ J/\psi
\end{pmatrix},~
\begin{pmatrix}
\pi^+\\ \pi^-\\h_c
\end{pmatrix},~
\begin{pmatrix}
\pi^+\\ D^{*-}\\D^0
\end{pmatrix},~
\begin{pmatrix}
\pi^+\\ D^{*-}\\D^{*0}
\end{pmatrix}.
\eeq
Then the Mandelstam variables read
\begin{equation}
\begin{aligned}
s&=(P^\mu-p_1^\mu)^2=(p_2^\mu+p_3^\mu)^2=m_{23}^2,\\
t&=(P^\mu-p_2^\mu)^2=(p_1^\mu+p_3^\mu)^2=m_{13}^2,\\
u&=(P^\mu-p_3^\mu)^2=(p_1^\mu+p_2^\mu)^2=m_{12}^2
\end{aligned}
\end{equation}
and satisfy the standard kinematic constraint, $s+t+u=m_1^2+m_2^2+m_3^2+P^2$. We remind the reader that the overall center-of-mass (c.m.) energy is denoted as $E=\sqrt{s_{\rm tot}}$, while the variable $s$ defined above refers to the invariant mass squared in the given two-body subsystem. In the c.m. frame with $P=(E,\vec{0})$, the three-momenta of the particles in the final state can be expressed as
\begin{equation}
\begin{aligned}
&p_1^2\equiv\vec{p}_1^2=\frac{1}{4E^2}{\lambda(E^2,m_1^2,s)},\\
&p_2^2\equiv\vec{p}_2^2=\frac{1}{4E^2}{\lambda(E^2,m_2^2,t)},\\
&p_3^2\equiv\vec{p}_3^2=\frac{1}{4E^2}{\lambda(E^2,m_3^2,u)},\\
&\vec{p}_1\cdot\vec{p}_2=\frac12{\left((\vec{p}_1+\vec{p}_2)^2-p_1^2-p_2^2\right)}=\frac12\left({p_3^2-p_1^2-p_2^2}\right).
\end{aligned}
\end{equation}

Thus, for the four observation channels in Eq.~\eqref{observation1} with three-body final states, the doubly differential cross section with respect to the Mandelstam variables $s$ and $t$ is given by
\beq
\frac{d\sigma_a}{ds\,dt}(E,s,t)=\frac{1}{(2\pi)^3}\frac{1}{32E^{4}}\overline{\left|\mathcal{M}_a\right|^2},\quad a=1\ldots4.
\eeq
The integration over $t$ yields the invariant-mass distribution in the $m_{23}$ subsystem,
\begin{align}
\frac{d\sigma_a}{ds}(E,s)=\frac{1}{(2\pi)^3}\frac{1}{32E^{4}}\int dt\ \overline{\left|\mathcal{M}_a\right|^2}
\label{eq:dsigds},
\end{align}
and the yield of the events in the $k$-th bin with $s\in (s_k,s_{k+1})$ is calculated as
\begin{align}
\Delta N_a^k(E)= 2\bar{\mathcal{L}}_a(E)\int_{s_k}^{s_{k+1}}ds\,\frac{d\sigma_a}{ds}(E,s).
\label{eq:DeltaJpsipi}
\end{align}
The prefactor 2 above implicitly accounts for the charge-conjugate channels in the given distributions (for example, $J/\psi\pi^+$ and $J/\psi\pi^-$ in the channel with $a=1$). At a fixed total energy $E$, the normalization factor $\bar{\mathcal{L}}_a(E)$ with a dimension of the inverse cross section absorbs experimental normalization factors such as the luminosity, branching fractions, ISR factors, and detection efficiency.
Then the total three-body cross section, obtained from the differential cross section in Eq.~\eqref{eq:dsigds}, reads
\beq
\sigma_a(E)=\int ds\ \frac{d\sigma_a}{ds}(E,s),\quad a=1\ldots4.
\label{xsec14}
\eeq

Finally, for the two-body observation channels in Eq.~\eqref{observation2}, the total cross section is
\beq
\sigma_a(E)=\frac{|\vec p|}{8\pi E^3} \overline{\left|\mathcal{M}_a\right|^2},\quad a=5,6,
\label{xsec56}
\eeq
with $\vec p$ the c.m. three-momentum of the final particles.

In the fits performed below, the formulas in Eqs.~\eqref{xsec14} and \eqref{xsec56} are evaluated at the beam energies $\{E_n\}$ from the discrete set studied at BESIII. The energies $E_n$ and the corresponding factors $\bar{\mathcal{L}}_1(E)$ for the channel $J/\psi\pi^+\pi^-$ are collected in Table~\ref{tab:jpsipiNormConstants}. The normalization constants $\bar{\mathcal{L}}$ for the other observation channels are not known and, therefore, are treated as fit parameters; see the discussion in Section~\ref{sec:parameters} below and Appendix~\ref{app:params}, where all fit parameters are listed explicitly, together with their values returned by the fits performed in different schemes.

\begin{table}
\caption{Energy-dependent normalization factors for the $J/\psi\pi^\pm$ invariant-mass distributions relevant for BESIII, constructed using the information from Ref.~\cite{BESIII:2022qal}.}\label{tab:jpsipiNormConstants}
\vspace{0.2cm}
\renewcommand{\arraystretch}{1.15}
\setlength{\tabcolsep}{0pt}
\begin{tabular*}
{\linewidth}{@{\extracolsep{\fill}} c c @{\hspace{0.8em}\vrule\hspace{0.8em}} c c @{\hspace{0.8em}\vrule\hspace{0.8em}} c c @{}}
\hline\hline
$E_n$, GeV & $\bar{\mathcal L}_1(E_n), {\rm pb}^{-1}$ &
$E_n$, GeV & $\bar{\mathcal L}_1(E_n), {\rm pb}^{-1}$ &
$E_n$, GeV & $\bar{\mathcal L}_1(E_n), {\rm pb}^{-1}$ \\
\hline
$4.1271$ & $41.3$ & $4.2187$ & $42.8$ & $4.2776$ & $15.7$ \\
$4.1567$ & $43.2$ & $4.2263$ & $81.2$ & $4.2866$ & $46.4$ \\
$4.1780$ & $280.0$ & $4.2357$ & $43.6$ & $4.3115$ & $52.9$ \\
$4.1888$ & $52.5$ & $4.2438$ & $49.5$ & $4.3370$ & $52.6$ \\
$4.1989$ & $41.2$ & $4.2580$ & $73.2$ & $4.3583$ & $55.5$ \\
$4.2091$ & $45.5$ & $4.2667$ & $48.8$ & & \\
\hline\hline
\end{tabular*}
\setlength{\tabcolsep}{6pt}
\renewcommand{\arraystretch}{1.0}
\end{table}
\subsection{FSIs in three-body observation channels}
\label{sec:fsi}

\begin{figure}[t!]
\includegraphics[width=\linewidth]{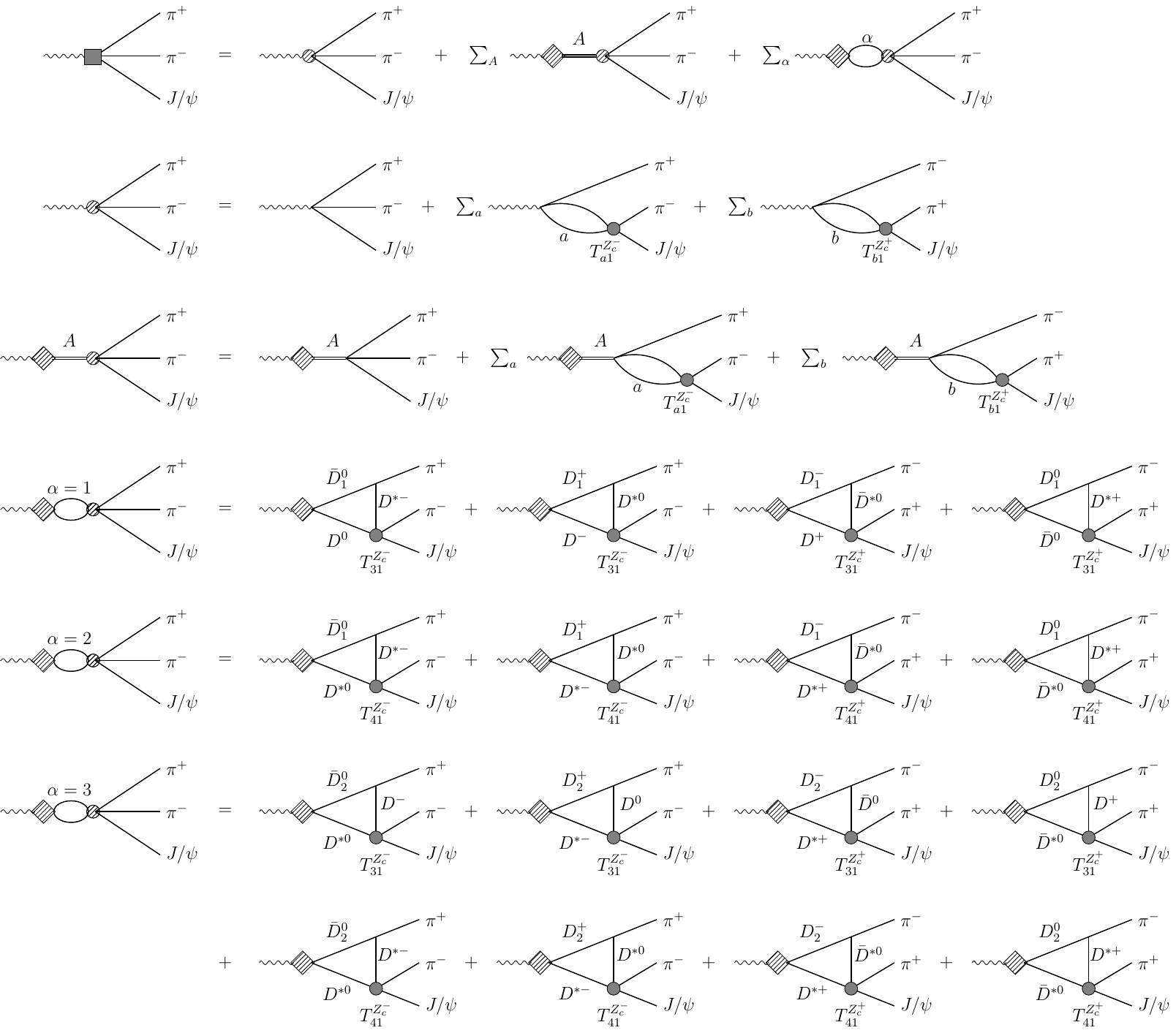}
\caption{Diagrams for the production of $J/\psi\pi^+\pi^-$, as a representative example for the observation channels with three-body final states in Eq.~\eqref{observation1}, from the virtual photon. The total amplitude is a sum of
the (i) direct production term, (ii) production through the bare charmonium poles, and (iii) production through the two-body elastic intermediate states, as given in Eq.~\eqref{M1ab}.
The internal sum in $a$ runs over two-body intermediate states enumerated in Eq.~\eqref{observation2body} and the sum in $b$ is for the respective charge-conjugate channels. The amplitudes $T^{Z_c^\pm}$'s with various lower indices are the corresponding components of the solution to the LSE in Eq.~\eqref{TZc}.}\label{fig:jpsipipi}
\end{figure}

Before performing combined fits to the experimental distributions, we need to address the effects of FSI in various two-body subsystems of the observation channels in Eq.~\eqref{observation1} with three-body final states.

\subsubsection{$Z_c$-related FSIs}
\label{sec:Tmatrix_Zc}

We start from the interactions in the two-body subsystems {of the three-body observation channels in Eq.~\eqref{observation1},}
\beq
\begin{split}
&J/\psi\pi~\mbox{($S$-wave) in observation channel 1},\\
&h_c\pi~\mbox{($P$-wave) in observation channel 2},\\
&(\bar{D}^*D)_-~\mbox{($S$-wave) in observation channel 3},\\
&\bar{D}^*D^*~\mbox{($S$-wave) in observation channel 4},
\end{split}
\label{observation2body}
\eeq
with $(\bar{D}^*D)_-\equiv\frac{1}{\sqrt{2}}(\bar{D}D^*-\bar{D}^*D)$ and $\bar{D}^*D^*$ that result in the formation of the charged exotic states $Z_c(3900)$~\cite{BESIII:2013ris,BESIII:2015cld,Belle:2013yex,BESIII:2013qmu} and $Z_c(4020)$ \cite{BESIII:2013ouc,BESIII:2014gnk,BESIII:2013mhi,BESIII:2015tix}, respectively. The corresponding interaction potential, a $4\times4$ matrix in the space of the channels in Eq.~\eqref{observation2body}, takes the form (see Appendix~\ref{app:Zc} for the derivation and further details)
\beq
V^{Z_c}=\begin{pmatrix}
0& V^{Z_c}_{(c\bar{c})\pi\leftrightarrow H\bar{H}}\\
V^{Z_c\dagger}_{(c\bar{c})\pi\leftrightarrow H\bar{H}}&
V^{Z_c}_{H\bar{H}}
\end{pmatrix},
\label{eq:VZc_phen}
\eeq
where, in the strict HQSS limit,
\beq
\left[V^{Z_c}_{(c\bar{c})\pi\leftrightarrow H\bar{H}}\right]_{\rm HQSS}=\frac 1{\sqrt2}
\begin{pmatrix}
F&F \\
-Gp_\pi&Gp_\pi
\end{pmatrix},\quad
\left[V^{Z_c}_{H\bar{H}}\right]_{\rm HQSS}=
\begin{pmatrix}
C_{1Z}&C_{34}\\
C_{34}& C_{1Z}
\label{VZcDDHQSS}
\end{pmatrix},
\eeq
The quantities $F$, $G$, $C_{1Z}$, and $C_{34}$ are four independent low-energy constants to be fitted to the data.
In the actual fitting procedure we employ an energy-dependent phenomenological parametrization that preserves the dominant channel structure above while allowing for HQSS breaking effects required by the data \cite{Chen:2023def}. Thus, instead of Eq.~\eqref{VZcDDHQSS}, we use
\beq
V^{Z_c}_{(c\bar{c})\pi\leftrightarrow H\bar{H}}=\frac 1{\sqrt2}\begin{pmatrix}
F&fF \\
-Gp_\pi&gGp_\pi
\end{pmatrix}
\eeq
and
\beq
V^{Z_c}_{H\bar{H}}=
\begin{pmatrix}
C_{1Z}+B\frac{s-M_{DD^*}^2}{2M_{DD^*}}&C_{34}\\
C_{34}& c_ZC_{1Z}+b\, B\frac{s-M_{D^*D^*}^2}{2M_{D^*D^*}}
\label{VZcDD}
\end{pmatrix},
\eeq
with
\beq
M_{DD^*}=m_{D}+m_{D^*},\quad M_{D^*D^*}=2m_{D^*}.
\eeq
Now, in addition to $F$, $G$, $C_{1Z}$, and $C_{34}$, the factors $B$, $f$, $g$, $c_Z$, and $b$ are also treated as fit parameters responsible for possible HQSS breaking effects, previously found necessary to explain the different manifestation of the $Z_c(3900)$ and $Z_c(4020)$ in different channels~\cite{Ye:2026azi}. 
We also note that, since the $Z_c$ states may possess further decay channels beyond the four channels included explicitly
here --- for example, the channel
$\eta_c\rho$~\cite{BESIII:2019rek} --- the parameter $C_{1Z}$ is allowed to acquire an imaginary part while all other parameters controlling the $Z_c$ states are adopted to be real (see also the discussion in Section~\ref{sec:parameters}).

The scattering amplitude is obtained from this potential by solving the LSE with the potential in Eq.~\eqref{eq:VZc_phen},
\begin{equation}
T^{Z_c}=\left({1-V^{Z_c}G^{Z_c}}\right)^{-1}{V^{Z_c}}.
\label{TZc}
\end{equation}
Here, $G^{Z_c}$ is a diagonal matrix of the two-point scalar loop functions for the channels in Eq.~\eqref{observation2body} evaluated in the DR scheme using the loop function defined in Eq.~\eqref{eq:GDR}. The subtraction constant $a(\mu)$ is set to $-2.8$, $-2.8$, $-3.0$, and $-3.0$ for these four channels, respectively~\cite{Du:2022jjv}. Different components of the amplitude $T^{Z_c}$ enter as building blocks in the production amplitudes for the observation channels in Eq.~\eqref{observation1} as shown in Figure~\ref{fig:jpsipipi}. Thus, the $Z_c$-related FSIs described above are included in the corresponding production amplitudes and, therefore, also in the cross sections in Eqs.~\eqref{eq:dsigds} and \eqref{xsec14}. To make this explicit, below we refer to these differential and total cross sections as $d\sigma_a^{(Z_c)}/ds$ and $\sigma_a^{(Z_c)}$, respectively.

\subsubsection{Treatment of the $\pi\pi$ FSI}\label{sec:pipiFSI}

In the observation channels $J/\psi\pi\pi$ and $h_c\pi\pi$ (channels 1 and 2 in Eq.~\eqref{observation1}), the $\pi\pi$ or $\pi\pi$-$K\bar K$ FSIs contain the scalar and tensor light mesons $f_0(500)$, $f_0(980)$ and $f_2(1270)$.
However, since we focus on the vector charmonium(-like) sector, a complete incorporation of this interaction into the present coupled-channel scheme goes beyond the scope of this work, so we resort to a simplified treatment.
In particular, the effects related to the formation of the $f_0(500)/f_0(980)$ mesons in both aforementioned observation channels are effectively included via a non-interfering term $\mathcal{B}^{\pi\pi}_a(E,s)$ in the differential cross section~\cite{Albaladejo:2015lob,Du:2022jjv}.
We used the calculation of Ref.~\cite{Chen:2023def} to verify that the interference term does not exceed approximately 5\%. This contribution is sufficiently small for the purposes of the present analysis, and its neglect is therefore justified.

We start from $J/\psi\pi^+\pi^-$ channel and represent the differential cross section as
\beq
\frac{d\sigma_1}{ds}(E,s)=
\frac{d\sigma_1^{(Z_c)}}{ds}(E,s)+\mathcal{B}_1^{\pi\pi}(E,s),
\label{BG1}
\eeq
where the contribution $d\sigma_1^{(Z_c)}/ds$ was
previously introduced in Eq.~\eqref{eq:dsigds} and the $\pi\pi$ term is parametrized as
\beq
\mathcal{B}_1^{\pi\pi}(E,s)=\frac{\sigma_1^{(f_0)}(E)}{N_{1,\pi\pi}^{\rm exp}}\int_{u_\downarrow(E,s)}^{u_\uparrow(E,s)}du \frac{\Delta N_{1,\pi\pi}^{\rm exp}(E,u)}{\Delta s(E,u)\Delta u}.
\label{eq:B1Es}
\eeq
For each fixed $E$ and $s$, the quantities $u_\downarrow(E,s)$ and $u_\uparrow(E,s)$ define the kinematical boundaries for $u$ and, similarly, $s_{\uparrow}(E,u)$ and $s_{\downarrow}(E,u)$ are the boundaries for $s$ at fixed $E$ and $u$; then $\Delta s(E,u)=s_{\uparrow}(E,u)-s_{\downarrow}(E,u)$. The experimental distribution of the numbers of $\pi\pi$ events $\Delta N^{\rm exp}_{1,\pi\pi}(E,u)$ in all $u$-bins with $\Delta \sqrt{u}=20~\text{MeV}$ and the total number of such events,
\beq
N_{1,\pi\pi}^{\exp}=\int d s\, d u\, \frac{\Delta N_{1,\pi\pi}^{\rm exp}(E,u)}{\Delta s(E,u)\Delta u},
\eeq
is taken from Ref.~\cite{BESIII:2025qkn}. Then it is easy to see that the quantity
\beq
\sigma_1^{(f_0)}(E)=\int d s\, \mathcal{B}_1^{\pi\pi}(E,s)
\eeq
admits interpretation as the contribution of the interacting pions in the $J/\psi \pi^+\pi^-$ final state, mainly through the formation of $f_0$'s. We parametrize this contribution in the form
\beq
\sigma_1^{(f_0)}(E)=\frac{|\mbox{Pol}(E)|^2}{|\mbox{det}(1-\bar VG)|^2} \ ,
\eeq
where the denominator captures the leading energy dependence provided by the $1^{--}$ states and the residual, slow dependence on the energy $E$ is encoded in a polynomial,
\beq
\mbox{Pol}(E)=\sum_{n=0}^{n_{\rm max}}\alpha_n(E-E_0)^n,
\label{eq:poly}
\eeq
with $E_0=4.2$~GeV and the $\alpha_n$'s as free parameters determined in the fit. {We have explored several values of $n_{\rm max}$ and found that, while $n_{\rm max}\geqslant 1$ is necessary to reproduce the gross features of the dipion distributions (which are not included in the combined fit), increasing $n_{\rm max}$ does not generally improve the fit quality already achieved for $n_{\rm max}=1$. We therefore choose to employ a linear function Pol$(E)$ with two fit parameters, $\alpha_0$ and $\alpha_1$.} 

There is no publicly available information on the $\pi\pi$ distributions for the channel $h_c\pi\pi$~\cite{BESIII:2025bce}. We therefore choose to neglect the $\pi\pi$ term in this channel. This is consistent with the analogous findings in the bottom sector where a prominent $\pi\pi$ FSI is observed
in the $\Upsilon(nS)\pi\pi$ ($n=1,2,3$) final states~\cite{Chen:2015jgl,Chen:2016mjn,Baru:2020ywb}, while the corresponding FSI effects in the $h_b(mP)\pi\pi$ ($m=1,2$) final states are found to be negligible~\cite{Baru:2020ywb}.

\subsection{Review of the parameters}
\label{sec:parameters}

Below we review the parameters of the coupled-channel scheme developed in this work. As was previously introduced in Section~\ref{sec:formalism}, the scheme includes $\Np$ bare poles, with $\Np$ varying between 0 and 2, $\Ne=3$ two-body elastic channels, and $\Nexp=6$ observation channels. For practical purposes, all fitted parameters can be grouped into four physically distinct classes:
\begin{itemize}
\item[(1)] short-distance production parameters introduced in Section~\ref{sec:amplitudes};
\item[(2)] effective interaction parameters in the $1^{--}$ coupled-channel sector introduced in Eq.~\eqref{eq:VY};
\item[(3)] parameters associated with the $Z_c$-related FSIs introduced in Eq.~\eqref{eq:VZc_phen};
\item[(4)] smooth {$\pi\pi$ FSI contribution} and normalization factors introduced in Eqs.~\eqref{eq:DeltaJpsipi} and \eqref{eq:poly}.
\end{itemize}
Let us briefly summarize these parameter classes and count the number of free parameters. 

We start with the short-distance production parameters. For the observation channels in Eq.~\eqref{observation1} and \eqref{observation2}, the production amplitudes contain a contribution from the direct virtual photon production terms, parametrized by $\Nexp$ real constants ${\cal P}_a^{\gamma^*}$, and the production through the bare poles, parametrized by $\Np\times\Nexp$ real constants ${\cal P}_a^A$.
The production of the two-body elastic channels in Eq.~\eqref{elastic} is parametrized by $\Ne$ constants $\Palpha$, all of which are allowed to be complex. The motivation is that, in the heavy-quark limit, the direct $S$-wave production of the two-body elastic channels from the virtual photon is suppressed~\cite{Li:2013yka} (see also Table VI in Ref.~\cite{Eichten:1978tg}) and is expected to proceed mainly through the mixing of the $D_1(2420)$ and $D_1(2430)$ states~\cite{vonDetten:2024eie}, the latter being dominated by the light-cloud configuration with $j_\ell=1/2$.\footnote{It has been suggested that the $D_1(2430)$ listed in the Review of Particle Physics (RPP)~\cite{ParticleDataGroup:2024cfk} should be replaced by two states~\cite{Guo:2006rp,Albaladejo:2016lbb}. Nevertheless, the light degrees of freedom in both states correspond to the same light-cloud configuration with $j_\ell=1/2$. As a matter of fact, the $B_1'\bar{B}$ and $B_0^*\bar{B}^*$ channels were suggested in Ref.~\cite{Wu:2018xaa} as a production mechanism for the $Z_b$ states in $\Upsilon(10860)$ decays, where $B_1'$ and $B_0^*$ are the $j_\ell^P=1/2^+$ bottom counterparts.}
Since the underlying short-distance production mechanism is not sufficiently constrained, we allow for nonvanishing phases in the corresponding source terms.
The production of the bare quark model states in Eq.~\eqref{bare} is parametrized through $\Np$ constants $\Pzero$. The magnitude of each source term can be constrained from the electronic widths of the corresponding bare $c\bar c$ state~\cite{ParticleDataGroup:2024cfk},
\beq
\begin{split}
|\Pzero |&=\sqrt{\frac{3M_{\psi}^3}{\alpha}\Gamma^{\psi}_{\rm tot}\cdot\text{BR}(\psi\to e^+e^-)}\approx\left\{\begin{array}{cc}
0.12\,{\rm GeV}^2 & \text{for the}~\psi(4160),\\
0.11\,{\rm GeV}^2 & \text{for the}~\psi(4415).
\end{array}\right.
\end{split}
\eeq
At the same time, given that the production of the $\psi(4160)$ may receive contributions from the nearby $\psi(4040)$ tail and in view of the generally complicated situation in the energy region under study, we allow all terms
$\mathcal P_{A}^{\gamma^*}$ to acquire nonvanishing imaginary parts.
Thus, altogether, the production sector of the model contains $\Nexp\times\Np+\Nexp+\Np+2\Ne=7\Np+12$ real free parameters.

We next turn to the $1^{--}$ coupled-channel interaction in the two-body elastic channels $D\bar D_1$, $D^*\bar D_1$ and $D^*\bar D_2^*$. As encoded in Eq.~\eqref{eq:VY}, the LO contact potential depends only on three independent combinations,
\begin{align}
F_{01}^c,\quad F_{01}^d,\quad \delta F_{02}\equiv F_{02}^d-F_{02}^c.
\label{eq:Fterms}
\end{align}
As discussed above, the net effect of the observation channels with lower-lying thresholds consists of a dispersive contribution to the real part of the potential and an absorptive contribution associated with the loss of flux into these channels. Since they are not included explicitly in the reduced LSE of Eq.~\eqref{LSEred}, the minimal phenomenological way to account for their effects within the reduced coupled-channel framework is to treat the parameters $F_{01}^c$, $F_{01}^d$, and $\delta F_{02}$ as complex-valued quantities. We therefore have 6 real parameters describing the effective elastic interaction. 
We do not include HQSS-violating corrections to the potential in Eq.~\eqref{eq:VY}, since the fits discussed below indicate that the currently available data do not require such corrections. 
Next, the transitions between the bare $c\bar c$ seeds and two-body elastic channels in Eq.~\eqref{elastic} are described by $\Np\times \Ne$ real couplings $V_{A\alpha}$. Two additional real parameters, $F_{J/\psi\eta}$ and $F_{\chi_{c0}\omega}$, describe the transitions from the two-body elastic channels to the two-body final states in Eq.~\eqref{observation2}, $J/\psi\eta$ and $\chi_{c0}\omega$. Thus the $1^{--}$ interaction sector contributes $3\Np+8$ real free parameters in total.

The $Z_c$-related sector of the framework introduces another set of phenomenological parameters. They control the couplings of the hidden-charm channels to the open-charm channels, the diagonal and off-diagonal open-charm interactions in the isovector $J^{PC}=1^{+-}$ subsystem, and the mild energy dependence required by the data and encoded in Eq.~\eqref{VZcDD}. As discussed in Section~\ref{sec:Tmatrix_Zc}, this sector contains the parameters $F$, $G$, $C_{1Z}$, $C_{34}$, $B$, $f$, $g$, $c_Z$, and $b$, with one of them, $C_{1Z}$, allowed to be complex to effectively account for the decay channels of the $Z_c$ that are not included explicitly in the present coupled-channel scheme. Thus, altogether, the $Z_c$-related sector contributes 10 real free parameters.

Next, we specify the normalization factors entering the invariant-mass distributions. The $J/\psi\pi^\pm$ spectrum has been reported by BESIII at 21 energy points~\cite{BESIII:2025qkn}. We include the first 17 energy points in the fit, for which the normalization factors $\bar{\mathcal L}_1(E_n)$ can be determined explicitly from the published experimental information. These values are listed in Table~\ref{tab:jpsipiNormConstants}. The $h_c\pi^\pm$ distribution~\cite{BESIII:2013ouc} is available at three energy points, with the 4.23 and 4.26~GeV samples combined. The $D^0D^{*-}$ distributions~\cite{BESIII:2015pqw} are reported at 4.23 and 4.26~GeV, and the $D^{*0}D^{*-}$ distribution~\cite{BESIII:2013mhi} is reported at 4.26~GeV. In the absence of complete information on the experimental setup in these channels, we introduce 5 independent normalization factors,
\beq
\bar{\mathcal L}_2(4.23),~
\bar{\mathcal L}_2(4.36),~
\bar{\mathcal L}_3(4.23),~
\bar{\mathcal L}_3(4.26),~
\bar{\mathcal L}_4(4.26),
\label{Ls}
\eeq
as real parameters of the fit. In Eq.~\eqref{Ls}, the subscript tags the observation channel according to Eq.~\eqref{observation1} and the beam energy in units of GeV is quoted in parentheses. In addition, to mimic the $\pi\pi$ FSI, the $J/\psi\pi\pi$ channel admits a smooth noninterfering term, $\mathcal{B}_1^{\pi\pi}$, defined in Eq.~\eqref{BG1} and parametrized by a first-order polynomial with {two} real parameters. Thus, this sector of the model contributes {7} free real parameters altogether. 

Finally, we allow the fit to vary the parameters (the mass and width) of the bare states around the initial values of the $\psi(4160)$ and $\psi(4415)$ inferred from the RPP~\cite{ParticleDataGroup:2024cfk}. This adds $2\Np$ real parameters to the scheme. Combining together the results of the above counting of the free parameters of the coupled-channel scheme, one arrives at
\beq
N_{\rm params}=12\Np+37
\eeq
as the total number of real parameters to be fitted to the data. For the versions of the model with $\Np=0,1,2$, this count gives 37, 49, and 61 parameters, respectively.

\section{Fit results and predictions}
\label{sec:results}

\subsection{Fit models and fitted line shapes}
\label{sec:models}

Within the quark model picture, the masses of the vector $2D$ and $4S$ states were calculated~\cite{Godfrey:1985xj} to be close to the observed $\psi(4160)$ and $\psi(4415)$ ones and thus reside in the studied energy region. However, as should be clear from the discussion above, the actual positions of the corresponding poles of the amplitude, and even their very appearance in the considered interval, cannot be inferred directly from na{\"i}ve quark model calculations. We therefore consider different fitting models that either include or exclude particular bare compact charmonium seeds. When included, the initial bare masses and widths of these seed states are assigned to values taken from the RPP~\cite{ParticleDataGroup:2024cfk} as an educated guess. Then the fit is allowed to adapt these values to the analyzed data as discussed in the end of Section~\ref{sec:parameters}. The properties of the corresponding physical states are extracted afterwards from the pole positions of the resulting amplitudes and their residues at these poles; see Section~\ref{sec:poles} below. Thus, to study the interplay between dynamical poles and explicit seed states, we consider four benchmark models for the reduced LSE in Eq.~\eqref{LSEred}:
\begin{itemize}
\item Model M0: a purely dynamical scenario with $\Ne=3$ two-body elastic channels in Eq.~\eqref{elastic} included and no bare poles, so $\Np=0$;
\item Model $\overline{\mbox{M0}}$: same as model M0 with $\Np=0$, however, the data are re-weighted as explained in the text below to demonstrate the potential of this ansatz, see Section~\ref{sec:poles} for details;
\item Model M1: obtained from model M0 by adding one bare pole associated with $\psi(4160)$, so $\Np=1$;
\item Model M2: obtained from model M1 by adding one more bare pole associated with $\psi(4415)$, so $\Np=2$.
\end{itemize}

\begin{table}[t]
\caption{Summary of the fit qualities for the benchmark models. Model $\overline{\mbox M0}$ is defined at the end of Section~\ref{sec:poles} through a particular re-weighting of the data in the $J/\psi\pi\pi$ and $D\bar D^*\pi$ channels, while the corresponding $\chi^2$ is evaluated then after removing the artificial weights.
}
\vspace{0.2cm}
\label{tab:chisq}
\centering
\begin{tabular}{cccc}
\hline\hline
\vphantom{{\Large A}}Model & $N_{\rm params}$ & $\chi^2$ & $\chi^2/{\rm dof}$ \\
\hline
\vphantom{{\Large A}}M0 & 37 &  2248& 1.87 \\
\vphantom{{\Large A}}M1 & 49 &1852 &  1.55 \\
\vphantom{{\Large A}}M2 & 61 & 1677 &  1.42 \\
\hline
\vphantom{{\huge A}}$\overline{\mbox M0}$ & 37 &  3913 & 3.25 \\
\hline\hline
\end{tabular}
\end{table}

\begin{figure}[t]
\centering
\includegraphics[width=\linewidth]{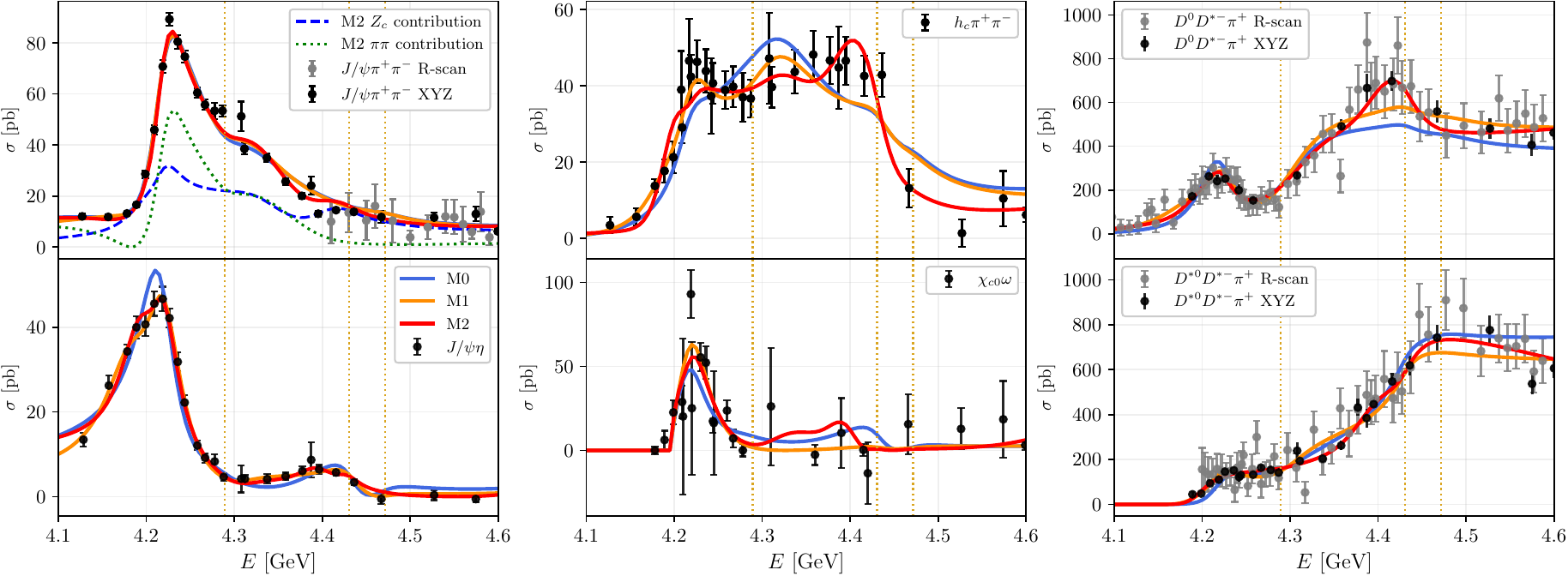}
\caption{Fitted cross sections of the $e^+e^-$ annihilation to the final states in Eqs.~\eqref{observation1} and \eqref{observation2} obtained in the three models, M0 (blue solid), M1 (orange solid), and M2 (red solid), introduced in Section~\ref{sec:models}. In the plot for $J/\psi\pi^+\pi^-$, the $Z_c$ and $\pi\pi$ individual contributions defined in Eq.~\eqref{BG1} are also shown for model M2.}
\label{fig:xsec-4model}
\end{figure}

For each model, the fit is performed simultaneously to all selected cross sections and invariant-mass distributions discussed in Section~\ref{sec:data}, comprising 1242 data points in total. For all three benchmark models, M0, M1, and M2, the qualities of the best fits are listed in Table~\ref{tab:chisq}, the corresponding values of all fitted parameters are collected in Appendix~\ref{app:params}, and the resulting cross sections are shown in Figure~\ref{fig:xsec-4model}.
From this figure, one can conclude that already the purely dynamical model M0 provides a decent overall description of the experimental line shapes, while the inclusion of the bare charmonium poles improves the fit especially in the higher-energy region of the open-charm final states.
In particular, it is instructive to observe a gradual enhancement of the fitted signal, bringing it closer to the data points in the hump region around 4.4 to 4.5~GeV in the $D\bar{D}^*\pi$ channel, as the fit acquires additional flexibility from the inclusion of extra bare poles. It is also instructive to note the different line shapes of the models in the 4.2\dash4.4~GeV region for the $h_c\pi\pi$ channel; thus improved data quality in this channel may allow for a better discrimination among the models
(this point becomes particularly clear in light of the results of model $\overline{\mbox{M0}}$, discussed below). 

\begin{figure}[th]
\centering
\includegraphics[width=\linewidth]{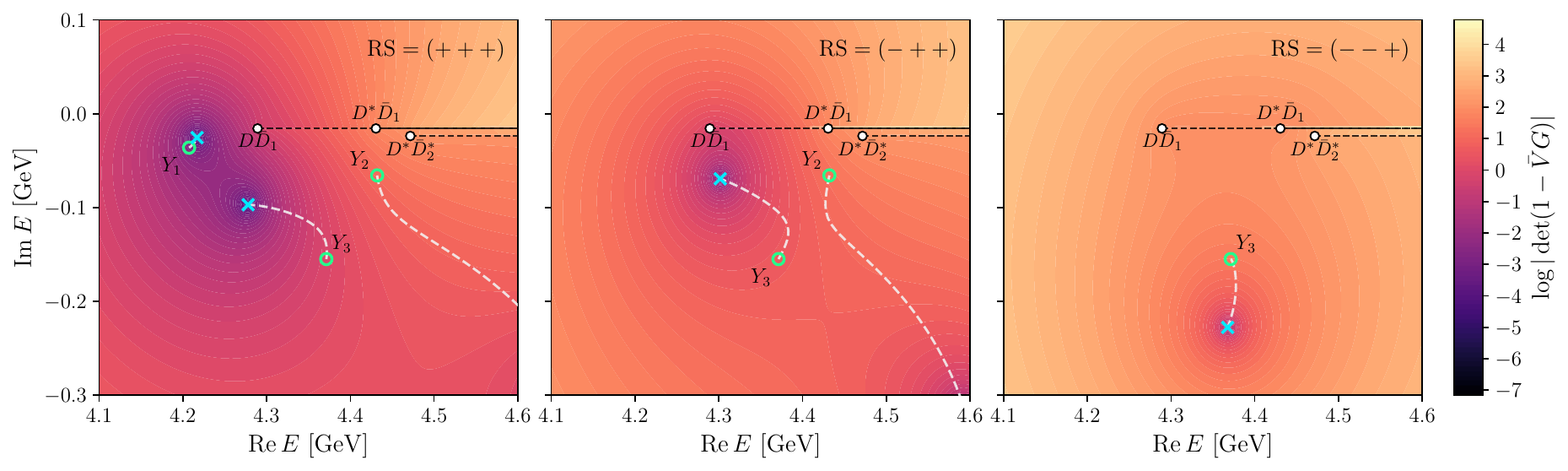}
\caption{Density plots for the fitting model M0. The pole positions in the no-coupling limit (see the main text for details) are shown as green open circles and assigned the tags $Y$ according to the serial number of the two-body elastic channel in Eq.~\eqref{elastic} in which they appear. The poles in the physical limit are shown as blue crosses; the positions of those relevant for the dynamics of the system are listed in Table~\ref{tab:four_model_poles}. The white dashed lines show the trajectories of the poles as the coupled-channel effects are gradually increased from zero to the physical value as explained in the text. }
\label{fig:YdensityplotM0variants}
\end{figure}

\subsection{Poles of the amplitude}
\label{sec:poles}
\begin{table}[t]
\caption{Poles of the amplitude and its residues at these poles for the three benchmark models introduced in Section~\ref{sec:models}. For the nomenclature of the Riemann sheets see the discussion in the main text. The uncertainties in the pole positions are statistical only.}
\label{tab:four_model_poles}
\vspace{0.2cm}
\resizebox{\linewidth}{!}{
\begin{tabular}{ccccccc}\hline\hline
Model & Pole & RS & $\sqrt{s_P}$ [MeV] & $g_1$ & $g_2$ & $g_3$ \\ \hline
\multirow{3}{*}{M0} & $\psi(4230)$ & $(+,+,+)$ & $4217(1)-25(1) i$ & $2.69(2)+0.53(3)\,i$ & $1.45(4)-1.44(4)\,i$ & $-0.11(6)+1.62(5)\,i$ \\[2mm]
 & $\psi(4320)$ & $(+,+,+)$ & $4278(3)-97(3) i$ & $1.17(5)-0.93(4)\,i$ & $-2.06(3)-1.05(4)\,i$ & $2.99(4)+0.40(5)\,i$ \\[0mm]
 & & $(-,+,+)$ & $4302(2)-69(2) i$ & $0.30(1)+0.37(1)\,i$ & $2.16(2)+0.43(2)\,i$ & $-2.23(2)-0.17(3)\,i$ \\
\hline
\multirow{4}{*}{M1} & $\psi(4160)$ & $(+,+,+)$ & $4185(3)-35(2) i$ & $1.95(12)-0.51(20)\,i$ & $0.03(17)-1.03(9)\,i$ & $-0.26(38)+0.59(8)\,i$ \\[2mm]
 & $\psi(4230)$ & $(+,+,+)$ & $4218(1)-25(1) i$ & $2.23(8)+0.83(13)\,i$ & $1.44(10)-1.13(8)\,i$ & $0.32(23)+1.25(10)\,i$ \\[2mm]
 & $\psi(4320)$ & $(+,+,+)$ & $4286(3)-85(3) i$ & $0.91(6)-0.89(7)\,i$ & $-2.12(8)-0.76(5)\,i$ & $2.85(6)+0.29(4)\,i$ \\[0mm]
 & & $(-,+,+)$ & $4304(2)-61(2) i$ & $0.27(1)+0.40(1)\,i$ & $2.16(5)+0.36(3)\,i$ & $-2.15(8)-0.13(4)\,i$ \\
\hline
\multirow{5}{*}{M2} & $\psi(4160)$ & $(+,+,+)$ & $4195(2)-26(2) i$ & $1.34(23)+0.06(12)\,i$ & $-0.13(13)-0.54(13)\,i$ & $-1.87(17)+0.49(9)\,i$ \\[2mm]
 & $\psi(4230)$ & $(+,+,+)$ & $4219(1)-20(1) i$ & $2.29(10)+0.36(6)\,i$ & $1.33(8)-1.10(6)\,i$ & $0.75(14)+1.07(11)\,i$ \\[2mm]
 & $\psi(4320)$ & $(+,+,+)$ & $4314(6)-101(6) i$ & $1.09(7)-0.58(6)\,i$ & $-2.21(8)-0.89(5)\,i$ & $2.35(7)+0.55(8)\,i$ \\[2mm]
 & & $(-,+,+)$ & $4321(3)-66(3) i$ & $0.28(1)+0.42(1)\,i$ & $2.18(7)+0.51(5)\,i$ & $-1.86(6)-0.23(4)\,i$ \\[0mm]
 & $\psi(4415)$ & $(-,+,+)$ & $4398(15)-46(5) i$ & $0.10(4)+0.19(5)\,i$ & $0.84(25)-0.28(16)\,i$ & $0.17(36)+0.21(10)\,i$ \\
 \hline\hline
\end{tabular}
}
\end{table}

\begin{table}[t]
\caption{Masses and widths of the compact charmonium states $\psi(4160)$ and $\psi(4415)$ in model M2. Here, $M_{\rm PDG}$ and $\Gamma_{\rm PDG}$ are taken from the RPP~\cite{ParticleDataGroup:2024cfk} and used as initial values for the fit, while $M_0$ and $\Gamma_0$ denote the bare values returned by the fit. Finally, $M_{\rm pole}$ and $\Gamma_{\rm pole}$ are obtained as the real part and minus twice the imaginary part of the corresponding pole of the amplitude, respectively. The latter quantities are interpreted as the physical mass and width and are also quoted in Table~\ref{tab:four_model_poles}. All values are given in MeV. }
\vspace{0.2cm}
\label{tab:masswidths}
\centering
\begin{tabular}{l|cc}
\hline\hline
charmonium & $\psi(4160)$ & $\psi(4415)$ \\
\hline
\vphantom{{\Large A}}$(M_{\rm PDG},\ \Gamma_{\rm PDG})$& $(4190,70)$ & $(4415,110)$ \\
$(M_0,\ \Gamma_0)$ & $(4204,53)$ & $(4406,80)$\\ 
$(M_{\rm pole},\ \Gamma_{\rm pole})$& $(4195,52)$ & $(4398,92)$ \\
\hline\hline
\end{tabular}
\end{table}

\begin{table}[t]
\caption{Masses and widths of the charmonium and charmonium-like states residing in the mass range from 4.1 to 4.6~GeV. For the present work, taken as a representative example, we cite the results for M2 collected in Table~\ref{tab:four_model_poles}.
The virtual state with the mass $4216.2\pm 0.5$~MeV and width $40.3\pm 1.0$~MeV found in Ref.~\cite{Nakamura:2023obk} in the channel $D_s^*\bar{D}_s^*$ is not quoted since the sector of charmed--strange mesons is not considered in this work. All masses and widths are given in MeV.}
\vspace{0.2cm}
\label{tab:compare}
\resizebox{\linewidth}{!}{
\begin{tabular}{l|c c c c c c}
\hline\hline
State & $M$ (this work) & $\Gamma$ (this work) &
$M$ \cite{Nakamura:2023obk} & $\Gamma$ \cite{Nakamura:2023obk} &
$M_{\rm PDG}$ \cite{ParticleDataGroup:2024cfk} & $\Gamma_{\rm PDG}$ \cite{ParticleDataGroup:2024cfk}
\\
\hline
$\psi(4160)$ & $4195\pm4$ & $52\pm4$ & $4192.2\pm 2.2$& $129.3\pm 4.2$ & $4191\pm 5$ & $69\pm 10$ \\
$\psi(4230)$ & $4219\pm1$ & $40\pm2$ & $4229.9\pm 0.9$ & $46.4\pm 2.6$ & $4222.2\pm 2.4$& $51\pm 8$ \\
$\psi(4320)$ & $4321\pm3$ & $132\pm6$ & $4308.1\pm 2.2$ & $138.2\pm 4.4$ & --- & --- \\
$\psi(4360)$ & --- & ---& $4346.2\pm 3.8$& $122.8\pm 6.7$& $4374\pm 7$ & $120\pm 12$ \\
$\psi(4415)$ & $4398\pm15$ & $92\pm10$ & $4390.1\pm 2.0$ & $106.5\pm 4.1$ & $4415\pm 5$ & $110\pm 13$ \\
\hline\hline
\end{tabular}
}
\end{table}

With the full multichannel amplitude constructed above, we are now in a position to compare the pole content of the models introduced in Section~\ref{sec:models}. The results are summarized in Table~\ref{tab:four_model_poles}. The sheets of the Riemann surface in the complex energy plane (we use RS as shorthand notation for them) are labeled according to the signs of the imaginary parts of the momenta in the two-body elastic channels in Eq.~\eqref{elastic}: the positive sign is denoted by ``$+$'', while the negative sign is denoted by ``$-$''. We therefore consider $2^3=8$ RSs in total. 
Notice that, in the presence of observation channels in Eqs.~\eqref{observation1} and \eqref{observation2} and inelastic light-hadron channels that introduce imaginary parts into the contact-term parameters in Eq.~\eqref{eq:Fterms}, the poles acquire imaginary parts even on the RS $(+++)$, which we hereafter refer to as the physical sheet, since it would become such in the absence of the aforementioned channels. For each found pole $s_P$, its couplings $g_\alpha$ to the two-body elastic channels are defined through the residues of the amplitude,
\beq
T_{\alpha\beta}(s)\mathop{=}^{s\to s_P}\frac{g_\alpha g_\beta}{s-s_P},
\label{gs}
\eeq
where the indices $\alpha,\beta=1,2,3$ enumerate the two-body elastic channels.

The purely dynamical model M0, which, as mentioned above, already provides a decent overall description of the data, generates two dynamical poles on the $(+++)$ and $(-++)$ sheets of the Riemann surface; see Table~\ref{tab:four_model_poles} for the pole positions and their couplings to the two-body elastic channels defined in Eq.~\eqref{gs}.
To gain insight into the physics of these dynamical poles, and in particular the role of coupled-channel effects, we fix the parameters of model M0 and then artificially rescale the off-diagonal terms in the potential $V_{\alpha\beta}$ ($\alpha,\beta=1,2,3$) (see Eq.~\eqref{eq:VY}) by a factor $\xi\in[0,1]$. 
This procedure provides a 
smooth interpolation (without refitting the other parameters) between a limit of decoupled two-body elastic channels and the physical coupled-channel scenario. In Figure~\ref{fig:YdensityplotM0variants}, we show the density plots for the representative Riemann sheets obtained 
in the dynamical model M0, overlaid with the pole trajectories as the parameter $\xi$ is varied from 0 to 1. The trajectories start from the decoupling limit, indicated by green open circles $(\xi=0)$, and evolve toward the physical situation, marked by blue crosses $(\xi=1)$.
In the decoupling limit, in agreement with the expectations in the hadronic molecular picture based on, for example, one-vector-meson-exchange model~\cite{Dong:2021juy},\footnote{Notice that the HQSS-friendly potential in Eq.~\eqref{eq:VY} for the $S$-wave contact interaction among $D^{(*)}\bar D_{1,2}$ channels~\cite{Guo:2017jvc} contains three independent parameters; therefore, additional dynamical arguments are required to infer the existence of other molecular states from a single observed one.} three states are found near the $D\bar D_1$, $D^*\bar D_1$, and $D^*\bar D^*_2$ thresholds. They are labeled as $Y_1$, $Y_2$, and $Y_3$, respectively, and represent bound states, that is, poles on the physical RS for each decoupled two-body elastic channel introduced above, although the real part of $Y_2$ is larger than that of $Y_3$. 
In Figure~\ref{fig:YdensityplotM0variants}, these correspond to $Y_1$ in the first plot, $Y_2$ in the second plot, and $Y_3$ in the third plot.
As the channels become coupled and the coupling strength increases, the poles move such that, in the physical case 
($\xi=1$):
\begin{itemize}
\item The pole $Y_1$ remains nearly stable and represents the physical state $\psi(4230)$.
\item The pole $Y_2$ follows a long trajectory and eventually leaves the near-threshold region, so that it no longer affects the system dynamics and therefore does not lead to any observable effects that can be assigned to a physical state.
\item The pole $Y_3$ on RS $(--+)$ moves deeper into the complex plane, away from the physical region. In contrast, the shadow poles associated with $Y_3$ on RSs $(+++)$ and $(-++)$, which emerge due to coupled-channel effects, undergo a strong downward shift in mass and, in the physical limit $\xi=1$, both correspond to the same physical state $\psi(4320)$.
\end{itemize}
Thus, in the physical limit, only three near-threshold poles remain relevant for the system dynamics, one of them representing $\psi(4230)$ and the other two corresponding to $\psi(4320)$, significantly shifted downward in mass from the $D^* \bar D_2^*$ threshold by strong coupled-channel effects. The positions of these poles and the
respective residues are quoted in Table~\ref{tab:four_model_poles}. 

While the appearance of shadow poles on different Riemann sheets is a generic feature of coupled-channel systems~\cite{Eden:1964zz,Zhang:2024qkg}, the pole positions, residues, and trajectories shown in Figure~\ref{fig:YdensityplotM0variants} suggest that both poles associated with the $\psi(4320)$ play equally important roles in observables. 
We therefore conclude that coupled-channel dynamics plays an essential role in understanding the nature of the enigmatic $Y$ states in the charmonium spectrum. 
In particular, we observe a nontrivial two-pole realization of the $\psi(4320)$ structure dynamically generated by coupled-channel effects. 
These two poles should not be interpreted as two independent resonances. 
Rather, they originate from the same dynamical state in the decoupling limit and become companion, or shadow, poles on different Riemann sheets once the elastic channels are coupled. 
Their similar distances to the physical region and sizable residues indicate that both poles can contribute to the same observed enhancement around 4.32~GeV.

This situation is very different from the well-known two-pole scenarios discussed for systems such as the $\Lambda(1405)$~\cite{Oller:2000fj,Jido:2003cb,Ikeda:2012au,Mai:2020ltx,Lu:2022hwm,Guo:2023wes,BaryonScatteringBaSc:2023zvt,He:2026mkf} and $D_0^*(2300)$~\cite{Kolomeitsev:2003ac,Guo:2006fu,Liu:2012zya,Guo:2015dha,Guo:2018tjx,Du:2020pui,Zhuang:2026lta}. 
In those cases,
the corresponding two poles are
associated with two different underlying dynamical components, for example different SU(3) multiplets or different dominant channel configurations, and therefore are regarded as two nearby physical states~\cite{Meissner:2020khl}. 
In the present case, by contrast, the two poles associated with $\psi(4320)$ are generated from the same underlying dynamical state and represent its coupled-channel shadow-pole manifestation rather than two distinct vector charmonium-like states.

Extending the fitting model from M0 to M1 by adding a single bare pole in the lower part of the spectrum predictably introduces an additional pole of the physical amplitude at about $4185-35i$~MeV corresponding to the physical (``dressed'') charmonium state $\psi(4160)$. The quality of the fit improves from $\chi^2/{\rm dof}=1.87$ for model M0 to $\chi^2/{\rm dof}=1.55$ for model M1. Importantly, the two dynamical physical states predicted by model M0 remain in model M1 and undergo only moderate shifts in both mass and width, as can be seen from Table~\ref{tab:four_model_poles}.

The best formal description of the data, with $\chi^2/{\rm dof}=1.42$, is achieved in model M2, in which both bare poles are included: one near the lower end of the studied energy interval, around 4.1~GeV, and the other near its upper end, around 4.4~GeV. Similarly to the case of model M1, both bare quark model states become dressed and produce two poles of the amplitude at approximately $4195-26i$~MeV and $4398-46i$~MeV, corresponding to the charmonia $\psi(4160)$ and $\psi(4415)$, respectively. Remarkably, the physical masses and widths of these charmonium states, extracted from the corresponding poles of the amplitude, deviate only mildly from both the PDG values used as initial input and the bare values returned by the fit; see Table~\ref{tab:masswidths}.
This finding, 
which indicates that the coupling of the additional states to the channels already included is weak,
is consistent with the above statement that the gross features of all data sets studied are already described by the two-body elastic channels; the two additional states only contribute to some finer details of the observed structures. 

Thus, the comparison among the three models employed, M0, M1, and M2, suggests the following physical picture:
\begin{itemize}
\item The structures associated with the $\psi(4230)$ and the$\psi(4320)$ are robust and are generated predominantly by the coupled-channel dynamics of the explicitly included two-body open-charm channels.
\item Incorporating the compact charmonium states $\psi(4160)$ and $\psi(4415)$ into the coupled-channel scheme provides a quantitatively better global description of the data, but does not qualitatively alter the above pattern of the dynamically generated poles.
\end{itemize}
In Table~\ref{tab:compare}, we compare the masses and widths of the vector charmonia obtained in this work and listed in Table~\ref{tab:four_model_poles} (for definiteness, we refer to model M2) with those reported in Ref.~\cite{Nakamura:2023obk} and in the RPP~\cite{ParticleDataGroup:2024cfk}. We find good agreement between our results and those of Ref.~\cite{Nakamura:2023obk} for the states $\psi(4160)$, $\psi(4230)$, and $\psi(4415)$, except that in our study the $\psi(4160)$ is much narrower than in~\cite{Nakamura:2023obk}, making the width consistent with the value reported by the PDG. At the same time, equipped with insights into the dynamics of the strongly coupled two-body elastic channels, we identify the two companion poles in the vicinity of 4.32~GeV as a single physical state, $\psi(4320)$. This contrasts with the interpretation in Ref.~\cite{Nakamura:2023obk}, where two distinct states, $\psi(4320)$ and $\psi(4360)$, were reported. We also note that comparison with the PDG values given in the last two columns of Table~\ref{tab:compare} should be made with caution, since these values are extracted independently in different experimental analyses that rely largely on Breit--Wigner fits. 

\begin{figure}[t!]
\centering
\includegraphics[width=\linewidth]{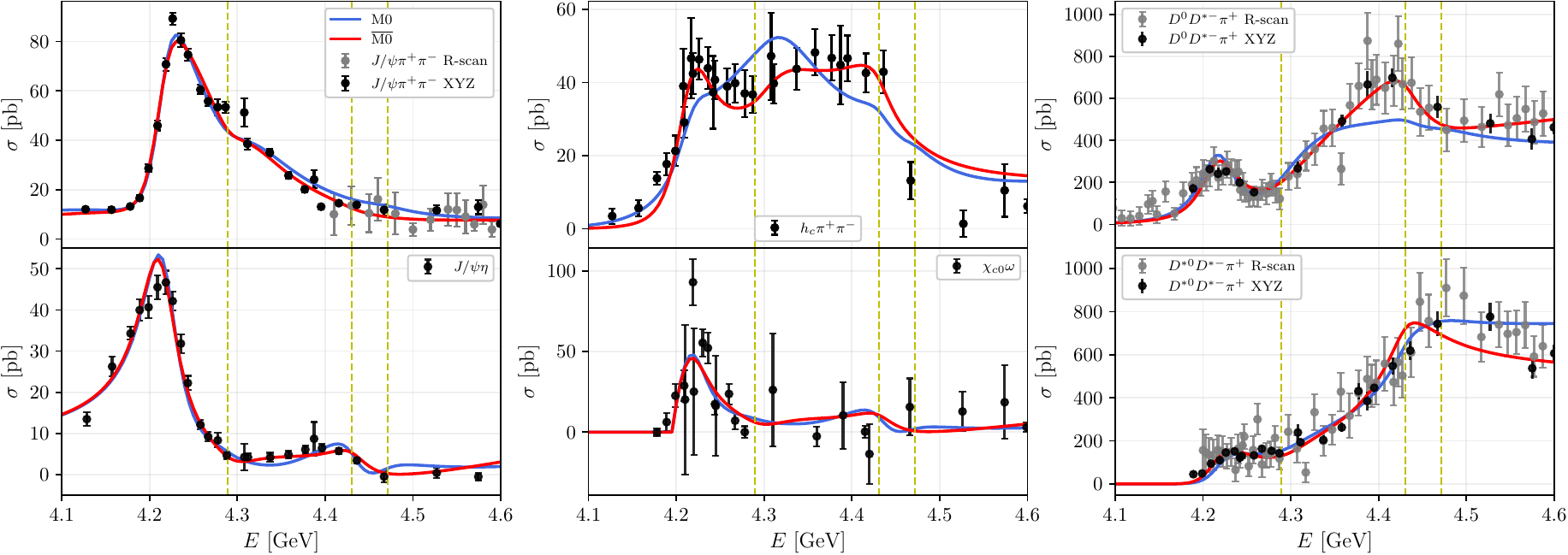}
\caption{Comparison of the fitted cross sections in the observation channels in Eqs.~\eqref{observation1} and \eqref{observation2} obtained in purely dynamical models M0 and $\overline{\mbox{M0}}$. {For the fit $\overline{\mbox{M0}}$, we introduced weights for selected data points, as discussed in the text, while the plot shows the original experimental data.
}}
\label{fig:crosssectionM0variants}
\end{figure}

\begin{table}[t]
\caption{The poles of the amplitude and its residues at these poles for the purely dynamical models M0 and $\overline{\mbox M0}$. Note that, to facilitate comparison of the models, here we quote all poles of the amplitude rather than only those relevant for the dynamics of the system listed in Table~\ref{tab:four_model_poles}.}
\label{tab:poles}
\vspace{0.2cm}
\resizebox{\linewidth}{!}{
\begin{tabular}{lccccc}
\hline\hline
Solution & RS & $\sqrt{s_p}$~[MeV] & $g_1$ & $g_2$ & $g_3$ \\
\hline
\multirow{4}{*}{M0} & \vphantom{{\Large A}}$(+,+,+)$ & $4217(1)-25(1) i$ & $2.69(2)+0.53(3)\,i$ & $1.45(4)-1.44(4)\,i$ & $-0.11(6)+1.62(5)\,i$ \\
 & \vphantom{{\Large A}}$(+,+,+)$ & $4278(3)-97(3) i$ & $1.17(5)-0.93(4)\,i$ & $-2.06(3)-1.05(4)\,i$ & $2.99(4)+0.40(5)\,i$ \\
 & \vphantom{{\Large A}}$(-,+,+)$ & $4302(2)-69(2) i$ & $0.30(1)+0.37(1)\,i$ & $2.16(2)+0.43(2)\,i$ & $-2.23(2)-0.17(3)\,i$ \\
 & \vphantom{{\Large A}}$(-,+,+)$ & $4594(43)-315(46) i$ & $0.19(5)+0.52(6)\,i$ & $1.75(15)+1.71(12)\,i$ & $1.88(11)+1.62(8)\,i$ \\
\hline
\multirow{4}{*}{$\overline{\mbox M0}$} & \vphantom{{\Large A}}$(+,+,+)$ & $4214(1)-26(1) i$ & $2.98(3)+0.42(3)\,i$ & $0.99(5)-1.76(5)\,i$ & $-0.01(6)+1.93(6)\,i$ \\
 & \vphantom{{\Large A}}$(+,+,+)$ & $4274(2)-77(2) i$ & $0.94(5)-1.32(5)\,i$ & $-2.46(4)-0.65(4)\,i$ & $3.07(5)+0.40(5)\,i$ \\
 & \vphantom{{\Large A}}$(-,+,+)$ & $4301(1)-58(1) i$ & $0.32(1)+0.39(1)\,i$ & $2.21(2)+0.18(1)\,i$ & $-2.19(3)-0.38(2)\,i$ \\
 & \vphantom{{\Large A}}$(-,+,+)$ & $4478(7)-84(9) i$ & $0.20(2)-0.37(2)\,i$ & $-1.05(5)-1.08(4)\,i$ & $-1.54(8)-0.66(5)\,i$ \\
\hline\hline
\end{tabular}
}
\end{table}

One last comment is in order here. Comparison of the fitted line shapes in the three-body open-charm channel $D\bar{D}^*\pi$ in Figure~\ref{fig:xsec-4model} may leave an impression that the visible enhancement around 4.4~GeV calls for the inclusion of a bare pole associated with the charmonium $\psi(4415)$ and cannot be described otherwise, in particular within the purely dynamical model M0. This conclusion; however, would be misleading and merely reflects the very different weight of the data sets analyzed simultaneously.
Actually, the $J/\psi\pi\pi$ channel carries a much larger weight in the global fit, since the 17 energy points are supplemented by the corresponding $J/\psi\pi$ invariant-mass distributions. By contrast, the $D\bar{D}^*\pi$ channel is constrained by only a small number of total-cross-section measurements and therefore has a much weaker impact on the fit.
To balance the relative impact of these two channels and emphasize an essential role played by the data 
weight in the various open-charm three-body channels in the studied energy range, we consider a modified purely dynamical model $\overline{\mbox{M0}}$. This fitting model is identical to model M0 discussed above but
\begin{itemize}
\item the contribution to $\chi^2$ from the channel $J/\psi\pi\pi$ is artificially suppressed by a factor of 0.025 and
\item the contribution to $\chi^2$ from the channel $D\bar D^*\pi$ above 4.35~GeV is artificially enhanced by a factor of 5.
\end{itemize}
In this way, we reduce the dominance of the $J/\psi\pi\pi$ channel in the global fit, driven mainly by the many additional $J/\psi\pi$ invariant-mass-distribution data points, and partially relax the constraints that this channel imposes on the fit.
At the same time, we enhance the weight of the data in the open-charm $D\bar D^*\pi$ channel around the peak structure at approximately 4.4~GeV clearly visible in this data set.
Although the overall fit quality for $\overline{\mbox{M0}}$ deteriorates, mainly driven by the $J/\psi\pi$ invariant mass distributions,
the fitted line shapes in the modified model $\overline{\mbox{M0}}$ still provide a good visual description of the data and, most importantly, the fit is now capable of reproducing the structure around 4.4~GeV in the $D\bar D^*\pi$ channel. This fit also captures the nontrivial energy dependence in the $h_c\pi\pi$ channel; see Figure~\ref{fig:crosssectionM0variants}. 
In Table~\ref{tab:poles}, the poles obtained in models M0 and $\overline{\rm M0}$ are compared. The first three poles and their residues are found to be consistent between the two models. As discussed above, in the decoupling limit these poles are associated with $D \bar D_1$ and $D^*\bar D_2^*$ quasibound states, respectively, with the
latter being shifted significantly towards lower energies by strong coupled-channel effects. 
However, model $\overline{\rm M0}$ features an additional dynamical state that can be traced back, in the decoupling limit, to a $D^*\bar D_1$ quasibound state and generates an additional 
pronounced bump structure near 4.4 GeV in the $h_c\pi\pi, D\bar D^*\pi,D^*\bar D^*\pi$ cross sections. In contrast, the corresponding $D^*\bar D_1$ pole in model M0 is located far from the physical axis and therefore 
 has only a minor impact on the line shapes. 

The results of model $\overline{\rm M0}$ 
highlight the need for more sufficiently precise $D\bar D^*\pi$ data before definitive conclusions can be drawn regarding the compact-state content of the charmonium(-like) spectrum in the studied energy range.

\subsection{Invariant-mass distributions}
\label{sec:invmasses}

\begin{figure}
\centering
\includegraphics[width=\linewidth]{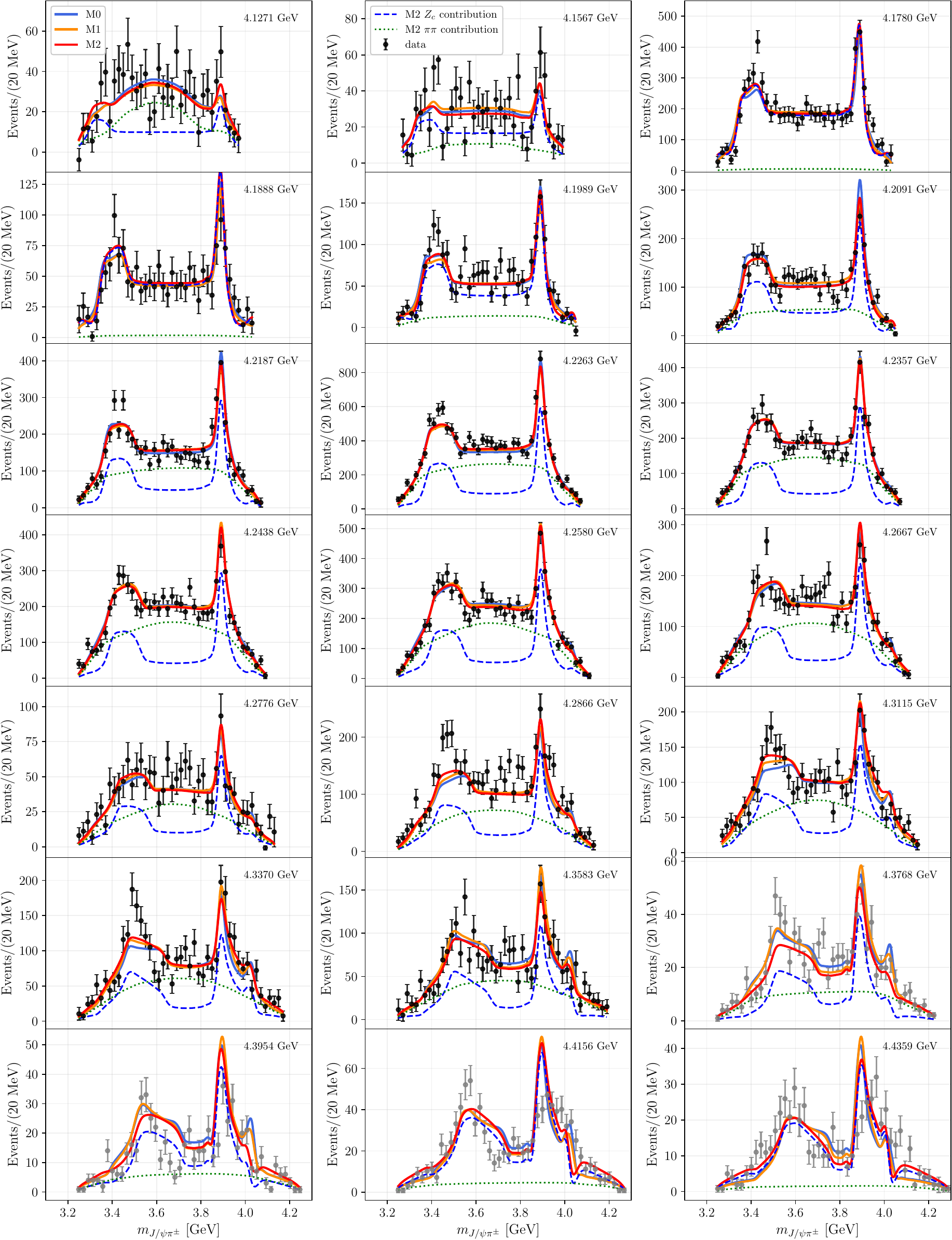}
\caption{{Invariant mass distributions at the BESIII energies listed in Table~\ref{tab:jpsipiNormConstants} for the $J/\psi\pi^\pm$ two-body subsystem in the observation channel 1 of Eq.~\eqref{observation1}, corresponding to the best fits obtained in M0 (blue solid), M1 (orange solid), and M2 (red solid), introduced in Section~\ref{sec:models}. In each plot, the full result is accompanied by the $Z_c$ and smooth $\pi\pi$ contributions defined in Eq.~\eqref{BG1} for Model M2. The last four data sets (shown in gray) were not included in the fit since they are not corrected for efficiency.}}
\label{fig:jpsipi}
\end{figure}

\begin{figure}[t]
\centering
\includegraphics[width=\linewidth]{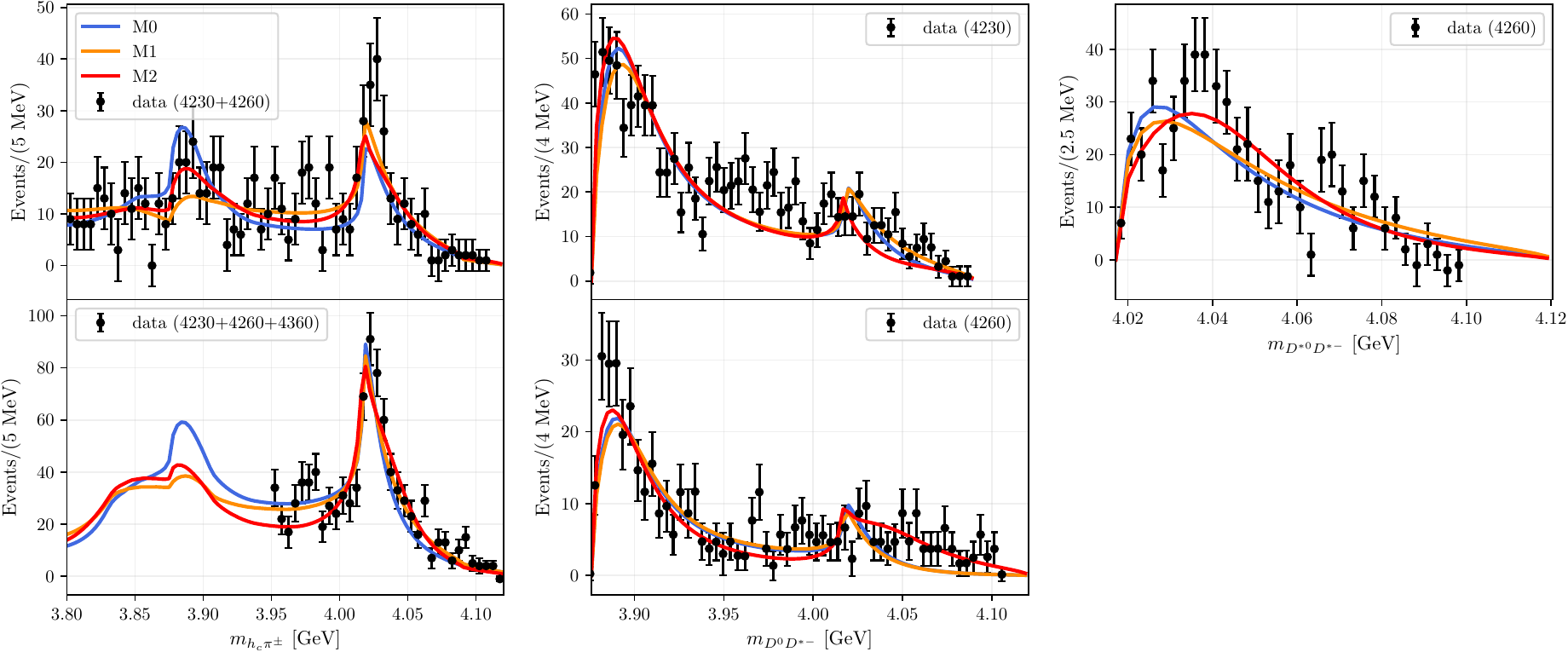}
\caption{Fitted invariant-mass distributions in the two-body subsystems $h_c\pi^\pm$, $D^0D^{*-}$, and $D^{*0}D^{*-}$ of the corresponding observation channels in Eq.~\eqref{observation1} at the c.m. energies 4.23, 4.26 and/or 4.36~GeV.}
\label{fig:zc2}
\end{figure}

The profiles of the invariant-mass distributions provide an important additional check of the validity of the coupled-channel scheme constructed in this work. In particular, the plots in Figure~\ref{fig:jpsipi} demonstrate that all benchmark models reproduce the $J/\psi\pi^\pm$ enhancement associated with the $Z_c(3900)$, while the dependence on the fitting model remains mild in this channel. This observation is further supported by Table~\ref{tab:zc_poles_four_models}, where the extracted poles corresponding to the $Z_c(3900)$ and $Z_c(4020)$ are shown to remain largely stable across all fitting models, with only a moderate effect on the imaginary parts in M2. Such stability indicates that the $Z_c$ sector is governed predominantly by near-threshold coupled-channel dynamics and is therefore much less sensitive to the detailed bare-pole content of the model. The invariant-mass distributions in the $h_c\pi^\pm$, $D^0D^{*-}$, and $D^{*0}D^{*-}$ channels shown in Figure~\ref{fig:zc2}
also confirm that the current data are generally well described by all three models and do not allow for a clear distinction among them.

\begin{table}[h]
\caption{The $Z_c$'s pole positions (in MeV) extracted in the three benchmark models and quoted from the RPP in the form of $(M-i\,\Gamma/2)$~\cite{ParticleDataGroup:2024cfk}.}
\vspace{0.2cm}
\label{tab:zc_poles_four_models}
\centering
\begin{tabular}{lcc}
\hline\hline
Model & $Z_c(3900)$ & $Z_c(4020)$\\
\hline
\vphantom{{\Large A}}M0 & $3875(1)-i\,28(2)$ & $4008(2)-i\,14(1)$ \\
M1 & $3876(2)-i\,34(3)$ & $3998(6)-i\,13(7)$ \\
M2 & $3868(4)-i\,43(6)$ & $4036(2)-i\,35(5)$ \\
PDG & $3887(3)-i\,14(1)$ & $4024(1)-i\,7(3)$ \\

\hline\hline
\end{tabular}
\end{table}

\begin{figure}
\centering \includegraphics[width=0.85\linewidth]{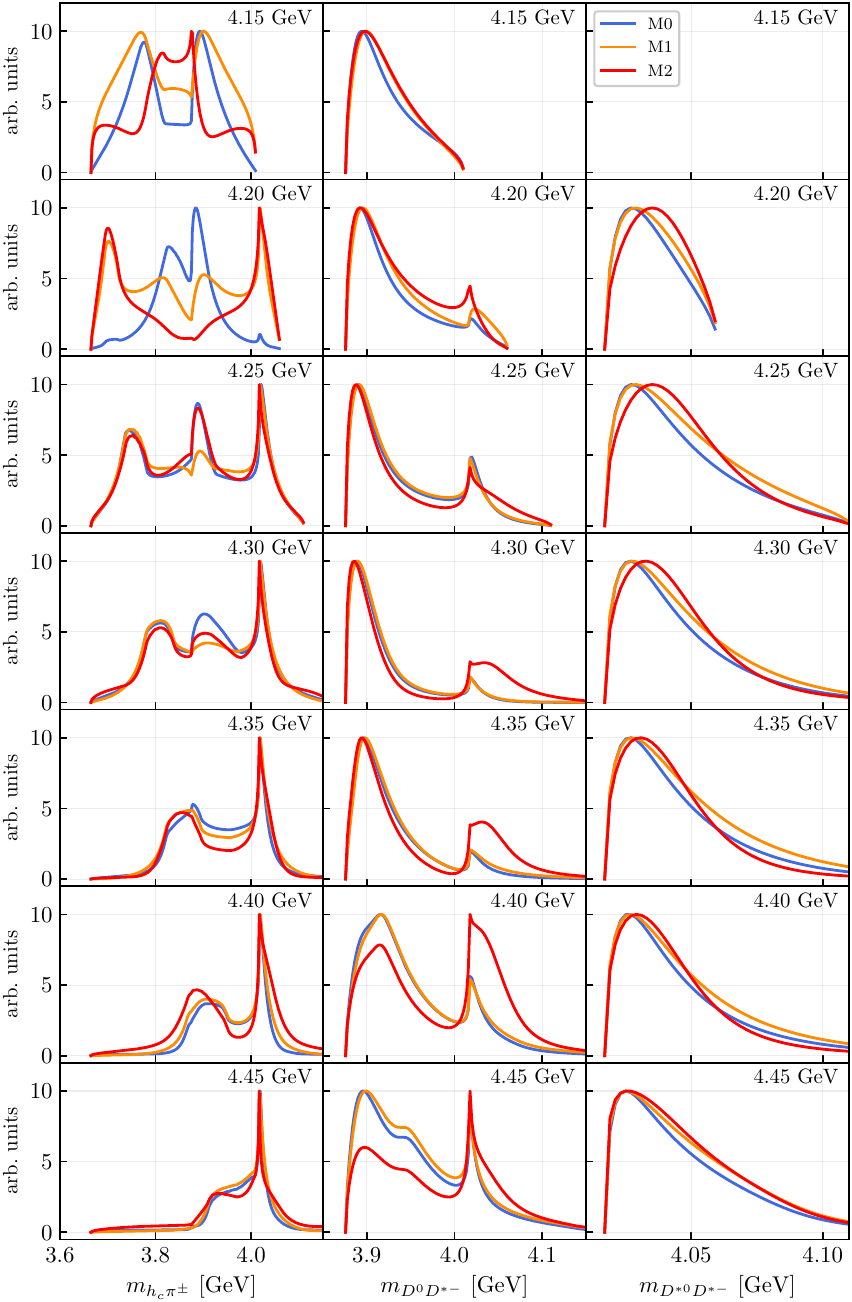}
\caption{Predicted invariant-mass distributions in the two-body subsystems $h_c\pi^\pm$, $D^0D^{*-}$, and $D^{*0}D^{*-}$ of the corresponding observation channels in Eq.~\eqref{observation1} obtained for selected values of the c.m. energy, explicitly indicated in each plot. Note that there is no phase space for the $D^*\bar D^*\pi$ final state at $E=4.15$~GeV, so the corresponding slot in the upper right corner of the figure is left blank. 
}
\label{fig:pre_selected_E}
\end{figure}

With all parameters fully fixed by the global fit, the present framework allows one to predict the invariant-mass distributions in channels and energy regions where data are sparse or not yet available. Figure~\ref{fig:pre_selected_E} shows the predictions at several selected energies, while Figure~\ref{fig:pre_all_E} in Appendix~\ref{app:predictions} presents the corresponding results for all 21 BESIII energy points, analogous to those shown in Figure~\ref{fig:jpsipi}. We note, in particular, that the low-mass enhancements associated with the $Z_c$ sector are present in all models, which is again consistent with the stability of the two $Z_c$ poles listed in Table~\ref{tab:zc_poles_four_models}. However, the strength and energy dependence of the higher-mass tails differ noticeably among the models. The comparison therefore shows that some features are robust, while others are clearly model-dependent and may thus provide a means to discriminate between the models in the future, once the missing data become available. These predictions are especially relevant and timely, since future measurements of the $h_c\pi^\pm$, $D^0D^{*-}$, and $D^{*0}D^{*-}$ invariant-mass distributions at higher energies should allow for a much sharper distinction between the different scenarios considered in this work. 

\subsection{Predictions for the HQSS partners}
\label{sec:HQSSprediction}

HQSS relates the short-range interactions in different open-charm channels. In addition to the three $1^{--}$ channels investigated in this work, the $\{D,D^*\}$--$\{\bar D_1,\bar D_2^*\}$ systems give rise to channels with quantum numbers $J^{PC}=0^{-\pm},1^{-+},2^{-\pm}$, and $3^{-\pm}$. The corresponding short-range interaction potentials can all be expressed in terms of the four contact interactions introduced in Eq.~\eqref{eq:FIjl}; see Ref.~\cite{Guo:2017jvc} for details.
The channels with $J^{PC}=0^{--}$, $1^{-+}$, and $3^{-+}$ are of particular interest, since these quantum numbers cannot be realised by conventional $c\bar c$ charmonia and are therefore manifestly exotic. In particular, a $0^{--}$ $D^*\bar D_1$ bound state was predicted in Ref.~\cite{Ji:2022blw}, while the $1^{-+}$ $D\bar D_1$ system has been studied in Refs.~\cite{Li:2013bca,Wang:2014wga,Dong:2019ofp,Zhang:2025gmm}. For a discussion of HQSS partners in the compact tetraquark model~\cite{Maiani:2014aja}, we refer to Ref.~\cite{Cleven:2015era}.  Experimental searches for such exotic states have recently been carried out by the BESIII Collaboration~\cite{BESIII:2025bez,BESIII:2026dku}. Although no statistically significant signal has been observed so far, the current data do not yet allow one to draw definitive conclusions regarding their existence.
Interestingly, in the hidden-strangeness sector, the corresponding $0^{--}$ molecular state may have manifested itself in the decay $J/\psi\to\phi\eta\eta'$~\cite{BESIII:2018zbm,Dong:2024fjk}.

In this work, we focus on three channels whose contact interactions are determined by three independent combinations of the contact terms introduced in Eq.~\eqref{eq:Fterms} and constrained by the fit,
\beq
\begin{split}
V(0^{--},D^*\bar D_1) &= F^d_{01}+F^c_{01},\\
V(0^{-+},D^*\bar D_1) &=F^d_{01}-F^c_{01},\\
V(3^{--},D^*\bar D_2^*) &=\delta F_{02}.
\label{Fs}
\end{split}
\eeq
Meanwhile, a disclaimer is in order here. The absorptive parts of the contact terms in Eq.~\eqref{eq:Fterms}, extracted from the analysis of the $1^{--}$ sector, effectively parameterize the inelastic channels relevant to that sector. In general, however, the inelastic channels associated with other $J^{PC}$ quantum numbers are different, so the corresponding absorptive contributions need not be the same.
By contrast, the real parts of the contact terms are expected to be more universal across HQSS-partner channels. We therefore use them to estimate the strength of the short-range interaction in the channels considered here. For each set of quantum numbers $J^{PC}$, we solve the corresponding single-channel pole equation and determine the binding energy\footnote{For virtual states, this is not a binding energy in the strict sense, but only a measure of the pole distance from threshold.} according to
\begin{equation}
E_B=M_{\rm th}-\operatorname{Re}E_{\rm pole}.
\end{equation}
The results are summarized in Table~\ref{tab:single-channel-poles}, where ``$+$'' and ``$-$'' denote the first and second Riemann sheets, respectively, and a brief discussion of the individual channels follows below.

\begin{itemize}
\item The prediction for the $3^{--}$ $D^*\bar D_2^*$ partner is remarkably stable across the different models. All  solutions yield a strongly attractive interaction and predict a pole on the first Riemann sheet with a binding energy of approximately $90$--$100~\mathrm{MeV}$. This makes the existence of the $3^{--}$ partner a robust consequence of the interaction constrained by the data.
\item The $0^{-+}$ $D^*\bar D_1$ partner is also found to be bound in all considered solutions. Although, varying from
$47~\mathrm{MeV}$ to $81~\mathrm{MeV}$,
its binding energy exhibits a larger spread than in the $3^{--}$ channel, all models nevertheless predict an attractive interaction strong enough to support a bound state.
\item The predictions for the $0^{--}$ $D^*\bar D_1$ channel are the least conclusive. In the M0, M1, and M2 solutions, the corresponding pole is located on the second Riemann sheet. In models M0 and M1, however, the attraction is relatively weak, resulting in poles residing more than $100~\mathrm{MeV}$ below the $D^*\bar D_1$ threshold; these poles are not included in Table~\ref{tab:single-channel-poles}. There are two reasons for disregarding them. First, they lie well outside the near-threshold region where the leading-order contact potential can be expected to provide a reliable description. Second, such remote poles on the second Riemann sheet have little impact on the physical observables. In model $\overline{\mathrm{M0}}$, on the other hand, the pole moves much closer to the threshold, approaching the unitary limit.
\end{itemize}

We also examined 
whether the current experimental data in the
$J^{PC}=1^{--}$ channel allow for the existence of a bound state in the exotic $0^{--}$ channel, as predicted in Ref.~\cite{Ji:2022blw}.
It turns out when adjusting $F_{01}^c+F_{01}^d$ such that a
bound state in the $0^{--}$ $D^*\bar D_1$ channel appears with
the binding energy of $E_B=30~\mathrm{MeV}$,
the overall fit gets only slightly worse.
Moreover, the pole locations found for  the 
quantum numbers $0^{-+}$ and $3^{--}$
as well as those in the $1^{--}$ channels
remain consistent with those found before. 
We therefore conclude that the currently available data do not exclude the existence of an exotic $0^{--}$ bound state of the type predicted in Ref.~\cite{Ji:2022blw}.

Clearly, improved data in the $1^{--}$ channel would allow us to also refine further the predictions for the spin partner states. But this connection goes both ways: Once 
sufficiently good data are available in the channels for the spin partner states, those will allow us to better understand the physics of the vector states. 

\begin{table}[t]
\caption{ Estimates for the pole positions of the HQSS partners. Only poles located within $100~\mathrm{MeV}$ of the corresponding threshold are quoted, since farther poles are not reliably constrained within the simplified single-channel framework employed in the present analysis.}
\label{tab:single-channel-poles}
\begin{center}
\begin{tabular}{cccccc}
\hline\hline
\vphantom{{\huge A}}$J^{PC}$ & Channel & Threshold [MeV] & Model &  RS & $E_B$ [MeV] \\
\hline
\multirow{2}{*}{$0^{--}$}
& \multirow{2}{*}{$D^*\bar{D}_1$}
& \multirow{2}{*}{4431}
& $\overline{\mathrm{M0}}$ & $-$ & $1$ \\
& & & $\mathrm{M2}$ &  $-$ & 44 \\
\hline
\multirow{4}{*}{$0^{-+}$}
& \multirow{4}{*}{$D^*\bar{D}_1$}
& \multirow{4}{*}{4431}
& $\mathrm{M0}$ &  $+$ & 66 \\
& & & $\overline{\mathrm{M0}}$ & $+$ & 80 \\
& & & $\mathrm{M1}$ &$+$ & 81 \\
& & & $\mathrm{M2}$ &  $+$ & 47 \\
\hline
\multirow{4}{*}{$3^{--}$}
& \multirow{4}{*}{$D^*\bar{D}_2^*$}
& \multirow{4}{*}{4472}
& $\mathrm{M0}$ &  $+$ & 94 \\
& & & $\overline{\mathrm{M0}}$ &  $+$ & 89 \\
& & & $\mathrm{M1}$ & $+$ & 98 \\
& & & $\mathrm{M2}$ &  $+$ & 93 \\
\hline\hline
\end{tabular}
\end{center}
\end{table}

\section{Summary and Outlook}
\label{sec:summary}

In this work, we developed a unified coupled-channel framework for the vector charmonium(-like) structures residing in the energy region from 4.1 to 4.6 GeV. It incorporates the three $S$-wave open-charm channels $D\bar{D}_1$, $D^*\bar{D}_1$, and $D^*\bar{D}_2^*$ as well as (optionally) the bare poles associated with the $\psi(4160)$ and $\psi(4415)$ treated as compact quark model states. Four benchmark models were introduced and employed in the combined data analysis: the purely dynamical model M0 (together with its variant $\overline{\mbox{M0}}$, in which the relative weights of the datasets were artificially modified), the one-seed-state model M1, and the two-seed-state model M2. Simultaneous fits to cross sections and invariant-mass distributions 
measured at BESIII demonstrate that the strongly channel-dependent line shapes in this energy region can be described reasonably well within the proposed common multichannel framework.
This illustrates very clearly the need
of simultaneous analyses of various channels with proper theoretical formalisms, if one aims at extracting
the physics content of a certain system.

Assessment of the results obtained in this work leads to the following conclusions:
\begin{itemize}
\item The coupled-channel formalism employed here is capable of describing the studied experimental distributions and provides an overall good fit quality.
In particular, already the version without any bare poles (model M0) can describe simultaneously the total cross sections in the $J/\psi \pi\pi$ and the $D\bar D^*\pi$ channels with reasonable accuracy. This observation is in conflict with the conclusion of Ref.~\cite{vonDetten:2024eie}, which called for the inclusion of the $\psi(4160)$. Thus, the coupled-channel dynamics not considered in
Ref.~\cite{vonDetten:2024eie} can produce additional structures. However, some detailed features of the observed channels, such as $h_c \pi\pi$ and $D\bar D^*\pi$, are better captured when additional charmonium poles are included.
\item In models M1 and M2, the physical masses and widths of the compact charmonium states extracted from the poles of the amplitude deviate only mildly from both the PDG values used as initial input and the bare values returned by the best fit.
This is consistent with the quarkonia largely decoupling from the two-meson states --- a scenario proposed in Refs.~\cite{Hammer:2016prh,Hanhart:2022qxq}.
It should be stressed, however, that the quark-model states play an important role in describing the data for the various $D^{(*)}\bar D^{(*)}$ channels with $J^{PC}=1^{--}$.
\item The two dynamical {states} found in all fitting models at around 4.23 and 4.32~GeV are robust, and {their properties} show only minor variations among the models. 
It is interesting to observe that, although three coupled two-body elastic channels were included in the analysis, only two near-threshold states were generated dynamically: 
one can be identified with the $\psi(4230)$, and the other corresponds to the $\psi(4320)$, which is significantly shifted downward from the $D^* \bar D_2^*$ threshold due to strong coupled-channel effects 
(see the results for model M0 in Figure~\ref{fig:YdensityplotM0variants} and Table~\ref{tab:four_model_poles}).
The pole that lies near the $D^* \bar D_1$ threshold in the decoupled-channel limit moves far away from the near-threshold region and does not affect observables.
However, we demonstrated by re-weighting some selected data sets that a third dynamical state could also be present and generate additional structure around 4.4~GeV (see the results for model $\overline{\rm M0}$ and their comparison with those for model M0 in Figure~\ref{fig:crosssectionM0variants} and Table~\ref{tab:poles}). Thus, improved data for different channels in this mass range are needed to draw firmer conclusions on the number of dynamically generated poles.
\item Contrary to the findings of Ref.~\cite{vonDetten:2024eie},
in all fits presented here, there is a dynamically generated pole close to 4.3~GeV. The coupled-channel dynamics driving the emergence of this pole was absent in Ref.~\cite{vonDetten:2024eie}, which might be the origin of this difference. However, as already pointed out above, more definite conclusions on the pole content in the mass range studied here can only be drawn once better data for all relevant channels are available.

\item The present analysis supports an alternative, theoretically motivated framework in which the experimentally observed vector enhancements above 4.2~GeV need not be interpreted as a na{\"i}ve sequence of isolated Breit--Wigner resonances. 
Instead, the results favor a mixed scenario in which
threshold-driven dynamical states may coexist and interfere with explicit seed states (generally consistent with the expectations of the quark model).
\item  Employing the contact interactions constrained by the coupled-channel fit in the $1^{--}$ sector, we also predicted the pole structure of several HQSS-partner channels. The $3^{--}$ $D^*\bar D_2^*$ and $0^{-+}$ $D^*\bar D_1$ partners are found to be robustly bound across all considered models. The predictions for the exotic $0^{--}$ $D^*\bar D_1$ channel are far less constrained; nevertheless, the current data were found compatible with the existence of a bound state in this channel, too, as predicted in Ref.~\cite{Ji:2022blw}.
\item The $Z_c$-related sector shows significantly greater stability across the benchmark models, suggesting that the two $Z_c$ poles are largely governed by near-threshold dynamics and are much less sensitive to the detailed bare-pole content of the vector sector.
\end{itemize}

The proposed framework is also capable of reproducing various measured invariant-mass distributions such as $J/\psi\pi^\pm$, $h_c\pi^\pm$, $D\bar{D}^*$, and $D^*\bar{D}^*$, and provides predictions for channels and energies for which data remain sparse. Further progress will come from more precise measurements of additional hidden-charm and open-charm final states, especially those capable of resolving the energy dependence of the $h_c\pi^\pm$, $D\bar{D}^*$, and $D^*\bar{D}^*$ subsystems. Such measurements will provide essential additional constraints on the interplay between explicit bare poles and dynamical poles in the vector charmonium-like sector.
Moreover, having demonstrated the connection between the system with $1^{--}$
and its spin partner channels, additional data in the latter will not only teach us a lot about the QCD spectrum but also improve our understanding of the vector channel.

\acknowledgments
We are grateful to Qi-Ming Li, Zhen-Tian Sun and Chang-Zheng Yuan for fruitful discussions and to Satoshi Nakamura and Chang-Zheng Yuan for useful comments on the manuscript. 
This work is supported in part by the National Key R\&D Program of China under Grant No. 2023YFA1606703; by the National Natural Science Foundation of China (NSFC) under Grants No. 12125507, No. 12361141819, and No. 12447101, by the Chinese Academy of Sciences (CAS) under Grant No.~YSBR-101, and by
Deutsche Forschungsgemeinschaft (DFG) under Grant No. 525056915
and under Germany's Excellence Strategy -- EXC 3107 -- Project-ID~533766364. In addition, U.-G.M., C.H., and A.N. thank the CAS President's International Fellowship Initiative (PIFI) under Grant Nos. 2025PD0022, 2025PD0087, and 2024PVA0004\_Y1, respectively, for partial support.

\bibliographystyle{JHEP}
\bibliography{refs}

\appendix

\section{HQSS-constrained interactions}
\label{app:HQSS}

Hadrons containing one or more heavy quarks exhibit an emergent symmetry known as heavy quark spin symmetry (HQSS) often employed in the low-energy effective field theory of QCD \cite{Isgur:1989ed,Isgur:1989vq}. This symmetry arises because the spin $s_Q$ of a heavy quark $Q$ effectively decouples from the momentum of the light degrees of freedom $j_\ell$. Specifically, in the limit $m_Q \to \infty$, the chromomagnetic interaction responsible for spin-dependent interactions between heavy and light quarks is suppressed as
${\mathbf{\sigma} \cdot \mathbf{B}}/{m_Q} \sim {\Lambda_{\rm QCD}}/{m_Q}\ll 1$, with
$\Lambda_{\rm QCD} \sim$ 300 MeV being the characteristic nonperturbative QCD scale. Therefore, strong interactions become independent of the heavy quark spin orientation. Note that, in this appendix, the low case Latin letters $a$ and $b$ are used for the SU(2) flavor indices.

\subsection{HQSS basis and LO potentials in the $1^{--}$ sector}
\label{sec:HbarT_potential}

For the ground states of charmed mesons, HQSS naturally organizes the spin-1 vector meson $P^{*(Q)}_\mu$ and the spin-0 pseudoscalar meson $P^{(Q)}$ into a heavy-light spin multiplet with $j_\ell = 1/2$. In the heavy quark effective theory formalism, these states are conveniently described by the superfield $H_{a}^{(Q)}$ \cite{Wise:1992hn,Casalbuoni:1992gi,Casalbuoni:1996pg},
\beq
H_{a}^{(Q)}=\frac{1+\slashed v}{2}\left[P_{a}^{*(Q) \mu} \gamma_{\mu}-P_{a}^{(Q)} \gamma_5\right],
\label{Hmeson}
\eeq
where the corresponding isospin doublets are
\beq
P^{(Q)}=(D^0,D^+), \quad
P^{*(Q)}_{\mu}=(D^{*0}_{\mu},D^{*+}_\mu),
\eeq
with $v^\mu=p^\mu/M$ the four-velocity of the heavy meson satisfying $v\cdot v=1$. The heavy field operators include a factor $\sqrt{M_H}$ (here $M_H$ is the heavy meson mass) and have the mass dimension 3/2.

Similarly, the $P$-wave ($L=1$) heavy-light mesons form two HQSS multiplets \cite{Falk:1991nq,Falk:1992cx}: one with the light-quark angular momentum $j_\ell=1/2$,
\beq
S_{a}^{(Q)}=\frac{1+\slashed v}{2}\left[P_{1 a}^{\prime(Q) \mu} \gamma_{\mu} \gamma_5-P_{0 a}^{*(Q)}\right],
\label{Smeson}
\eeq
and the other with $j_\ell=3/2$,
\beq
T_{a}^{(Q)\mu}=\frac{1+\slashed v}{2}\bigg[P_{2 a}^{*(Q) \mu \nu} \gamma_{\nu}-\sqrt{\frac{3}{2}} P_{1 a \nu}^{(Q)} \gamma_5\left(g^{\mu \nu}-\frac{1}{3} \gamma^{\nu}\left(\gamma^{\mu}-v^{\mu}\right)\right)\bigg].
\label{Tmeson}
\eeq
The states in the multiplet $S$ are generally broad due to their predominantly $S$-wave decays. They are not included explicitly in the current analysis but are expected to contribute to nonvanishing phases of some production parameters as discussed in Section~\ref{sec:parameters}. The mesons in the multiplet $T$,
\begin{align}
P_1^{(Q)}=(D_1(2420)^0,D_1(2420)^+),\quad P_2^{(Q)}=( D_2^*(2460)^0, D_2^*(2460)^+),
\end{align}
can only couple to $D^{(*)}\pi$ in the $D$ wave in the heavy quark limit. These $D_1(2420)$ and $ D_2^*(2460)$ mesons are the key ingredients of the molecular states studied in this work.

The coupled-channel potential is derived from the LO effective Lagrangian for the $H\bar{T}+\text{c.c.}$ interaction. To this end we first decompose the wave functions of the relevant two-meson states in terms of the HQSS basis states $\ket{s_Q,j_\ell,J}$, where $s_Q$, $j_\ell$, and $J$ are the heavy quark spin, angular momentum of the light degrees of freedom, and the total angular momentum, respectively. The physical $1^{--}$ channels relevant to our analysis are projected onto the HQSS basis as
\beq
\begin{split}
&\ket{D\bar{D}_1}=\frac1{2}\ket{0,1,1}_1-\sqrt{\frac{1}{8}}\ket{1,1,1}_1+\sqrt{\frac{5}{8}}\ket{1,2,1}_1,\\
&\ket{D^*\bar{D}_1}=-\sqrt{\frac{1}{8}}\ket{0,1,1}_1+\frac{3}{4}\ket{1,1,1}_1+\frac{\sqrt5}{4}\ket{1,2,1}_1,\\
&\ket{D^*\bar{D}_2^*}=\sqrt{\frac{5}{8}}\ket{0,1,1}_1+\frac{\sqrt5}{4}\ket{1,1,1}_1-\frac{1}{4}\ket{1,2,1}_1,
\label{setalpha}
\end{split}
\eeq
while for their charge conjugates one has
\beq
\begin{split}
&\ket{D_1\bar{D}}=\frac1{2}\ket{0,1,1}_2+\sqrt{\frac{1}{8}}\ket{1,1,1}_2+\sqrt{\frac{5}{8}}\ket{1,2,1}_2,\\
&\ket{D_1\bar{D}^*}=-\sqrt{\frac{1}{8}}\ket{0,1,1}_2-\frac{3}{4}\ket{1,1,1}_2+\frac{\sqrt5}{4}\ket{1,2,1}_2,\\
&\ket{D_2^*\bar{D}^*}=\sqrt{\frac{5}{8}}\ket{0,1,1}_2-\frac{\sqrt5}{4}\ket{1,1,1}_2-\frac{1}{4}\ket{1,2,1}_2,
\end{split}
\label{setalphabar}
\eeq
where the subscripts $1$ and $2$ distinguish the $H\bar T$ and $T\bar H$ ordering, respectively. We choose the relative phase of the $T\bar H$ basis such that the $D_1\bar{D}^*$ channel carries an extra overall minus sign as shown above. If we denote the states from the sets in Eqs.~\eqref{setalpha} and \eqref{setalphabar} as $\ket{\alpha}$ and $\ket{\bar{\alpha}}$, respectively, and employ the charge conjugation operation in Eq.~\eqref{M1M2C}, then
\beq
\ket{\alpha;C=-}=\frac{1}{\sqrt{2}}(\ket{\alpha}-\ket{\bar{\alpha}}),
\eeq
and the potential in Eq.~\eqref{eq:VY} follows from the projection
\begin{equation}
V_{\alpha\beta}=
\braket{\alpha;C=-|\mathcal H_{\rm int}|\beta;C=-}=
\frac12\Bigl(\langle\alpha|\mathcal H_{\rm int}|\beta\rangle+\langle\bar{\alpha}|\mathcal H_{\rm int}|\bar{\beta}\rangle-\langle\alpha|\mathcal H_{\rm int}|\bar{\beta}\rangle-\langle\bar{\alpha}|\mathcal H_{\rm int}|\beta\rangle\Bigr).
\end{equation}
Assuming that the interaction Hamiltonian $\mathcal H_{\rm int}$ conserves HQSS and, therefore, is diagonal in the HQSS basis, we define the following nonvanishing matrix elements for the direct ($d$) and cross ($c$) interactions:
\beq\label{eq:FIjl}
\begin{split}
&\bra{s_Q,j_\ell,J}\mathcal H_{\rm int}\ket{s_Q',j_\ell',J'}= \bra{\overline{s_Q,j_\ell,J}}\mathcal H_{\rm int}\ket{\overline{s_Q',j_\ell',J'}}
=\delta_{s_Qs_Q'}\delta_{j_\ell j_\ell'}\delta_{JJ'}F_{Ij_\ell}^d,\\
&\bra{s_Q,j_\ell,J}\mathcal H_{\rm int}\ket{\overline{s_Q',j_\ell',J'}}=
\bra{\overline{s_Q,j_\ell,J}}\mathcal H_{\rm int}\ket{s_Q',j_\ell',J'}=
\delta_{s_Qs_Q'}\delta_{j_\ell j_\ell'}\delta_{JJ'}F_{Ij_\ell}^c,
\end{split}
\eeq
where $I$ is the isospin quantum number. In the present work, we set $I=0$ throughout, since all channels considered here are isoscalar. By projecting the interaction strength defined above onto the physical channel basis, we obtain the coupled-channel potential $V$ for the $H\bar T$ interaction, as quoted in Eq.~\eqref{eq:VY}.

\subsection{$TH\Phi$ Lagrangian and tensor decay amplitudes}
\label{sec:THPhi_coupling}

The nonrelativistic form of the HQSS multiplets in Eqs.~\eqref{Hmeson} and \eqref{Tmeson} read
\beq
H_a =\vec{D}_a^* \cdot \vec{\sigma}+D_a\quad \text{and}\quad
T_a^i =D_{2 a}^{i j} \sigma^j+\sqrt{\frac{2}{3}} D_{1 a}^i+i \sqrt{\frac{1}{6}} \epsilon_{i j k} D_{1 a}^j \sigma^k,
\label{HTmesonnr}
\eeq
respectively, while the matrix of the pions is
\beq
\Phi=\left(\begin{array}{cc}
\pi^0 / \sqrt{2} & \pi^{+} \\
\pi^{-} & -\pi^0 / \sqrt{2}
\end{array}\right).
\label{Phimatrix}
\eeq
Then the LO effective Lagrangian for the transition of a $P$-wave heavy meson $T$ into a ground-state heavy meson $H$ and a light pseudoscalar $\Phi$, collected in Eqs.~(\ref{HTmesonnr},\ \ref{Phimatrix}), that satisfies HQSS and corresponds to a $D$-wave coupling, reads
\beq
\mathcal{L}_D=\frac{h_D}{2F_\pi}\left\langle T_b^i\sigma^jH_a^\dagger\right\rangle\partial^i\partial^j\Phi_{ba},
\label{THPhi}
\eeq
where $F_{\pi}=92.4$~MeV is the pion decay constant. Meanwhile, the width of the decay $D_1\to D^*\pi$
derived from the Lagrangian in Eq.~\eqref{THPhi} for the physical $D_1(2420)$ meson fails to saturate its full experimental decay width \cite{Falk:1992cx,Guo:2020oqk}. It suggests sizable HQSS-breaking effects and the necessity to take into account $S$-wave contribution to this decay that stems from the Lagrangian
\beq
\mathcal L_S= i\frac{h_S}{\sqrt6 F_{\pi}}D_{1b}^i D_a^{*\dagger i}\partial^0\Phi_{ba}.
\label{LagS}
\eeq
The absolute values of the couplings $h_S$ and $h_D$ can be determined by fitting the experimental decay widths of the $ D_2^*(2460)$ and $D_1(2420)$ mesons, yielding $|h_S|=0.57$ and $|h_D|=1.17$~GeV$^{-1}$ \cite{Guo:2020oqk}. Then, from the effective Lagrangians $\mathcal{L}_D$ and $\mathcal{L}_S$ in Eqs.~\eqref{THPhi} and \eqref{LagS}, respectively, one can extract
\begin{equation}
\begin{aligned}
\mathcal L_{D_1D^*\pi}&=\frac{h_D}{\sqrt6F_\pi}\left(3D_{1b}^iD_a^{*\dagger j}\partial^i\partial^j\Phi_{ba}-D_{1b}^i D_a^{*\dagger i}\partial^2\Phi_{ba}\right)+i\frac{h_S}{\sqrt6 F_{\pi}}D_{1b}^i D_a^{*\dagger i}\partial^0\Phi_{ba},\\
\mathcal L_{ D_2^*D^*\pi}&=i\frac{h_D}{F_\pi}\epsilon^{ijk} D_{2b}^{il}D_a^{*\dagger k}\partial^j\partial^l\Phi_{ba},\\
\mathcal L_{ D_2^*D\pi}&=\frac{h_D}{F_\pi} D_{2b}^{ij}D_a^{\dagger}\partial^i\partial^j\Phi_{ba}.
\end{aligned}
\end{equation}
These interaction terms lead to the decay amplitudes
\beq
\begin{aligned}
\mathcal M_{D_1\to D^*\pi}&=-\frac{\sqrt{m_{D_1}m_{D^*}}}{\sqrt6F_\pi}\epsilon_{D_1}^i\epsilon_{D^*}^{*j}\left(h_S p_\pi^0\delta^{ij}+h_D\left(3p_{\pi}^i p_{\pi}^j-p_\pi^2\delta^{ij}\right)\right)=-c_1\epsilon_{D_1}^i\epsilon_{D^*}^{*j}\mathcal M_{TH1}^{ij},\\
\mathcal M_{ D_2^*\to D^*\pi}&=-i\frac{\sqrt{m_{ D_2^*}m_{D^*}}}{F_\pi}\epsilon_{ D_2^*}^{il}\epsilon_{D^*}^{*k}\left(h_D\epsilon^{ijk}p_{\pi}^j p_{\pi}^l\right)=-ic_2\epsilon_{ D_2^*}^{il}\epsilon_{D^*}^{*k}\mathcal M_{TH2}^{ilk},\\
\mathcal M_{ D_2^*\to D\pi}&=-\frac{\sqrt{m_{ D_2^*}m_{D}}}{F_\pi}\epsilon_{ D_2^*}^{ij}(h_Dp_{\pi}^i p_{\pi}^j)=-c_3\epsilon_{ D_2^*}^{ij}\mathcal M_{TH3}^{ij},
\end{aligned}
\eeq
where
\beq
c_1=\frac{\sqrt{m_{D_1}m_{D^*}}}{\sqrt6F_\pi},\quad c_2=\frac{\sqrt{m_{ D_2^*}m_{D^*}}}{F_\pi},\quad c_3=\frac{\sqrt{m_{ D_2^*}m_{D}}}{F_\pi}\label{eq:ci}
\eeq
and
\beq
\mathcal M_{TH1}^{ij}=h_S p_\pi^0\delta^{ij}+h_D\left(3p_{\pi}^i p_{\pi}^j-p_\pi^2\delta^{ij}\right),\quad
\mathcal M_{TH2}^{ilk}=h_D{\epsilon^{ijk}}p_{\pi}^j p_{\pi}^l,\quad
\mathcal M_{TH3}^{ij}=h_Dp_{\pi}^i p_{\pi}^j.\label{eq:MTH}
\eeq

\subsection{Transitions from two-body elastic channels to $J/\psi\eta$ and $\chi_{c0}\omega$}

In this appendix we provide some details concerning production of the observation channels in Eq.~\eqref{observation2} with two-body final states. 
To proceed we decompose the corresponding states in the basis $\{\ket{a;s_Q,j_\ell,J}\}$, with $a=5$ for $J/\psi\eta$ and $a=6$ for $\chi_{c0}\omega$, as defined in Eq.~\eqref{observation2}. Then we arrive at
\beq
\begin{split}
\ket{J/\psi\eta,L=1}&=\ket{J/\psi\eta;1,1,1},\\
\ket{\chi_{c0}\omega,L=0}&=\frac13\ket{\chi_{c0}\omega;1,0,1}-\frac{1}{\sqrt3}\ket{\chi_{c0}\omega;1,1,1}+\frac{\sqrt5}{3}\ket{\chi_{c0}\omega;1,2,1}
\end{split}
\eeq
so the transition amplitudes can be parametrized as
\beq
\begin{split}
\bra{1,1,1}\mathcal H_{\rm int}\ket{J/\psi\eta,L=1}
&=-\bra{\overline{1,1,1}}\mathcal H_{\rm int}\ket{J/\psi\eta,L=1}=F_{J/\psi\eta}p_{\eta},\\[1mm]
\bra{1,1,1}\mathcal H_{\rm int}\ket{\chi_{c0}\omega,L=0}
&=-\bra{\overline{1,1,1}}\mathcal H_{\rm int}\ket{\chi_{c0}\omega,L=0}=F_{\chi_{c0}\omega},
\end{split}
\eeq
with $F_{J/\psi\eta}$ and $F_{\chi_{c0}\omega}$ two free parameters and $p_\eta$ the momentum of $\eta$ in the c.m. frame.
Since the $l_\ell =1/2$ component in $D_1(2420)$ is small, we do not consider the mixing here. Then it is straightforward to obtain for the LO transitions from a two-body elastic channel $\alpha$,
\beq
\begin{aligned}
V_{\alpha5}&=\bra{\alpha}\mathcal H_{\rm int}\ket{J/\psi\eta,L=1}=\left(-\frac12,{\frac3{\sqrt8}},\sqrt{\frac58}\right)F_{J/\psi\eta}p_{\eta},
\label{eq:Valpha}\\
V_{\alpha6}&=\bra{\alpha}\mathcal H_{\rm int}\ket{\chi_{c0}\omega,L=0}=\left(-\frac12,{\frac3{\sqrt8}},\sqrt{\frac58}\right)F_{\chi_{c0}\omega},
\end{aligned}
\eeq
with $\alpha=1,2,3$ enumerating the two-body elastic channels in Eq.~\eqref{elastic}.

\subsection{HQSS basis and the potentials in the $Z_c$-related sector}
\label{app:Zc}

We consider the two-body subsystems of the three-body observation channels with
$Z_c$-related interactions; see Eq.~\eqref{observation2body}---and decompose the corresponding states in the basis $\ket{s_Q,j_\ell,J}$,
\begin{equation}
\begin{aligned}
&\left|{J/\psi \pi}\right\rangle=\left|{a=1;1,0,1}\right\rangle,\\[2mm]
&\left|{h_c \pi}\right\rangle=\left|a=2;{0,1,1}\right\rangle,\\
&\left|{\bar{D}D^{*}}\right\rangle=\frac{1}{2}\left|{a=3;1,0,1}\right\rangle-\frac{1}{2}\left|{a=3;0,1,1}\right\rangle+\frac{1}{\sqrt2}\left|{a=3;1,1,1}\right\rangle,\\
&\left|{\bar{D}^*D}\right\rangle=-\frac{1}{2}\left|{a=3;1,0,1}\right\rangle+\frac{1}{2}\left|{a=3;0,1,1}\right\rangle+\frac{1}{\sqrt2}\left|{a=3;1,1,1}\right\rangle,\\
&\left|{\bar{D}^{*}D^{*}}\right\rangle=\frac{1}{\sqrt2}\left|{a=4;1,0,1}\right\rangle+\frac{1}{\sqrt2}\left|{a=4;0,1,1}\right\rangle,
\end{aligned}
\end{equation}
which gives for the relevant linear combination
\begin{align}
&\left|{\bar{D}^*D}\right\rangle_+
\equiv\frac{1}{\sqrt2}\left(\left|{\bar{D}D^*}\right\rangle+\left|{\bar{D}^*D}\right\rangle\right)
=\left|{a=3;1,1,1}\right\rangle,\\
&\left|{\bar{D}^*D}\right\rangle_-
\equiv\frac{1}{\sqrt2}\left(\left|{\bar{D}D^*}\right\rangle-\left|{\bar{D}^*D}\right\rangle\right)
=\frac{1}{\sqrt2}\left(\left|{a=3;1,0,1}\right\rangle-\left|{a=3;0,1,1}\right\rangle\right).
\end{align}
Then the LO HQSS-motivated transition potential in the basis of physical states
\begin{equation}
\{\ket{J/\psi\pi}, \ket{h_c\pi}, \left|\bar{D}^*D\right\rangle_-,
\left|\bar{D}^*D^*\right\rangle\}
\end{equation}
reads
\begin{align}
\label{eq:VZc1}
V^{Z_c}_{\rm HQSS}=\left(\begin{matrix}
0&0&\,\frac 1{\sqrt2}F&\,\frac 1{\sqrt2}F\\
0&0&-\frac 1{\sqrt2}G\,p_\pi&\,\frac 1{\sqrt2}G\,p_\pi\\
\frac 1{\sqrt2}F&-\frac 1{\sqrt2}G\,p_\pi&\,\frac12(C_0+C_1)&\,\frac12(C_0-C_1)\\
\frac 1{\sqrt2}F&\,\frac 1{\sqrt2}G\,p_\pi&\,\frac12(C_0-C_1)&\,\frac12(C_0+C_1)
\end{matrix}\right),
\end{align}
where the four independent low-energy constants appearing at this level, $F$, $G$, $C_0$, and $C_1$, are defined through
\begin{align}
F=\left\langle{a;1,0,1}\right|\mathcal H_{\rm int}\left|{J/\psi \pi}\right\rangle, \quad G\,p_\pi=\left\langle{a;0,1,1}\right|\mathcal H_{\rm int}\left|{h_c \pi}\right\rangle,
\end{align}
for $a=1,2$, and
\begin{align}
\left\langle{a;s_Q,j_\ell,J}\right|\mathcal H_{\rm int}\left|a';s_Q',j_\ell',J'\right\rangle=\delta_{s_Q s_Q'}\delta_{j_\ell j_\ell'}\delta_{JJ'}C_{j_\ell},
\end{align}
for $a,a'=3,4$. It proves convenient to define the linear combinations $C_{1Z}\equiv (C_0+C_1)/2$ and $C_{34}\equiv (C_0-C_1)/2$.

Note that, for the transition $a\to b$ ($a,b=1\ldots4$), the corresponding matrix elements in Eq.~\eqref{eq:VZc1} should be multiplied by the mass factor $4\sqrt{m_{a1}m_{a2}m_{b1}m_{b2}}$ to account for the normalization of fields, where $\{m_{a1},m_{a2}\}$ and $\{m_{b1},m_{b2}\}$ are the masses of the two particles in the given initial- and final-state observation channel, respectively.

\section{Production amplitudes in the observation channels}\label{app:prodi}

In this appendix we collect explicit channel-dependent formulas for the production of the observation channels in Eqs.~\eqref{observation1} and \eqref{observation2} used to evaluate the amplitudes in Eq.~\eqref{Mas}.

\subsection{Observation channel $J/\psi\pi^+\pi^-$ ($a=1$)}

The set of the diagrams relevant for building the total production amplitude in the channel $J/\psi\pi^+\pi^-$ is shown in Figure~\ref{fig:jpsipipi}. The components of this amplitude contributing to the decomposition in Eq.~\eqref{M1ab} read (summations over repeated indices of the observation channels $b$ and $c$ as well as the three-dimensional Cartesian index $k$ are implied; no summation over $A$ is understood)
\begin{align}
\mathcal M^{\gamma^*,ij}_1=&\left(\mathcal P^{\gamma^*}_{a=1} + 
\mathcal P^{\gamma^*}_bG^{Z_c}_{bc}(s)T^{Z_c^-}_{c1}(s) + 
\mathcal P^{\gamma^*}_bG^{Z_c}_{bc}(t)T^{Z_c^+}_{c1}(t)\right)\delta^{ij},
\label{M1gij}\\
\mathcal M^{A,ij}_1=&-U_AG_{AA}\left(\mathcal P^A_{a=1} + 
\mathcal P^A_bG^{Z_c}_{bc}(s)T^{Z_c^-}_{c1}(s) + 
\mathcal P^A_bG^{Z_c}_{bc}(t)T^{Z_c^+}_{c1}(t)\right)\delta^{ij},\\
\mathcal M^{\alpha=1,ij}_1=&\,\frac{1}{\sqrt2}c_1U_{\alpha=1}T_{D_1DD^*}
\left(\mathcal M^{ij}_{TH1}(p_1)T_{31}^{Z_c^-}(s) + 
\mathcal M^{ij}_{TH1}(p_2)T_{31}^{Z_c^+}(t)\right),\\
\mathcal M^{\alpha=2,ij}_1=&\,\frac{1}2c_1U_{\alpha=2}T_{D_1D^*D^*}
\left[
\left(\mathcal M^{ij}_{TH1}(p_1)-\delta^{ij}\mathcal M^{kk}_{TH1}(p_1)\right)T_{41}^{Z_c^-}(s)\right.\notag\\
&\hspace{3.6cm}\left.
+\left(\mathcal M^{ij}_{TH1}(p_2)-\delta^{ij}\mathcal M^{kk}_{TH1}(p_2)\right)T_{41}^{Z_c^+}(t)
\right],\\
\mathcal M^{\alpha=3,ij}_1=&\,\frac{\sqrt{3}}{\sqrt{10}}c_3U_{\alpha=3}T_{ D_2^*DD^*}
\left[
\left(\mathcal M^{ij}_{TH3}(p_1)- \delta^{ij}\mathcal M^{kk}_{TH3}(p_1)/3\right)T_{31}^{Z_c^-}(s)\right.\notag\\
&\hspace{3.6cm}\left.
+\left(\mathcal M^{ij}_{TH3}(p_2)- \delta^{ij}\mathcal M^{kk}_{TH3}(p_2)/3\right)T_{31}^{Z_c^+}(t)
\right]\notag\\
&\,+\frac{\sqrt3 }{2\sqrt{10}}c_2U_{\alpha=3}T_{ D_2^*D^*D^*}\epsilon^{jbk}
\left[
\left(\mathcal M_{TH2}^{ibk}(p_1)+\mathcal M_{TH2}^{bik}(p_1)\right)T_{41}^{Z_c^-}(s)\right.\notag\\
&\hspace{3.6cm}\left.
+\left(\mathcal M^{ibk}_{TH2}(p_2)+\mathcal M^{aik}_{TH2}(p_2)\right)T_{41}^{Z_c^+}(t)
\right],
\label{M1alpha3ij}
\end{align}
where the triangle loop function denoted by $T_{P_1P_2P_3}$ is approximated as
\begin{align}
T_{P_1P_2P_3}\simeq -I_{P_1P_2P_3},
\end{align}
with the scalar function $I_{P_1P_2P_3}$ provided explicitly in Ref.~\cite{Chen:2023def}.
The constants $c_i$ and the amplitude $\mathcal M_{TH}$ are related to the transition of $T\to H\pi$ and the corresponding expressions are given in Eqs.~\eqref{eq:ci} and \eqref{eq:MTH}, respectively. The tensor amplitude in Eq.~\eqref{Mas} given by the sum of the contributions in Eqs.~\eqref{M1gij}-\eqref{M1alpha3ij} can be decomposed in the independent tensor structures as
\beq
\mathcal{M}^{ij}_1=M_{11}(s,t)\delta^{ij}+M_{12}(s)p_1^i p_1^j+M_{12}(t)p_2^i p_2^j,
\label{eq:M1tensor0}
\eeq
where the coefficients of this decomposition take the form
\begin{align}
M_{11}(s,t)=&\bigl(\mathcal P^{\gamma^*}_{a=1}
+\mathcal P^{\gamma^*}_bG^{Z_c}_{bc}(s)T^{Z_c^-}_{c1}(s)
+\mathcal P^{\gamma^*}_bG^{Z_c}_{bc}(t)T^{Z_c^+}_{c1}(t)\bigr)\notag\\
&-\sum_{A=1}^2U_{A}G_{AA}\bigl(\mathcal P^A_1
+\mathcal P^A_bG^{Z_c}_{bc}(s)T^{Z_c^-}_{c1}(s)
+\mathcal P^A_bG^{Z_c}_{bc}(t)T^{Z_c^+}_{c1}(t)\bigr)\notag\\
&+\frac{1}{\sqrt 2}c_1 U_{\alpha=1}\bigl[
T_{D_1\bar{D} D^*}(s)T^{Z_c^-}_{31}(s)\left(h_Sp_1^0-h_Dp_1^2\right)\notag\\
&\hspace{3.2cm}
+T_{D_1\bar{D} D^*}(t)T^{Z_c^-}_{31}(t)\left(h_Sp_2^0-h_Dp_2^2\right)\bigr]\notag\\
&-c_1 U_{\alpha=2}\bigl[
T_{D_1\bar{D}^* D^*}(s)T^{Z_c^-}_{41}(s)\left(h_Sp_1^0+h_Dp_1^2/2\right)\notag\\
&\hspace{3.2cm}
+T_{D_1\bar{D}^* D^*}(t)T^{Z_c^+}_{41}(t)\left(h_Sp_2^0+h_Dp_2^2/2\right)\bigr]\notag\\
&-\frac{1}{\sqrt{30}}c_2 U_{\alpha=3}\bigl[
T_{{ D_2^*\bar{D}^*D}}(s)T^{Z_c^-}_{31}(s)h_Dp_1^2
+T_{{ D_2^*\bar{D}^*D}}(t)T^{Z_c^+}_{31}(t)h_Dp_2^2\bigr]\notag\\
&-\frac{\sqrt3 }{2\sqrt{10}}c_2U_{\alpha=3}\bigl[
T_{{ D_2^*\bar{D}^*D^*}}(s)T^{Z_c^-}_{41}(s)h_Dp_1^2
+T_{{ D_2^*\bar{D}^*D^*}}(t)T^{Z_c^+}_{41}(t)h_Dp_2^2\bigr],\\
M_{12}(s)=&\,\frac{3}{\sqrt 2}c_1U_{\alpha=1}
T_{D_1\bar{D} D^*}(s)T^{Z_c^-}_{31}(s)h_D+\frac{3}{2}c_1U_{\alpha=2}
T^{Z_c^+}_{{D_1\bar{D}^*D^*}}(s)T^{Z_c^-}_{41}(s)h_D\notag\\
&+\sqrt{\frac{3}{10}} c_3U_{\alpha=3}
T_{{ D_2^*\bar{D}^*D}}(s)T^{Z_c^-}_{31}(s)h_D+\frac{3\sqrt3}{2\sqrt{10}}c_2U_{\alpha=3}
T_{{ D_2^*\bar{D}^*D^*}}(s)T^{Z_c^-}_{41}(s)h_D.
\end{align}
Then the vector $\bar{M}_1(s,t)$ introduced in Eq.~\eqref{barMdef} reads
\beq
\bar{M}_1(s,t)=\Bigl(M_{11}(s,t),
M_{12}(s),M_{12}(t)\Bigr)^T
\eeq
and the $3\times 3$ kinematical matrix $Q_1$ is
\beq
Q_1=\frac13\begin{pmatrix}
3&p_1^2&p_2^2\\
p_1^2&p_1^4&(\vec p_1\cdot\vec p_2)^2\\
p_2^2&(\vec p_1\cdot\vec p_2)^2&p_2^4
\end{pmatrix}.
\label{kinmatr}
\eeq

\subsection{Observation channel $h_c\pi^+\pi^-$ ($a=2$)}
\begin{figure}[t]
\includegraphics[width=\linewidth]{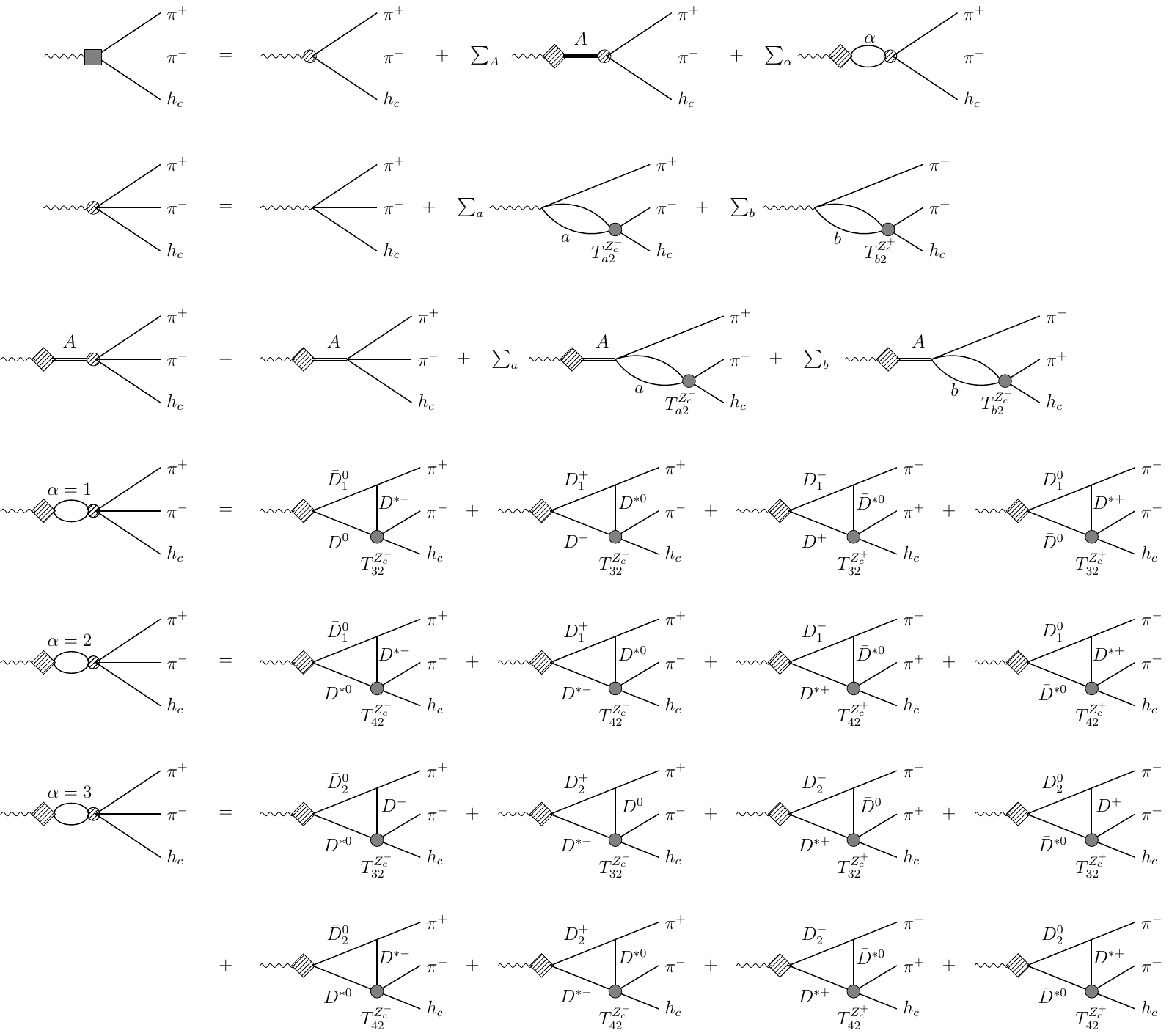}
\caption{Diagrams for the production of $h_c\pi^+\pi^-$ from the virtual photon. }\label{fig:hcpipi}
\end{figure}

The production of the observation channel $h_c\pi^+\pi^-$ is illustrated in Figure~\ref{fig:hcpipi} and described by the tensor $\mathcal M_2^{ij}$ introduced in Eq.~\eqref{Mas}.
Following the decomposition explained in Eq.~\eqref{M1ab}, we write the individual contributions to this tensor as (summations over repeated indices of the observation channels $b$ and $c$ as well as the three-dimensional Cartesian indices $k$ and $l$ are implied; no summation over $A$ is understood)
\begin{align}
 \mathcal M^{\gamma,ij}_2=&\bigl(\mathcal P^{\gamma^*}_{a=2}
 +\mathcal P^{\gamma^*}_bG^{Z_c}_{bc}(s)T^{Z_c^-}_{c2}(s)
 +\mathcal P^{\gamma^*}_bG^{Z_c}_{bc}(t)T^{Z_c^+}_{c2}(t)\bigr)\delta^{ij},\\
 \mathcal M^{A,ij}_2=&-U_AG_{AA}\bigl(\mathcal P^A_{a=2}
 +\mathcal P^A_bG^{Z_c}_{bc}(s)T^{Z_c^-}_{c2}(s)
 +\mathcal P^A_bG^{Z_c}_{bc}(t)T^{Z_c^+}_{c2}(t)\bigr)\delta^{ij},\\
 \mathcal M^{\alpha=1,ij}_2=&\,\frac{1}{\sqrt 2}c_1U_{\alpha=1}T_{D_1DD^*}
 \bigl(\mathcal M_{TH1}^{ij}(p_1)T_{32}^{Z_c^-}(s)
 +\mathcal M_{TH1}^{ij}(p_2)T_{32}^{Z_c^+}(t)\bigr),\\
 \mathcal M^{\alpha=2,ij}_2=&-\frac{1}{2}c_1U_{\alpha=2}T_{D_1D^*D^*}
 \bigl[\bigl(\delta^{ij}\mathcal M^{ll}_{TH1}(p_1)-\mathcal M^{ij}_{TH1}(p_1)\bigr)T_{42}^{Z_c^-}(s)\notag\\
 &\hspace{3.5cm}
 +\bigl(\delta^{ij}\mathcal M^{ll}_{TH1}(p_2)-\mathcal M^{ij}_{TH1}(p_2)\bigr)T_{42}^{Z_c^+}(t)\bigr],\\
 \mathcal M^{\alpha=3,ij}_2=&\,\sqrt{\frac{3}{10}}c_3U_{\alpha=3}T_{ D_2^*D^*D^*}
 \bigl[\bigl(\mathcal M^{ij}_{TH3}(p_1)- \frac13\delta^{ij}\mathcal M^{ll}_{TH3}(p_1)\bigr)T_{32}^{Z_c^-}(s)\notag\\
 &\hspace{3.5cm}
 +\bigl(\mathcal M^{ij}_{TH3}(p_2)- \frac13\delta^{ij}\mathcal M^{ll}_{TH3}(p_2)\bigr)T_{32}^{Z_c^+}(t)\bigr]\notag\\
 &+\frac{\sqrt3}{4\sqrt{10}}c_2U_{\alpha=3}T_{ D_2^*D^*D^*}\epsilon^{jkl}\notag\\
 &\quad \times\left[\bigl(\mathcal M_{TH2}^{ikl}(p_1)-\mathcal M_{TH2}^{ilk}(p_1)
 +\mathcal M_{TH2}^{kil}(p_1)-\mathcal M_{TH2}^{lik}(p_1)\bigr)T_{42}^{Z_c^-}(s)\right.\notag\\
 &
 \quad\quad\left.+\bigl(\mathcal M_{TH2}^{ikl}(p_2)-\mathcal M_{TH2}^{ilk}(p_2)
 +\mathcal M_{TH2}^{kil}(p_2)-\mathcal M_{TH2}^{lik}(p_2)\bigr)T_{42}^{Z_c^+}(t)\right].
\end{align}
After summing up the contributions above, the full amplitude $\mathcal M_2^{ij}$ can be expressed as
\beq
\mathcal{M}^{ij}_2=M_{21}(s,t)\delta^{ij}+M_{22}(s)p_1^i p_1^j+M_{22}(t)p_2^i p_2^j,
\label{eq:M2tensor}
\eeq
with the coefficients
\begin{align}
M_{21}(s,t)=&\left(\mathcal P^{\gamma^*}_{a=2}+\mathcal P^{\gamma^*}_bG^{Z_c}_{bc}(s)T^{Z_c^-}_{c2}(s)+\mathcal P^{\gamma^*}_bG^{Z_c}_{bc}(t)T^{Z_c^+}_{c2}(t)\right)\notag\\
 &-\sum_{A=1}^2U_AG_{AA}\left(\mathcal P^A_{a=2}+\mathcal P^A_bG^{Z_c}_{bc}(s)T^{Z_c^-}_{c2}(s)+\mathcal P^A_bG^{Z_c}_{bc}(t)T^{Z_c^+}_{c2}(t)\right)\notag\\
 &+\frac{1}{\sqrt 2}c_1U_{\alpha=1}T_{D_1\bar{D} D^*}
\left[
T^{Z_c^-}_{32}(s)\left(h_S p_1^0-h_D p_1^2\right)
+T^{Z_c^+}_{32}(t)\left(h_S p_2^0-h_D p_2^2\right)
\right]\notag\\
&+c_1U_{\alpha=2}T_{D_1\bar{D}^* D^*}
\left[
T^{Z_c^-}_{42}(s)\left(h_S p_1^0+\tfrac{1}{2}h_D p_1^2\right)
+T^{Z_c^+}_{42}(t)\left(h_S p_2^0+\tfrac{1}{2}h_D p_2^2\right)
\right]\notag\\
&-\frac{1}{\sqrt{30}}c_3U_{\alpha=3}T_{ D_2^*\bar{D}^* D}\,
h_D\left[
T^{Z_c^-}_{32}(s)p_1^2+T^{Z_c^+}_{32}(t)p_2^2
\right]\notag\\
&-\frac{\sqrt{3}}{2\sqrt{10}}c_2U_{\alpha=3}T_{ D_2^*\bar{D}^* D^*}\,
h_D\left[
T^{Z_c^-}_{42}(s)p_1^2+T^{Z_c^+}_{42}(t)p_2^2
\right]\\
M_{22}(s)=&\,\frac{3}{\sqrt 2}c_1U_{\alpha=1}T_{D_1\bar{D} D^*}T^{Z_c^-}_{32}(s)h_D
-\frac{3}{2}c_1U_{\alpha=2}T_{D_1\bar{D}^* D^*}T^{Z_c^-}_{42}(s)h_D\notag\\
&+\sqrt{\frac{3}{10}}c_3U_{\alpha=3}T_{ D_2^*\bar{D}^* D}T^{Z_c^-}_{32}(s)h_D
+\frac{3\sqrt3}{2\sqrt{10}}c_2U_{\alpha=3}T_{ D_2^*\bar{D}^* D^*}T^{Z_c^-}_{42}(s)h_D.
\end{align}
Then the vector $\bar M_2$ in Eq.~\eqref{barMdef} reads
\begin{align}
\bar M_2=\Bigl(M_{21}(s,t),M_{22}(s),M_{22}(t)\Bigr)^{\rm T}
\end{align}
and the kinematical matrix $Q_2$ takes the same form as $Q_1$ in Eq.~\eqref{kinmatr}.

\subsection{Observation channel $D^{*-}D^0\pi^+$ ($a=3$)}

\begin{figure}[t]
 \includegraphics[width=\linewidth]{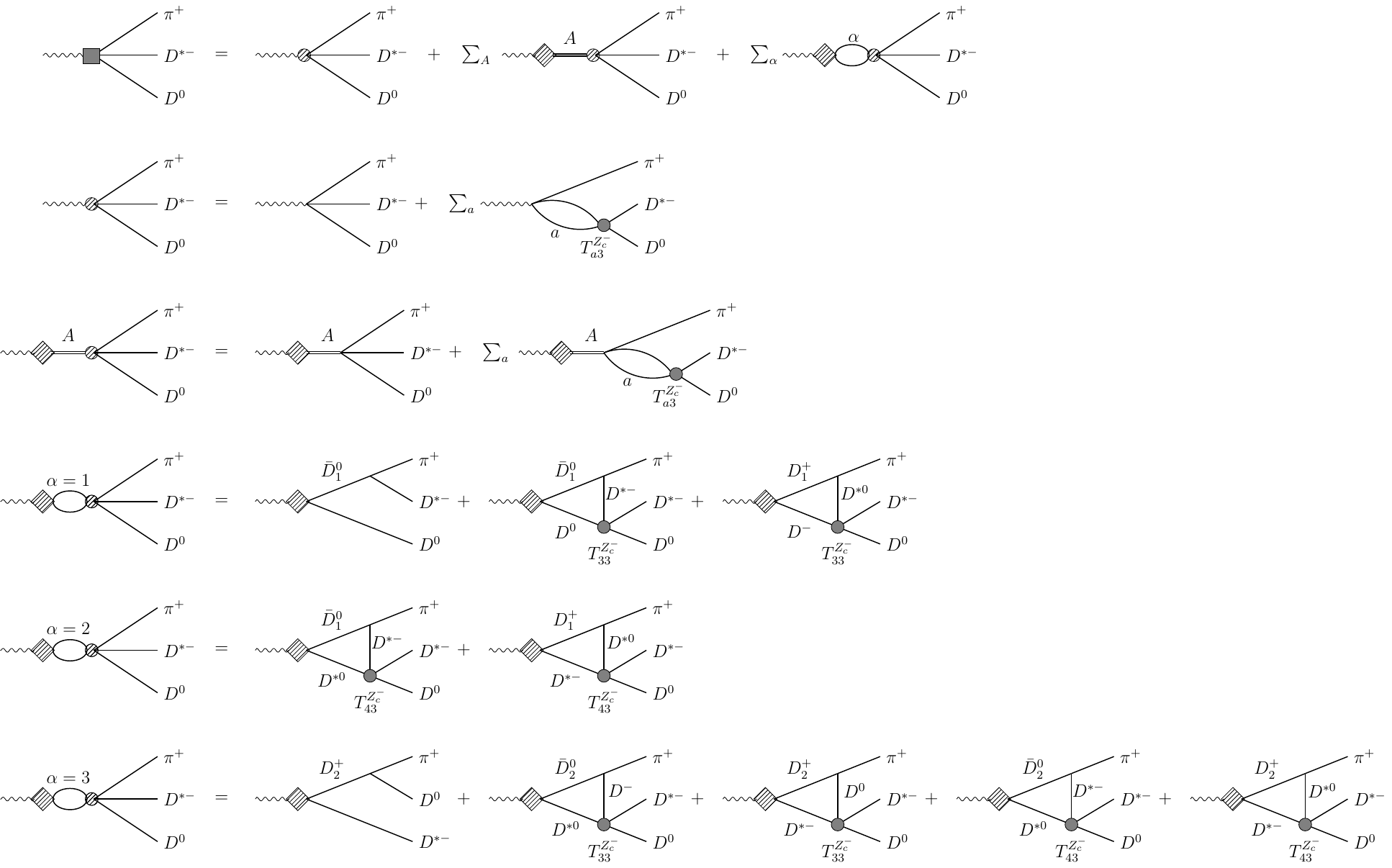}
 \caption{Diagrams for the production of \chIII from the virtual photon. }\label{fig:DDspi}
\end{figure}

The production of the $D^{*-}D^0\pi^+$ final state is illustrated in Figure~\ref{fig:DDspi}. The total tensor amplitude $\mathcal M_3^{ij}$ in Eq.~\eqref{Mas} admits various contributions as given in Eq.~\eqref{M1ab} which read (summations over repeated indices of the observation channels $b$ and $c$ as well as the three-dimensional Cartesian index $k$ are implied; no summation over $A$ is understood)
\begin{align}
 \mathcal M^{\gamma^*,ij}_3=&\,\frac1{\sqrt2}\left(\mathcal P^{\gamma^*}_{a=3}+\mathcal P^{\gamma^*}_bG^{Z_c}_{bc}(s)T^{Z_c^-}_{c3}(s)\right)\delta^{ij},\\
 \mathcal M^{A,ij}_3=&-\frac 1{\sqrt2}U_AG_{AA}\left(\mathcal P^A_{a=3}+\mathcal P^A_bG^{Z_c}_{bc}(s)T^{Z_c^-}_{c3}(s)\right)\delta^{ij},\\
 \mathcal M^{\alpha=1,ij}_3=&\,\frac{1}{2}c_1U_{\alpha=1}\mathcal M^{ij}_{TH1}(p_1)\left(-G_{D_1}(u)+T_{D_1DD^*}T_{33}^{Z_c^-}(s)\right),\\
 \mathcal M^{\alpha=2,ij}_3=&-\frac{1}{2\sqrt2}c_1U_{\alpha=2}T_{D_1D^*D^*}
 \left(\mathcal M^{ij}_{TH1}(p_1)-\delta^{ij}\mathcal M^{kk}_{TH1}(p_1)\right)T_{43}^{Z_c^-}(s),\\
 \mathcal M^{\alpha=3,ij}_3=&\,\frac {\sqrt3}{2\sqrt5}c_3U_{\alpha=3}\left(-G_{ D_2^*}(t)+T_{ D_2^*DD^*}T_{33}^{Z_c^-}(s)\right)
 \left[\mathcal M^{ij}_{TH3}(p_1)- \delta^{ij}\mathcal M^{kk}_{TH3}(p_1)/3\right]\notag\\
 &+\frac{\sqrt3}{4\sqrt5}c_2U_{\alpha=3}
 T_{ D_2^*D^*D^*}
 \left(\epsilon^{jkl}\mathcal M_{TH2}^{ikl}(p_1)-\epsilon^{kjl}\mathcal M_{TH2}^{kil}(p_1)\right)T_{43}^{Z_c^-}(s).
\end{align} 

Then the total amplitude can be written as
\begin{align}
\mathcal{M}^{ij}_3=M_{31}(s,t)\delta^{ij}+M_{32}(s,t)p_1^{i}p_1^{j},
\end{align}
with the coefficients
\begin{align}
M_{31}(s,t)=&\left(\mathcal P^{\gamma^*}_{a=3}+\mathcal P^{\gamma^*}_bG_{bc}(s)T^{Z_c^-}_{c3}(s)\right)-\sum_{A=1}^2U_AG_{AA}
\left(\mathcal P^A_{a=3}+\mathcal P^A_bG^{Z_c}_{bc}(s)T^{Z_c^-}_{c3}(s)\right)\notag\\
&+\frac{1}{2}c_1U_{\alpha=1}\left(h_Sp_1^0-h_Dp_1^2\right)
\left(G_{D_1}(u)-T_{D_1\bar{D} D^*}T^{Z_c^-}_{33}(s)\right)\notag\\
&+\frac{1}{\sqrt2}c_1U_{\alpha=2}T_{D_1\bar{D}^* D^*}T^{Z_c^-}_{43}(s)
\left(h_Sp_1^0+\frac12h_Dp_1^2\right)\notag\\
&+\frac{1}{2\sqrt{15}}c_3U_{\alpha=3}
\left(G_{ D_2^*}(t)-T_{{ D_2^*\bar{D}^*D}}T^{Z_c^-}_{33}(s)\right)h_Dp_1^2\notag\\
&-\frac{\sqrt3}{4\sqrt{5}}c_2U_{\alpha=3}T_{{ D_2^*\bar{D}^*D^*}}T^{Z_c^-}_{43}(s)h_Dp_1^2,\\
M_{32}(s,t)=&-\frac{3}{2}c_1U_{\alpha=1}
\left(G_{D_1}(u)-T_{D_1\bar{D} D^*}T^{Z_c^-}_{33}(s)\right)h_D+\frac{3}{2\sqrt 2}c_1U_{\alpha=2}T_{{D_1\bar{D}^*D^*}}T^{Z_c^-}_{43}(s)h_D\notag\\
&-\frac{\sqrt3}{2\sqrt5}c_3U_{\alpha=3}
\left(G_{ D_2^*}(t)-T^{Z_c^-}_{33}(s)T_{ D_2^*\bar{D}^*D}\right)h_D\notag\\
&+\frac{3\sqrt3}{4\sqrt5}c_2U_{\alpha=3}T_{ D_2^*\bar{D}^*D^*}T^{Z_c^-}_{43}(s)h_D.
\end{align}
Then the vector $\bar{M}_3$ and the kinematical matrix $Q_3$ in Eq.~\eqref{barMdef} read
\begin{equation}
\bar{M}_3=\Bigl(M_{31}(s,t),M_{32}(s,t)\Bigr)^{\rm T}
\end{equation}
and
\begin{equation}
Q_3=\frac13\left(\begin{array}{cc}
3&p_1^2\\
p_1^2&p_1^4
\end{array}\right),
\label{K3matr}
\end{equation}
respectively.

\subsection{Observation channel $D^{*-}D^{*0}\pi^+$ ($a=4$)}

\begin{figure}[t]
\includegraphics[width=\linewidth]{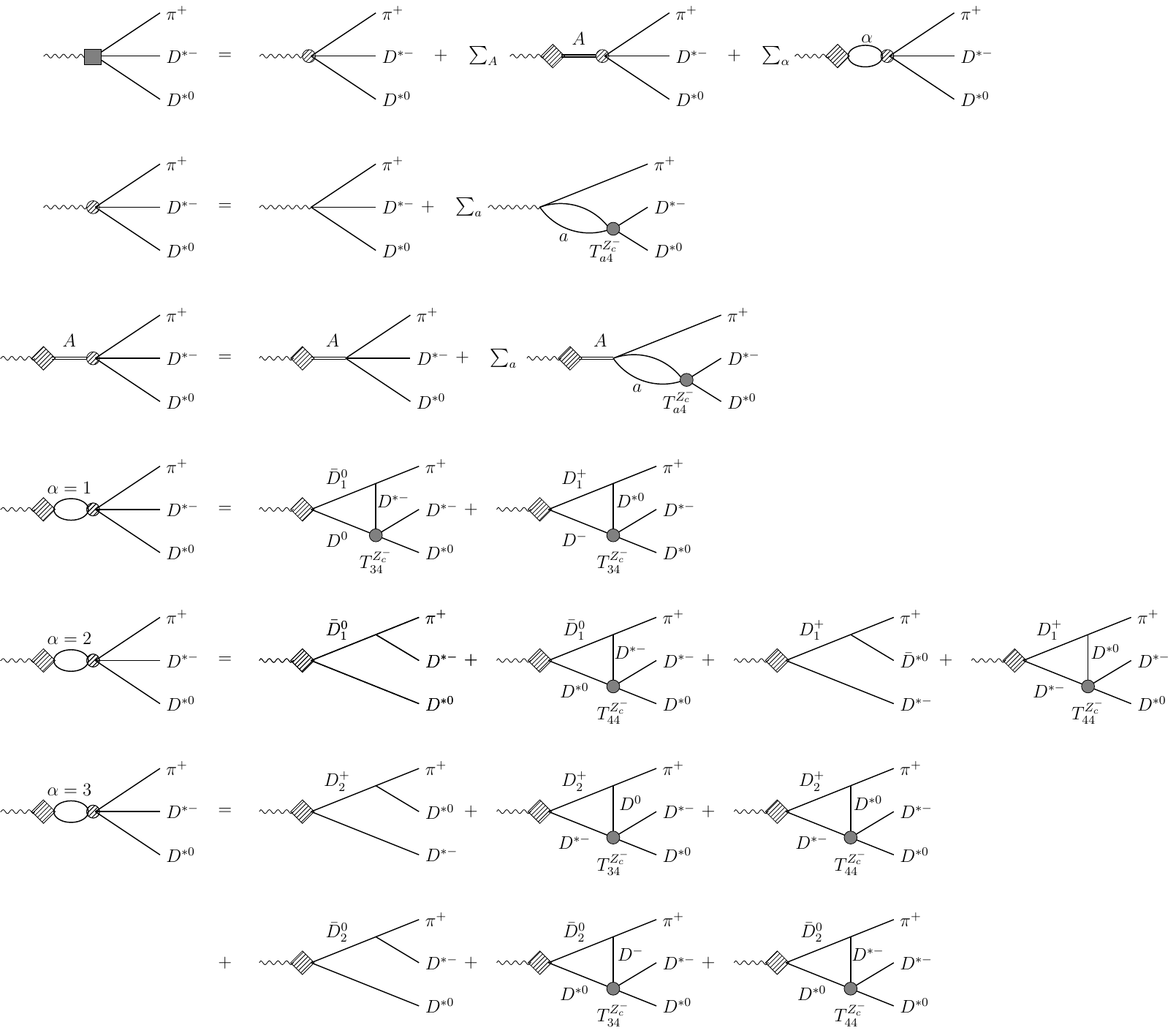}
\caption{Diagrams for the production of $D^{*-}D^{*0}\pi^+$ from the virtual photon. }\label{fig:DsDspi}
\end{figure}

The production of the $D^{*-}D^{*0}\pi^+$ final state is illustrated in Figure~\ref{fig:DsDspi}. The total tensor amplitude $\mathcal M_4^{ij}$ in Eq.~\eqref{Mas} admits various contributions as given in Eq.~\eqref{M1ab} which read (summations over repeated indices of the observation channels $b$ and $c$ as well as the three-dimensional Cartesian indices $k$ and $l$ are implied; no summation over $A$ is understood)
\begin{equation}
\begin{aligned}
\mathcal M^{\gamma^*,ij}_4=&\left(\mathcal P^{\gamma^*}_{a=4}+\mathcal P^{\gamma^*}_bG^{Z_c}_{bc}(s)T^{Z_c^-}_{c4}(s)\right)\delta^{ij},\\
\mathcal M^{A,ij}_4=&-U_AG_{AA}\left(\mathcal P^A_{a=4}+\mathcal P^A_bG_{bc}(s)T^{Z_c^-}_{c4}(s)\right)\delta^{ij},\\
\mathcal M^{\alpha=1,ij}_4=&\,\frac 1{\sqrt 2} c_1 U_{\alpha=1} T_{D_1DD^*}\mathcal M_{TH1}^{ij}(p_1)T_{34}^{Z_c^-}(s),\\
\mathcal M^{\alpha=2,ij}_4=&\,\frac{1}{2}c_1U_{\alpha=2}
\left[-(G_{D_1}(u)+G_{D_1}(t))
\left(\delta^{ij}\mathcal M_{TH1}^{ll}(p_1)-M_{TH1}^{ij}(p_1)\right)\right.\notag\\
&\quad\quad\left.
+T_{D_1D^*D^*}\left(\delta^{ij}\mathcal M_{TH1}^{ll}(p_1)-\mathcal M_{TH1}^{ij}(p_1)\right)T_{44}^{Z_c^-}(s)\right],\\
\mathcal M^{\alpha=3,ij}_4=&\,\frac {\sqrt3}{\sqrt{10}} c_3 U_{\alpha=3} T_{ D_2^*DD^*}
\left(\mathcal M_{TH3}^{ij}(p_1)-\delta^{ij}\mathcal M_{TH3}^{ll}(p_1)/3\right)T_{34}^{Z_c^-}(s)\notag\\
&+\frac{\sqrt3}{4\sqrt{10}}c_2 U_{\alpha=3}\, \epsilon^{jkl}
\left[-\left(\mathcal M_{TH2}^{ilk}(p_1)+\mathcal M_{TH2}^{lik}(p_1)\right)
\left(G_{ D_2^*}(t)+G_{ D_2^*}(u)\right)\right.\notag\\
&\quad\quad\left.
+ \left(\mathcal M_{TH2}^{ilk}(p_1)+\mathcal M_{TH2}^{lik}(p_1)
-\mathcal M_{TH2}^{ikl}(p_1)-\mathcal M_{TH2}^{kil}(p_1)\right)T_{ D_2^*DD^*}T_{34}^{Z_c^-}(s)\right].
\end{aligned}
\end{equation}
The total amplitude can be expressed as
\beq
\mathcal{M}^{ij}_4=M_{41}(s,t)\delta^{ij}+M_{42}(s,t)p_1^i p_1^j,
\label{eq:M4tensor}
\eeq
with the coefficients
\begin{align}
M_{41}(s,t)=&\left(\mathcal P^{\gamma^*}_{a=4}+\mathcal P^{\gamma^*}_bG^{Z_c}_{bc}(s)T^{Z_c^-}_{c4}(s)\right)-\sum_{A=1}^2U_AG_{AA}\left(\mathcal P^A_{a=4}+\mathcal P^A_bG_{bc}(s)T^{Z_c^-}_{c4}(s)\right)\notag\\
&+\frac{1}{\sqrt 2}c_1U_{\alpha=1}T_{D_1\bar{D} D^*}
T^{Z_c^-}_{34}(s)\left(h_S p_1^0-h_D p_1^2\right)\notag\\
&+c_1U_{\alpha=2}\left(T_{D_1\bar{D}^* D^*}T^{Z_c^-}_{44}(s)
- \frac12\left(G_{D_1}(u)+G_{D_1}(t)\right)\right)\left(h_S p_1^0+\tfrac{1}{2}h_D p_1^2\right)
\notag\\
&+\frac{\sqrt3}{2\sqrt{10}}c_2U_{\alpha=3}
\left(\frac12\left(G_{ D_2^*}(t)+G_{ D_2^*}(u)\right)-T_{ D_2^*\bar{D}^* D} T^{Z_c^-}_{34}(s)\right)h_D p_1^2\notag\\
&-\frac{1}{\sqrt{30}}c_3U_{\alpha=3}T_{ D_2^*\bar{D}^* D} T^{Z_c^-}_{34}(s)\,h_D p_1^2\\
M_{42}(s,t)=&\,\frac{3}{\sqrt 2}c_1U_{\alpha=1}T_{D_1\bar{D} D^*}T^{Z_c^-}_{34}(s)h_D\notag\\
&+\frac{3}{2}c_1U_{\alpha=2}
\left(\frac12\left(G_{D_1}(u)+G_{D_1}(t)\right)-T_{D_1\bar{D}^* D^*}T^{Z_c^-}_{44}(s)\right)h_D\notag\\
&-\frac{3\sqrt3}{2\sqrt{10}}c_2U_{\alpha=3}
\left(\frac12(G_{ D_2^*}(u)+G_{ D_2^*}(t))-T_{ D_2^*\bar{D}^* D^*}T^{Z_c^-}_{44}(s)\right)h_D\notag\\
&+\sqrt{\frac{3}{{10}}}c_3U_{\alpha=3}T_{ D_2^*\bar{D}^* D}T^{Z_c^-}_{34}(s)h_D
\end{align}
Then the vector $\bar{M}_4$ in Eq.~\eqref{barMdef} reads
\begin{align}
\bar M_4=\Bigl(M_{41}(s,t),M_{42}(s,t)\Bigr)^{\rm T}
\end{align}
while the kinematical matrix $Q_4=Q_3$, with the explicit form of $K_3$ given in Eq.~\eqref{K3matr}.

\subsection{Observation channels $J/\psi\eta$ and $\chi_{c0}\omega$ ($a=5,6$)}

\begin{figure}[t]
\centering
\includegraphics[width=\linewidth]{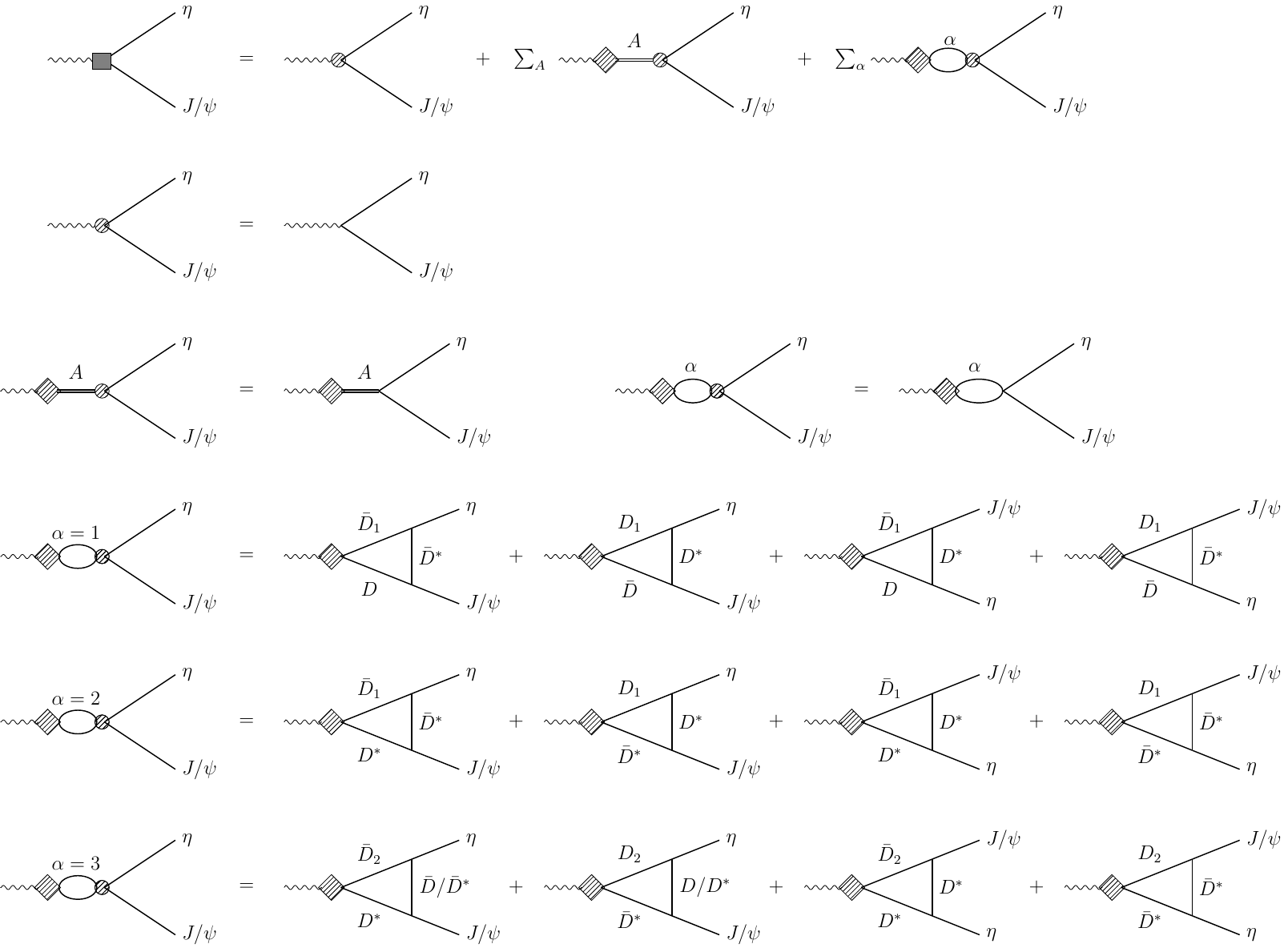}
\caption{Diagrams relevant for the production of $J/\psi\eta$. The meson-exchange mechanism for the transition of a two-body elastic channel to $J/\psi\eta$, as shown by the diagrams in the last rows, is represented by short-range contact terms, as illustrated by the last diagram in the third row. Diagrams for the production of $\chi_{c0}\omega$ look similar.}
\label{fig:VJpsieta}
\end{figure}

The production of the $J/\psi\eta$ final state is illustrated in Figure~\ref{fig:VJpsieta}, and it looks similar for the $\chi_{c0}\omega$ and $h_c\eta$ channels.
We employ the short-range contact terms in Eq.~\eqref{eq:Valpha} for the triangle-loop contributions to the transition from a two-body elastic channels to $J/\psi\eta$, $\chi_{c0}\omega$ and $h_c\eta$, respectively.
The individual contributions to the tensor amplitudes
$\mathcal M_{a=5,6}^{ij}$ (for $J/\psi\eta$, and $\chi_{c0}\omega$ channels, respectively), as defined in Eqs.~\eqref{Mas} and \eqref{M1ab}, read (no summation over $A$ and $\alpha$ is understood)
\begin{equation}
\mathcal M_a^{\gamma^*,ij}=\mathcal P^{\gamma^*}_{a}\,\delta^{ij},\quad
\mathcal M_a^{A,ij}=-U_AG_{AA}\mathcal P_{a}^A\,\delta^{ij},\quad
\mathcal M_a^{\alpha,ij}=-U_\alpha G_{\alpha\alpha} V_{\alpha a}\,\delta^{ij},
\end{equation}
so that
\begin{align}
\mathcal M^{ij}_a=M_{a1}\delta^{ij},
\end{align}
where
\begin{align}
M_{a1}=\mathcal P^{\gamma^*}_{a}-(U_A,U_\alpha)\left(\begin{array}{cc}
G_{AB}&\\
&G_{\alpha\beta}
\end{array}\right)\left(\begin{array}{c}
\mathcal P_{a}^B\\
V_{\beta a}
\end{array}\right).
\end{align}
In this way, one ends up with single-component coefficients in Eq.~\eqref{barMdef} that read
\begin{align}
\bar{M}_a=M_{a1},\quad Q_a=1.
\end{align}
Note that in the fit we use the reduced parameter 
$\bar{\mathcal P}^{\gamma^*/A}_{a=5}$ defined as 
$\mathcal P^{\gamma^*/A}_{a=5}
=
\bar{\mathcal P}^{\gamma^*/A}_{a=5}\, p_\eta$.

\newpage
\section{Predicted distributions in the channels $h_c\pi$, $D\bar D^*$, and $D^*\bar D^*$}
\label{app:predictions}

\begin{figure}[h]
\centering
\includegraphics[width=\linewidth]{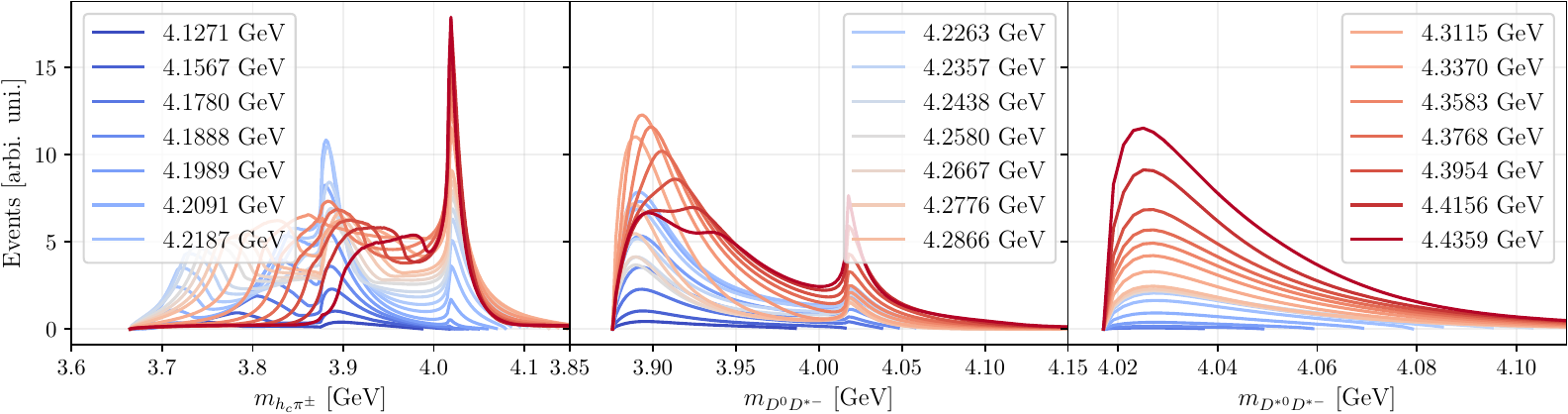}
\includegraphics[width=\linewidth]{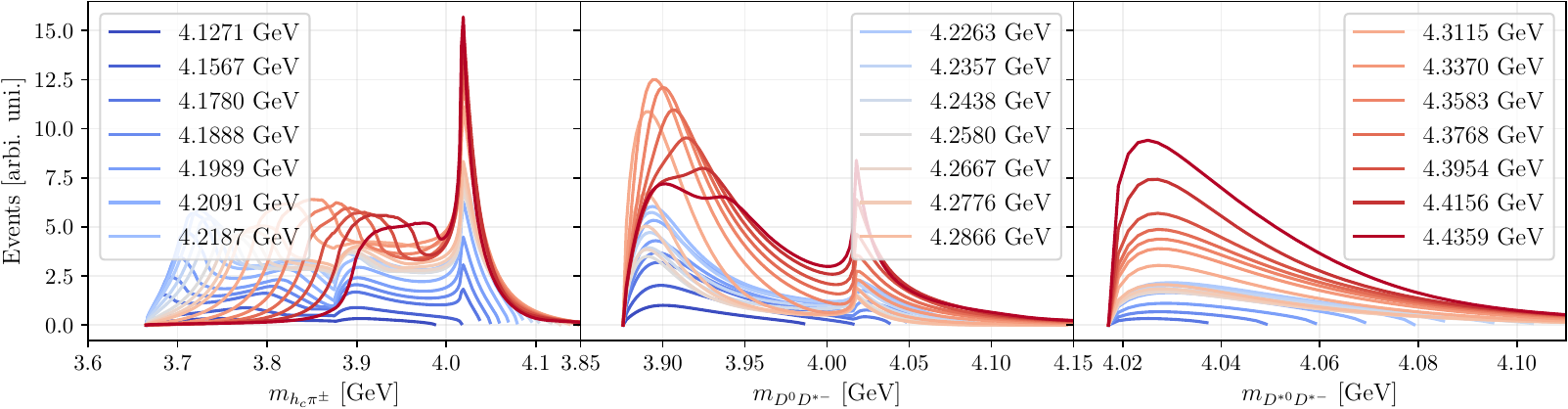}
\includegraphics[width=\linewidth]{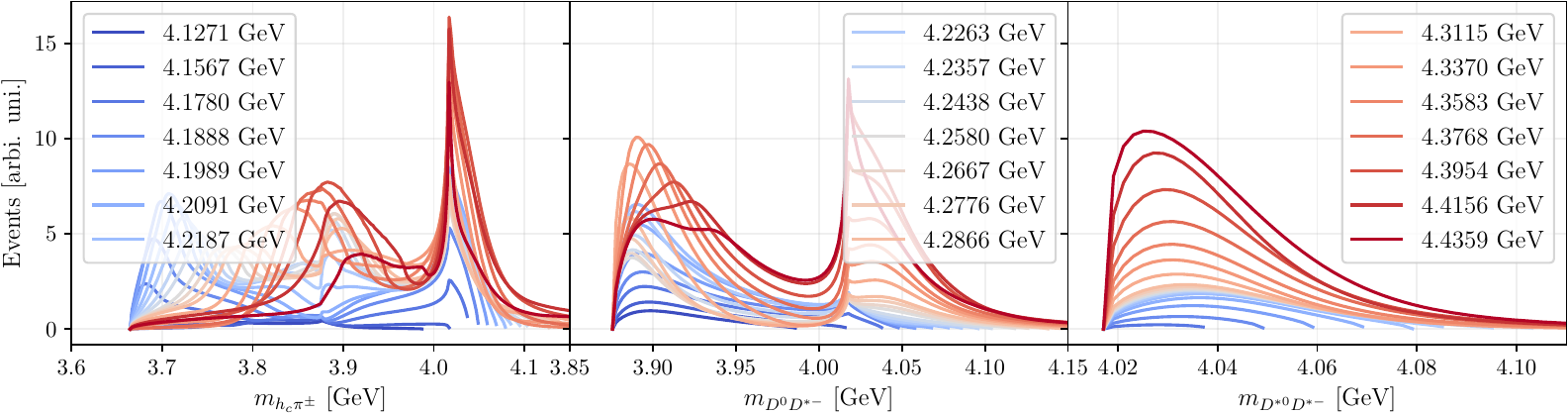}
\caption{Predicted invariant-mass distributions in the two-body subsystems $h_c\pi^\pm$, $D^0D^{*-}$, and $D^{*0}D^{*-}$ of the corresponding observation channels in Eq.~\eqref{observation1} obtained for c.m. energies performed in BESIII experiments, analogous to those shown in Figure~\ref{fig:jpsipi}. The upper, middle, and lower rows correspond to the models M0, M1, and M2, respectively.}
\label{fig:pre_all_E}
\end{figure}
\newpage
\section{Parameters of the best fits for models $\mathrm M0$, $\overline{\mathrm M0}$, $\mathrm M1$, and $\mathrm M2$}
\label{app:params}

\begin{table}[H]
\centering
\caption{The values and errors of the best fit parameters.}\label{tab:fitparams}
\vspace{0.2cm}
\resizebox{\linewidth}{!}{
\renewcommand{\arraystretch}{1.05}
\begin{tabular}{llcccc}
\hline\hline
\vphantom{\Large A}parameters & units & M0 & $\overline{\mbox M0}$ & M1 & M2 \\
\hline
$\mathcal P^{\gamma^*}_{a=1}$ & $1$ & $-0.414(62)$ & $-1.200(76)$ & $-0.329(97)$ & $0.037(107)$ \\
$\mathcal P^{\gamma^*}_{a=2}$ & $1$ & $-2.173(212)$ & $-0.914(170)$ & $-1.134(221)$ & $-0.148(421)$ \\
$\mathcal P^{\gamma^*}_{a=3}$ & $1$ & $1.052(209)$ & $-7.036(282)$ & $1.771(218)$ & $4.682(349)$ \\
$\mathcal P^{\gamma^*}_{a=4}$ & $1$ & $-3.44(18)$ & $-4.78(26)$ & $-2.07(38)$ & $2.13(88)$ \\
$\bar{\mathcal P}^{\gamma^*}_{a=5}$ & $10^{-1}$ & $-0.871(60)$ & $-1.141(44)$ & $-0.583(87)$ & $-1.493(134)$ \\
$\mathcal P^{\gamma^*}_{a=6}$ & $10^{-1}~\mathrm{GeV}$ & $1.489(199)$ & $2.042(200)$ & $1.292(264)$ & $3.984(668)$ \\
$\mathcal P^{\gamma^*}_{\alpha=1}$ & $\mathrm{GeV}$ & $2.26(7)\exp(1.63(4)\,i)$ & $3.65(8)\exp(1.23(3)\,i)$ & $2.97(11)\exp(1.96(7)\,i)$ & $2.36(13)\exp(1.66(6)\,i)$ \\
$\mathcal P^{\gamma^*}_{\alpha=2}$ & $\mathrm{GeV}$ & $3.24(8)\exp(-1.44(5)\,i)$ & $2.65(10)\exp(-1.45(5)\,i)$ & $3.22(13)\exp(-1.18(5)\,i)$ & $3.24(17)\exp(-1.54(6)\,i)$ \\
$\mathcal P^{\gamma^*}_{\alpha=3}$ & $\mathrm{GeV}$ & $5.83(15)\exp(1.52(4)\,i)$ & $4.95(13)\exp(1.11(4)\,i)$ & $5.95(17)\exp(1.82(6)\,i)$ & $4.62(21)\exp(1.71(10)\,i)$ \\
$F$ & $\mathrm{GeV}^{-2}$ & $1.105(28)$ & $0.803(33)$ & $1.111(42)$ & $1.034(38)$ \\
$f$ & $10^{-1}$ & $-1.954(392)$ & $-1.741(668)$ & $-2.476(658)$ & $-1.168(819)$ \\
$G$ & $\mathrm{GeV}^{-2}$ & $4.57(27)$ & $0.32(3)$ & $1.54(53)$ & $1.78(49)$ \\
$g$ & $1$ & $1.32(10)$ & $19.93(174)$ & $4.67(162)$ & $5.56(151)$ \\
$C_{1Z}$ & $\mathrm{GeV}^{-2}$ & $-5.14(4)+0.94(3)\,i$ & $-1.20(3)+0.50(3)\,i$ & $-5.06(5)+0.94(4)\,i$ & $-4.89(7)+0.47(5)\,i$ \\
$c_Z$ & $1$ & $0.972(11)$ & $3.387(81)$ & $0.926(18)$ & $0.945(33)$ \\
$C_{34}$ & $\mathrm{GeV}^{-2}$ & $0.737(35)$ & $2.336(54)$ & $0.844(46)$ & $1.541(82)$ \\
$B$ & $\mathrm{GeV}^{-3}$ & $5.390(786)$ & $0.605(85)$ & $2.268(1328)$ & $-8.358(1568)$ \\
$b$ & $1$ & $1.97(38)$ & $36.29(514)$ & $3.56(205)$ & $5.91(127)$ \\
$\alpha_0$ & $10^{-1}~{\mathrm{pb}^{1/2}}$ & $3.75(27)$ & $4.53(40)$ & $5.09(56)$ & $6.40(93)$ \\
$\alpha_1$ & $10~\mathrm{pb}^{1/2}\mathrm{GeV}^{-1}$ & $2.74(9)$ & $2.33(9)$ & $2.56(18)$ & $3.82(30)$ \\
$\bar{\mathcal{L}}_2(4.23)$ & $10^9\,\mathrm{pb}^{-1}$ & $2.26(20)$ & $2.30(18)$ & $2.42(19)$ & $2.44(20)$ \\
$\bar{\mathcal{L}}_2(4.36)$ & $10^9\,\mathrm{pb}^{-1}$ & $4.63(64)$ & $4.70(60)$ & $4.09(57)$ & $3.84(60)$ \\
$\bar{\mathcal{L}}_3(4.23)$ & $10^9\,\mathrm{pb}^{-1}$ & $1.35(6)$ & $1.33(6)$ & $1.46(7)$ & $1.54(9)$ \\
$\bar{\mathcal{L}}_3(4.26)$ & $10^9\,\mathrm{pb}^{-1}$ & $0.941(69)$ & $0.822(59)$ & $0.957(77)$ & $0.981(73)$ \\
$\bar{\mathcal{L}}_4(4.26)$ & $10^9\,\mathrm{pb}^{-1}$ & $1.32(9)$ & $1.45(10)$ & $1.56(11)$ & $1.31(9)$ \\
$F_{01}^c$ & $1$ & $171(11)+168(10)\,i$ & $129(6)+210(8)\,i$ & $177(12)+200(15)\,i$ & $106(26)+248(17)\,i$ \\
$F_{01}^d$ & $1$ & $-207(6)-330(7)\,i$ & $-293(5)-314(5)\,i$ & $-248(12)-285(12)\,i$ & $-220(14)-226(11)\,i$ \\
$\delta F_{02}$ & $1$ & $-480(3)+137(4)\,i$ & $-462(3)+157(3)\,i$ & $-495(6)+170(8)\,i$ & $476(5)-141(7)\,i$ \\
$F_{J/\psi \eta}$ & $1$ & $36.9(12)$ & $17.4(4)$ & $33.1(20)$ & $31.8(18)$ \\
$F_{\chi_{c0}\omega}$ & $\mathrm{GeV}$ & $-68.6(40)$ & $-32.5(17)$ & $-22.9(108)$ & $-74.9(82)$ \\
$\mathcal P^{A=1}_{a=1}$ & $1$ & $-$ & $-$ & $-1.88(30)$ & $-2.24(50)$ \\
$\mathcal P^{A=1}_{a=2}$ & $1$ & $-$ & $-$ & $-2.08(117)$ & $-3.79(163)$ \\
$\mathcal P^{A=1}_{a=3}$ & $1$ & $-$ & $-$ & $4.28(119)$ & $7.93(139)$ \\
$\mathcal P^{A=1}_{a=4}$ & $1$ & $-$ & $-$ & $17.6(21)$ & $32.7(38)$ \\
$\bar{\mathcal P}^{A=1}_{a=5}$ & $10^{-2}$ & $-$ & $-$ & $1.51(172)$ & $-3.03(191)$ \\
$\mathcal P^{A=1}_{a=6}$ & $\mathrm{GeV}$ & $-$ & $-$ & $1.01(21)$ & $0.225(234)$ \\
$\mathcal P^{\gamma^*}_{A=1}$ & $\mathrm{GeV}^{2}$ & $-$ & $-$ & $0.12\exp(0.23(13)\,i)$ & $0.12\exp(2.05(13)\,i)$ \\
$V_{A=1,\alpha=1}$ & $\mathrm{GeV}$ & $-$ & $-$ & $3.12(53)$ & $-2.08(53)$ \\
$V_{A=1,\alpha=2}$ & $\mathrm{GeV}$ & $-$ & $-$ & $-1.24(158)$ & $5.29(180)$ \\
$V_{A=1,\alpha=3}$ & $\mathrm{GeV}$ & $-$ & $-$ & $-8.45(328)$ & $22.21(194)$ \\
$m_{0,A=1}$ & $\mathrm{GeV}$ & $-$ & $-$ & $4.197(3)$ & $4.204(3)$ \\
$\Gamma_{0,A=1}$ & $\mathrm{GeV}$ & $-$ & $-$ & $0.085(10)$ & $0.053(7)$ \\
$\mathcal P^{A=2}_{a=1}$ & $1$ & $-$ & $-$ & $-$ & $-13.31(181)$ \\
$\mathcal P^{A=2}_{a=2}$ & $1$ & $-$ & $-$ & $-$ & $42.90(780)$ \\
$\mathcal P^{A=2}_{a=3}$ & $1$ & $-$ & $-$ & $-$ & $-4.72(517)$ \\
$\mathcal P^{A=2}_{a=4}$ & $1$ & $-$ & $-$ & $-$ & $59.6(93)$ \\
$\bar{\mathcal P}^{A=2}_{a=5}$ & $1$ & $-$ & $-$ & $-$ & $0.781(198)$ \\
$\mathcal P^{A=2}_{a=6}$ & $\mathrm{GeV}$ & $-$ & $-$ & $-$ & $-3.63(102)$ \\
$\mathcal P^{\gamma^*}_{A=2}$ & $\mathrm{GeV}^{2}$ & $-$ & $-$ & $-$ & $0.11\exp(-0.11(6)\,i)$ \\
$V_{A=2,\alpha=1}$ & $\mathrm{GeV}$ & $-$ & $-$ & $-$ & $-2.14(40)$ \\
$V_{A=2,\alpha=2}$ & $\mathrm{GeV}$ & $-$ & $-$ & $-$ & $-0.94(94)$ \\
$V_{A=2,\alpha=3}$ & $\mathrm{GeV}$ & $-$ & $-$ & $-$ & $-12.1(13)$ \\
$m_{0,A=2}$ & $\mathrm{GeV}$ & $-$ & $-$ & $-$ & $4.406(5)$ \\
$\Gamma_{0,A=2}$ & $\mathrm{GeV}$ & $-$ & $-$ & $-$ & $0.080(11)$ \\
\hline\hline
\end{tabular}}
\end{table}

\end{document}